%% file: main.tex
\documentclass[a4paper,11pt]{article}
\usepackage{jheppub} 
\usepackage{comment}
\usepackage{lineno}
\usepackage[dvipsnames]{xcolor}
\usepackage[utf8]{inputenc}  
\usepackage{pdflscape} 
\usepackage{fix-cm}  
\usepackage{tikz}    
\usepackage{float}
\usepackage{longtable}
\usepackage{array}
\usetikzlibrary{arrows}
\usepackage{tikz-cd} 
\usepackage{tkz-euclide}
\usepackage{circuitikz}
\usepgfmodule{shapes}
\usetikzlibrary{decorations.pathmorphing}
\usetikzlibrary{backgrounds, arrows,calc,shapes,decorations.pathreplacing, automata,positioning}
\usepackage{cancel} 
\usepackage{graphicx}
\usepackage{mathdots}
\usepackage{slashed}
\usetikzlibrary{decorations.markings,backgrounds}
\usetikzlibrary{matrix}
\usepackage{ytableau}
\usetikzlibrary{shapes.misc}
\usepackage[most]{tcolorbox}
\usepackage{hyperref}
\usepackage{amsmath} 
\usepackage{amssymb}
\usepackage{graphicx}
\usepackage{caption}
\usepackage{subcaption}
\usepackage{mathrsfs}
\usepackage{mathtools}
\usepackage{multirow}

\graphicspath{ {images/} }

\input{custom_commands}

\author[a]{Sergio Benvenuti}
\author[a,b]{Vittorio Cagioni}
\author[a,b]{Simone Rota}
\author[a,b]{Anant Shri}

\affiliation[a]{INFN, Sezione di Trieste, Via Valerio 2, I-34127 Trieste, Italy}
\affiliation[b]{SISSA, Via Bonomea 265, I-34136 Trieste, Italy}

\emailAdd{benve79@gmail.com}  
\emailAdd{vcagioni@sissa.it}  
\emailAdd{srota@sissa.it}  
\emailAdd{ashri@sissa.it}

\title{\boldmath Universal Planar Abelian Duals for 3d $\mathcal{N}=2$ Symplectic CS-SQCD} 

\abstract{
We propose a new class of infrared dualities relating three-dimensional $\mathcal N=2$ $USp(2N)$ Chern--Simons SQCD to planar Abelian quiver gauge theories. These dual descriptions are constructed via real mass deformations of established $\mathcal N=4$ mirror dualities between $\mathcal N=4$ $USp(2N)$ SQCD and unitary $D$-type quiver gauge theories. The resulting $\mathcal N=2$ dual pairs exhibit the characteristic exchange of topological and flavor symmetries. We provide nontrivial evidence for these dualities by matching $\mathbf S^3_b$ partition functions, superconformal indices, and gauge-invariant operator spectra. Furthermore, we systematically incorporate additional real mass deformations on both sides of the duality, allowing us to extend the construction to $\mathcal N=2$ symplectic SQCD with generic ranks, flavors, and Chern--Simons levels.
}

\begin{document}
\maketitle
\flushbottom


\section{Introduction and Summary}\label{sec: introduction}
Recent work has established that a broad class of three-dimensional $\mathcal N=2$ Chern--Simons--matter theories admit Abelian dual descriptions at their infrared fixed points \cite{Benvenuti:2024seb,Benvenuti:2025a}. These dualities are characterized by an \emph{exchange of topological and flavor symmetries} and can be viewed as the $\mathcal N=2$ avatar of mirror symmetry \cite{Hanany:1996ie,Intriligator_1996}.

A salient feature of these Abelian duals is that their gauge and matter content is encoded in a \textit{quiver}: an oriented diagram in which nodes correspond to symmetry groups and arrows encode matter fields. Remarkably, the quivers that arise in this framework are \emph{planar}, in the sense that they can be embedded in the plane such that each face corresponds to an interaction term in the action. Planarity plays a structural role in the construction, organizing both the matter content and the allowed superpotential couplings in a highly constrained manner. This notion closely parallels the planar quivers familiar from four-dimensional gauge theories with minimal supersymmetry, particularly in the context of the $\text{AdS}_5/\text{CFT}_4$ correspondence \cite{Hanany:2005ve,Franco:2005rj,Benvenuti:2005cz,Benvenuti:2005ja,Franco:2005sm,Butti:2005sw}.

\vspace{.5cm}
The paradigm of planar Abelianization is especially compelling for several reasons:
\begin{itemize}
    \item The resulting $\mathcal N=2$ mirror duals are Abelian \emph{and} admit a Lagrangian description — two highly nontrivial and \textit{a priori} independent properties.
    \item The construction admits straightforward generalizations to linear $A_N$ quivers, theories with special unitary gauge groups, circular quivers, and models with rich matter content, including adjoint fields as well as the presence of both fundamental and anti-fundamental representations.
    \item The framework enables systematic deformations, yielding theories with more general Chern--Simons levels and flavor content \cite{Benvenuti:2026a}. For special choices of parameters, the resulting theories display a rich infrared phenomenology, such as $s$-confinement, chiral symmetry breaking, and the emergence of non-Abelian topological quantum field theories (TQFTs).
    \item The paradigm further admits extensions beyond supersymmetry, encompassing non-supersymmetric dualities \cite{Benvenuti:2025qnq}.
\end{itemize}
The constructions of \cite{Benvenuti:2024seb,Benvenuti:2025a} focused on theories with unitary gauge groups. In this work, we extend the paradigm of planar Abelianization to theories with symplectic gauge symmetries. Three-dimensional $\mathcal N=4$ symplectic gauge theories are distinguished by the existence of two distinct mirror descriptions. These mirrors typically take the form of either unitary $D$-type quiver gauge theories \cite{Porrati:1996xi,Hori:1997zj,Kapustin:1998fa,Giveon:1998sr,Hanany:1999sj} or quivers involving orthosymplectic gauge groups \cite{Feng:2000eq}, which can be engineered via orientifolding Hanany--Witten brane setups.

While both classes of mirrors capture the same infrared physics \cite{ Nawata:2021nse,Nawata:2023rdx}, incorporating orthosymplectic quivers within the paradigm of planar Abelianization presents a significant challenge. In such descriptions, not all properties of the infrared SCFT are manifest in the ultraviolet Lagrangian, notably hidden Fayet--Iliopoulos (FI) parameters and the associated symmetry enhancement \cite{Kapustin:1998fa,Cremonesi_2014,Assel:2018exy}. By contrast, unitary $D$-type quivers provide a fully Lagrangian and transparent framework in which these issues do not arise.

Accordingly, in this paper we focus on planar Abelian duals of $USp$ gauge theories, obtained from $\mathcal N=4$ $USp$ SQCD and their unitary $D$-type quiver mirrors. This setting allows us to construct explicit $\mathcal N=2$ Abelian mirror duals for symplectic gauge theories and to extend the paradigm of planar Abelianization beyond the unitary case in a controlled and systematic manner.\newpage

We summarize the central result of this paper - three-dimensional $\mathcal{N}=2$ $USp(2N)_k$ SQCD theories with $F$ fundamental multiplets can be organized in a two-dimensional parameter space spanned by $(k,F)$ for fixed $N$. Each point in the plane corresponds to a distinct theory. We display the $(k,F)$-plane below, and defer a detailed discussion of the same to Section \ref{sec: mass flow}.
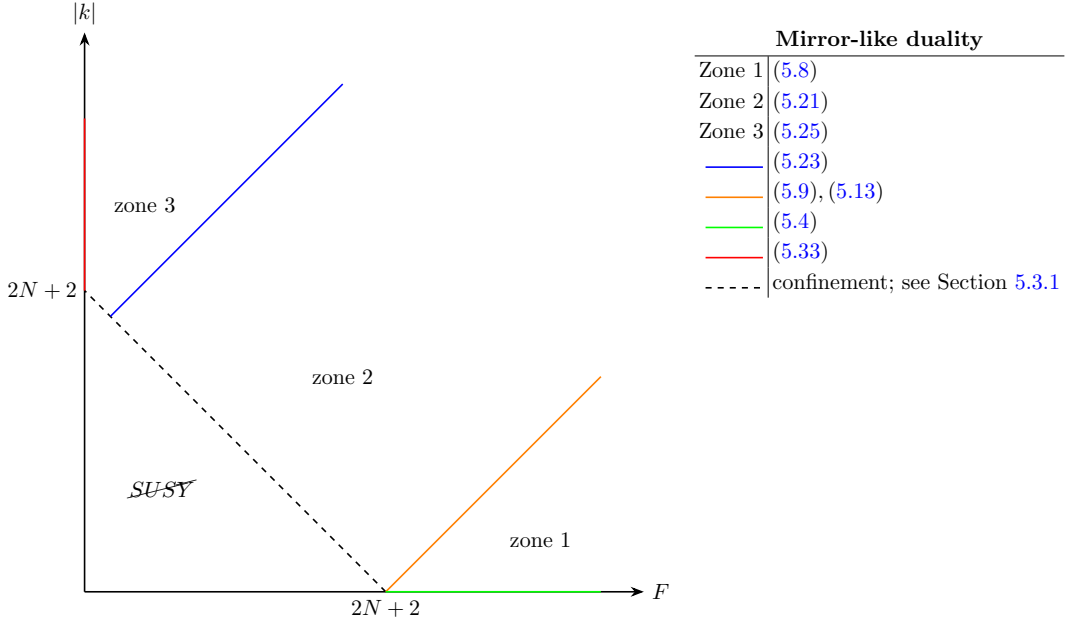
\begin{figure}[H]
    \centering
\resizebox{.95\hsize}{!}{ 

\begin{tikzpicture}[scale=1.5,>=Stealth]

\draw[->,thick] (0,0) -- (6.5,0) node[right] {$F$};
\draw[->,thick] (0,0) -- (0,6.5) node[above] {$|k|$};

\node[left] at (0,3.5) {$2N+2$};
\node[below] at (3.5,0) {$2N+2$};


\draw[dashed,thick] (3.5,0) -- (0,3.5);

\draw[thick,blue] (.3,3.2) -- (3,5.9);  
\draw[thick,orange] (3.5,0) -- (6,2.5); 
\draw[thick,green] (3.5,0) -- (6,0);
\draw[thick,red] (0,3.5) -- (0,5.5);


\node at (5.3,0.6) {zone $1$};
\node at (3,2.5) { zone $2$};
\node at (.7,4.5) { zone $3$};

\node[rotate=0] at (0.9,1.2) { $\cancel{SUSY}$};

\begin{scope}[shift={(7,5)}]
    \node[anchor=west] at (0,0) {
    $
    \begin{array}{r|l}
     
        \multicolumn{2}{c}{\textbf{Mirror-like duality}}\\
        \hline
       \text{Zone 1}  & \eqref{eq: 3d_USp_gen_k}  \\
       \text{Zone 2}  & \eqref{eq: Planar_N=4D_kl2NNegM_Fin}  \\
       \text{Zone 3}  & \eqref{eq: Planar_N=4D_k_BF_Col_NegM_Fin}   \\
      
       \tikz{\path[draw,thick,blue](0,0)--(1,0);} & \eqref{eq: Planar_N=4D_k_2N+1NegM_Fin} \\
       \tikz{\path[draw,thick,orange](0,0)--(1,0);} & \eqref{eq: 3d_USp_gen_k_N=4}, \eqref{eq: 3d_USp_N=4_D} \\
       \tikz{\path[draw,thick,green](0,0)--(1,0);} & \eqref{eq: 3d_USp_k=0} \\
       \tikz{\path[draw,thick,red](0,0)--(1,0);} & \eqref{eq: Planar_N=4D_TQFT_BF_Col_NegM_Fin} \\
       \tikz{\path[draw,thick,dashed](0,0)--(1,0);} & \text{confinement; see Section \ref{sec: confinement}} 
    \end{array}
    $
    };

\end{scope}

\end{tikzpicture}
}
    \caption{The landscape of $USp(2N)_k$ SQCD with $F$ fundamental fields and the corresponding planar Abelian duals organized in the $(F,\,|k|)$-plane for fixed $N$. Each point in the plane with integer $F$ and $k$ corresponds to a different SQCD theory. The diagram is partitioned into various zones, within which the planar Abelian duals take qualitatively distinct forms. 
    }
\end{figure}

\subsection*{Scope and Future Directions}
In this work, we have restricted attention to $USp(2N)$ CS-SQCD$_3$ with fundamental
matter and their real mass deformations. Within this class, we were able to track
how real mass deformations shift the Chern--Simons levels and follow the induced
RG flows in both the electric and planar Abelian mirror descriptions.
Understanding these RG flows also allows us to track analogous flows in
$U(2N)$ CS-SQCD$_3$ with fundamental matter and a rank-2 antisymmetric tensor and
their mirror duals, since these two families of theories share many qualitative
features. 

A natural extension of this framework is to generalize the planar Abelianization
prescription to $SO(N)$ CS--SQCD$_3$. For $\mathcal N=4$ theories, these gauge
groups can be realized as $\mathbb Z_2$ quotients of a $U(2N)$ gauge symmetry
\cite{Feng:2000eq,Gaiotto_2009}, and we anticipate that a similar mechanism
operates in the $\mathcal N=2$ setting considered here. Another promising
direction is the investigation of Aharony duality \cite{Benini_2011a} in the
\emph{symplectic} setting. In the \emph{unitary} case,
\cite{Benvenuti:2026a} showed that Aharony duality can be realized as a
consequence of mirror duality, and it would be interesting to uncover how an
analogous structure manifests in $USp(2N)$ CS-SQCD$_3$. 

A further direction is to explore generalizations of planar Abelianization to theories with more general tensor matter content. 

Finally, we anticipate that combining the results presented in this paper with the strategies employed in \cite{Benvenuti:2025qnq} would enable us to extend the scope of planar Abelianization to non-supersymmetric $USp(2N)_k$ CS-QCD$_3$ with generic fermionic and bosonic flavors.

\subsection*{Organization of the Paper}
\begin{itemize}
    \item We conclude Section \ref{sec: introduction} with a brief recap of planar Abelianization for unitary SQCD, which serves as the conceptual and technical template for all subsequent generalizations.
    \item In Section \ref{sec: antisymmetric}, we extend the paradigm of planar Abelianization to unitary SQCD with a rank-two antisymmetric tensor. This case plays a pivotal role, as it interpolates between unitary SQCD without tensor matter and symplectic SQCD, and should be regarded as an ancestor for both families. Furthermore, this is an interesting duality in its own right since it generalizes the notion of planar Abelanization to more general tensor matter. 
    \item Section \ref{sec: symplectic} is devoted to the study of real mass deformations of $\mathcal N=4$ symplectic SQCD and its unitary $D$-type mirror dual. This analysis allows us to construct planar Abelian duals for $\mathcal N=2$ symplectic SQCD
    \item In Section \ref{sec: d quiver}, we construct planar mirror duals of $D_F$ quiver gauge theories by considering the planar limit of $\NN=2$ symplectic SQCD. 
    \item Finally, Section \ref{sec: mass flow} focuses on the mapping of real mass deformations across mirror descriptions, where a representative example is analyzed in detail. 
    \item We summarize our conventions for quiver diagrams in Appendix \ref{app: notation}, and discuss some properties of monopoles in Appendix \ref{app: monopole}. We discuss a possible extension of the planar Abelianization paradigm to orthogonal gauge groups in Appendix \ref{app: planar bd}.

\end{itemize}

\newpage


\subsection*{Lightning Recap: General Yoga of Planar Abelianization} \label{sec: planar_ab}

We review the results of \cite{Benvenuti:2024seb, Benvenuti:2025a} in this section regarding the chiral-planar mirror duality of unitary CS-SQCD. 

Starting from the $\mathcal{N}=4$ mirror duality that relates an $\mathcal{N}=4$ $U(N)$ SQCD theory with $F \geq 2N$ flavors to a linear $A$-type quiver gauge theory with $F-1$ gauge nodes \cite{Intriligator_1996, Hanany:1996ie}, one can introduce a suitable real mass deformation on the electric side which breaks supersymmetry from $\mathcal{N}=4$ to $\mathcal{N}=2$. In the vacuum where the real scalar of the vector multiplet has no expectation value, this deformation leaves the $F$ fundamental chiral fields massless, while the adjoint and the $F$ antifundamental chiral fields acquire masses and are integrated out. As a result, a non-zero Chern-Simons (CS) level is generated, and one obtains the \textit{electric} $\mathcal{N}=2$ $U(N)$ CS-SQCD theory with $[F,0]$ chiral fields. We remind the reader that the global symmetries of this theory are $G_F = U(1)_{Top}\,\times SU(F)$, and we will now elaborate on how these are mapped to the planar dual theory. 

\begin{figure}[H]
     \centering
     \includegraphics[width=.9\linewidth]{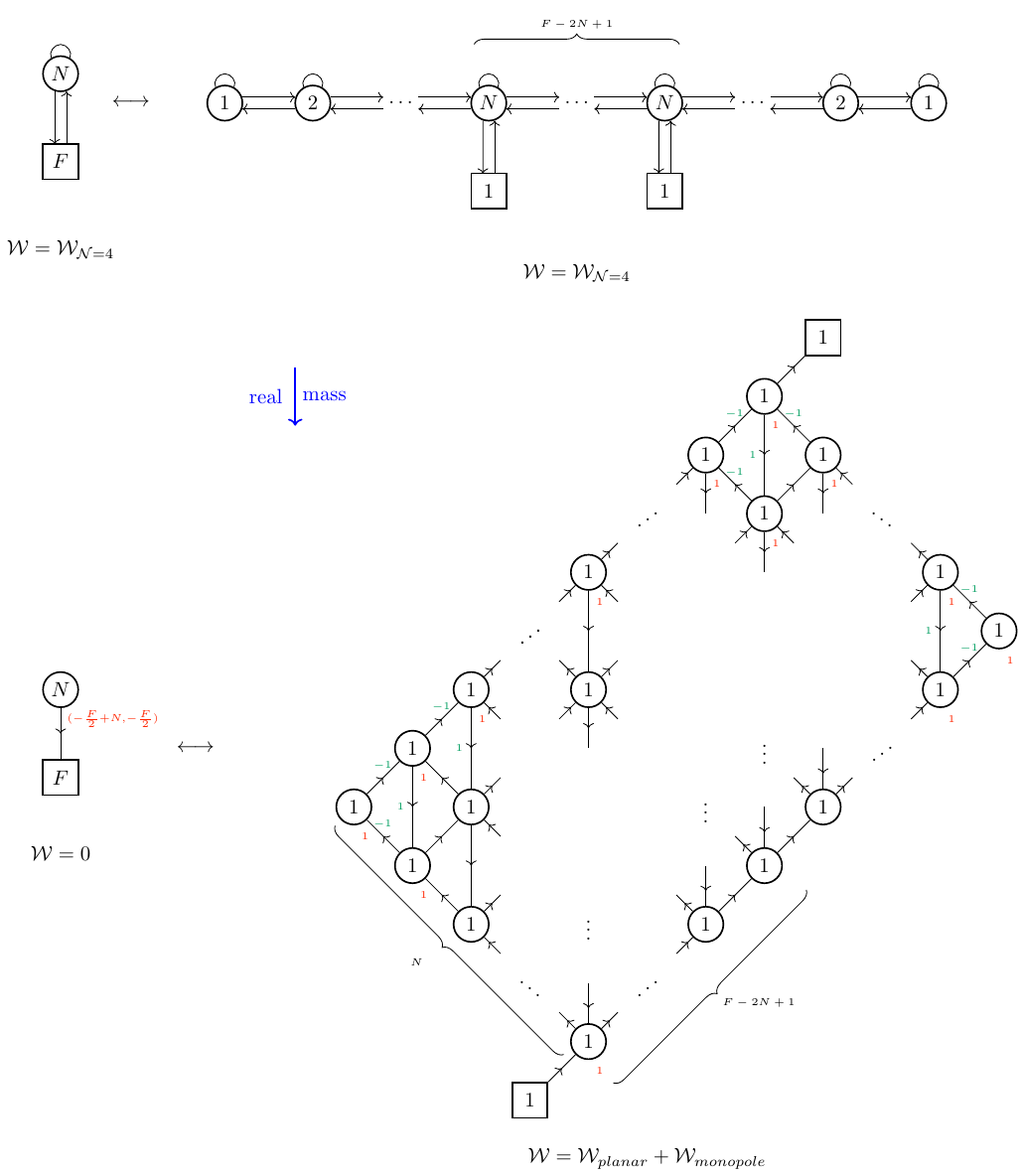}
     \caption{Schematic depiction of the $\mathcal{N}=2$ mirror-like duality (bottom row) between $U(N)$ CS-SQCD$_3$ with $F$ fundamental fields, and a planar Abelian quiver gauge theory obtained from an $\mathcal{N}=4$ mirror pair (top row). All quivers are drawn in $\mathcal{N}=2$ notation,
     with red and green labels indicating CS and mixed CS levels, respectively.
     FI terms have been suppressed for brevity.
     }
     \label{fig: 3d_N=4_N=2_Unit_Web}
 \end{figure}

The mirror description of this deformation is more subtle, due to the rich structure of the Coulomb branch in the dual theory. Recent works \cite{Benvenuti:2024seb, Benvenuti:2025a} have streamlined this analysis by examining the effect of the mass deformation directly on the $\mathbf{S}^3_b$ partition function \cite{vandebult, Benini_2011a} to arrive at an algorithmic prescription. On the mirror side, the vacuum corresponding to the electric theory is characterized by the Higgsing of each $U(k)$ gauge node to its maximal torus $U(1)^k$, giving rise to a column of $k$ abelian gauge nodes. This picture is supported by several complementary checks: analyses of the F- and D-term equations \cite{Intriligator_2013}, the asymptotic behavior of the $\mathbf{S}^3_b$ partition function under large mass deformations, which produces highly oscillatory phases that must match across the duality \cite{Aharony:2013dha}, and computations of the superconformal index \cite{Imamura:2011su, Kapustin:2011jm}. The resulting duality web is schematically depicted in Figure \ref{fig: 3d_N=4_N=2_Unit_Web}. 

\newpage

 We briefly highlight some salient features of the planar mirror theory:

\begin{itemize}
    \item Each triangular loop in the planar mirror corresponds to a cubic superpotential, with a relative sign depending on its orientation (clockwise or counterclockwise). These interactions descend from the cubic $\mathcal{N}=4$ superpotential and survive the real mass flow.  

    \item The chiral ring of the planar theory is generated by a mesonic operator built by contracting the $F$ diagonal bifundamentals connecting the two $U(1)$ flavor nodes. Although many paths can generate such an operator, F-term relations from the planar superpotential $\mathcal{W}_{planar}$ identify all of them. This meson maps to the dressed monopole operator of the electric theory, thereby identifying the mesonic $U(1)_{mes}$ symmetry of the mirror with the topological $U(1)_T$ symmetry of the electric theory.  

    \item Monopole superpotentials ($\mathcal{W}_{monopole}$) arise due to the Higgsing $U(k) \to U(1)^k$, following the Polyakov mechanism \cite{Polyakov1977}. Each vertical arrow in the planar quiver corresponds to a linear superpotential term coupling monopole operators with Goddard-Nuyts-Olive (GNO) flux \cite{Goddard:1976qe} $\pm1$ under adjacent nodes (from top to bottom).  

    These monopole terms break the topological symmetry to $U(1)^{F-1}$, which is expected to enhance in the deep IR to $SU(F)$—in agreement with the flavor symmetry of the electric theory. The FI parameters of the $U(1)$ gauge nodes are consistent with this symmetry enhancement. Specifically, the $\alpha^{\text{th}}$ gauge node in the $I^{\text{th}}$ column ($\alpha=1,\ldots, |G(I)|$, where $|G(I)|$ is the number of $U(1)$ nodes in that column) has FI term  
    \[
    X_{I+1}-X_I+\frac{iQ}{4}(\delta_{\alpha,1}-\delta_{\alpha,|G(I)|}).
    \]

    \item Each $U(1)$ gauge node carries a Chern–Simons term at level $+1$, while nodes connected by diagonal or vertical links possess mixed CS interactions at levels $-1$ or $+1$, respectively.  
\end{itemize}

\textit{As such, we observe the exchange of topological and flavor symmetries characteristic of} $3d$ \textit{mirror symmetry}. Having identified the key features of the $\mathcal{N}=2$ mirror symmetry between a linear chiral and planar quiver gauge theory, we will now extend these findings to theories with $USp(2N)$ gauge symmetry. 

\section{A Planar Abelian Dual for \texorpdfstring{U(2N) CS-SQCD$_3$ with a $\overline{\scalebox{0.6}{$\ydiagram{1,1}$}}$
}{U(N) CS-SQCD3} Tensor}\label{sec: antisymmetric}

We begin from a somewhat unconventional vantage point: the mirror dual of an $\mathcal N=4$ $U(2N)$ gauge theory with $F$ hypermultiplets and a rank-2 antisymmetric tensor \cite{Hanany:1999sj,Dey:2014dwa,Dey:2014tka}. This setup provides a natural and instructive arena in which to explore how the paradigm of planar Abelianization extends to gauge theories with tensor matter.

Beyond its intrinsic interest, this theory furnishes a nontrivial consistency check across different gauge groups. Indeed, \textbf{both unitary and symplectic theories} can be recovered as appropriate limits of this construction: the former by integrating out the antisymmetric tensor, and the latter by giving it a vacuum expectation value\footnote{The full analysis of these RG flows at the level of the $\mathbf S^3_b$ partition function is beyond the scope of this work. However, it is intuitively evident that if the antisymmetric field acquires a VEV proportional to an antisymmetric matrix $J$, the residual gauge group is made of those matrices satisfying $U^TJU=J$, which is the defining feature of $USp(2N)$ symmetries.}. From this perspective, the $\mathcal N=2$ chiral-planar mirror duals obtained in this section should be viewed as common ancestors of the unitary chiral-planar mirror duals studied in \cite{Benvenuti:2024seb,Benvenuti:2025a,Benvenuti:2026a}, as well as of the symplectic chiral-planar mirror duals constructed in later sections of this work.

\subsection{\texorpdfstring{The $\mathcal N=4$ Mirror Pair}{N=4 Pair}}
We show the mirror dual pair of an $\mathcal N=4$ $U(2N)$ gauge theory with $F\, (\geq 2N+2)$ hypermultiplets and a rank-2 antisymmetric tensor in Equation \ref{eq: 3d_N=4_Unitary_Antisymmetric}.

\begin{equation}
\label{eq: 3d_N=4_Unitary_Antisymmetric}
    \includegraphics[width=\textwidth]{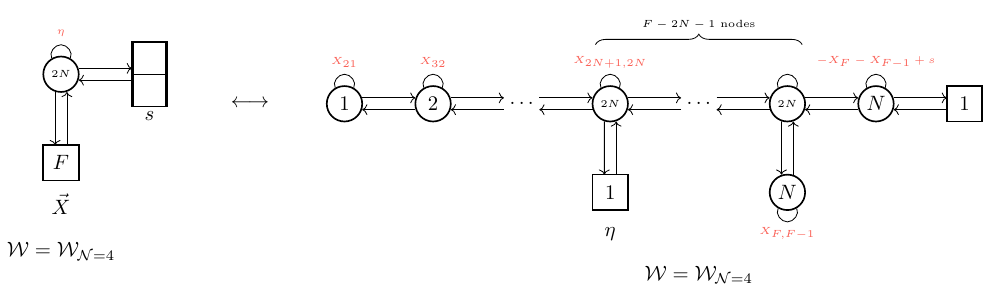}
\end{equation}

The FI terms are shown in orange, with the shorthand $X_{ij} := X_i - X_j$ in Equation \ref{eq: 3d_N=4_Unitary_Antisymmetric}. 

Although both theories preserve
$\mathcal N=4$ supersymmetry, it is convenient to describe them in the
$\mathcal N=2^*$ setup \cite{Tong:2000ky}. In this notation, 
we only keep track of the Cartan of
the $\mathcal N=4$ R-symmetry $SU(2)_C \times SU(2)_H$, parametrized as:
\begin{equation*}
U(1)_R \;=\; U(1)_{C+H}, \qquad
U(1)_\tau \;=\; U(1)_{C-H},
\end{equation*}
where $U(1)_C \subset SU(2)_C$ and $U(1)_H \subset SU(2)_H$. From the
$\mathcal N=2$ perspective, $U(1)_R$ is the R-symmetry,
while $U(1)_\tau$ should be regarded as a global symmetry.

However, we have the freedom to redefine $U(1)_R$ up to abelian flavor symmetries,
for example we can shift the R-charge by multiples of $U(1)_\tau$ charges. Using this freedom
we pick a convention in which, on the electric side, we assign trial R-charge $1-R$ to the chiral and anti-chiral multiplets
and R-charge $2R$ to the adjoint chiral multiplet.
The trial R-charges of the fields in the mirror dual follow and are summarized below.
On the electric side, we denote by $\Phi$, $Q$, $\tilde Q$, $\lambda$, and $\tilde\lambda$ the adjoint, fundamental/anti-fundamental chiral multiplets and the chiral/anti-chiral components of the rank-2 antisymmetric field of SQCD. On the mirror
side, we use the notation $\Phi_J$, $b_{J,J+1}$, and
$\widetilde b_{J+1,J}$ for the adjoint and bifundamental chiral
multiplets of the linear quiver, where $J$ denotes the position of the corresponding node(s), together with fundamental chiral
multiplets $q,\tilde q$ and $p,\tilde p$.
The charges of the various fields under the global symmetry are summarized in \eqref{eq: fields_n=4}.

\begin{equation}\label{eq: fields_n=4}
\begin{array}{c|c|c|c|c|c|c}
	 & U(2N) & SU(F)& U(1)_s & U(1)_{\tau} & U(1)_R & U(1)_\eta \\
	\hline
	\Phi & Adj & 1 & 0 & 1 & 2R & 0 \\
	Q & \square & \overline{\square} & 0 & -\frac{1}{2} & 1-R & 0 \\
	\tilde{Q} & \overline{\square} & {\square} & 0 & - \frac{1}{2} & 1-R & 0  \\
    \lambda & \scalebox{0.4}{$\ydiagram{1,1}$} & 1 &-1  & -\frac{1}{2} & 1-R & 0 \\
    \tilde\lambda & \overline{\scalebox{0.4}{$\ydiagram{1,1}$}} & 1 & 1 & -\frac{1}{2} & 1-R & 0
\end{array}
\qquad 
\begin{array}{c|c|c|c|c|c}
	  & SU(F)& U(1)_s & U(1)_\tau & U(1)_R & U(1)_{\eta} \\
	\hline
	\Phi_J  & 1 & 0 & -1 & 2-2R & 0 \\
	b_{J,J+1},\,\widetilde{b}_{J,J+1} & 1 & 0& \frac{1}{2} & R & 0 \\
	q,\, \widetilde{q} & 1 & 0& \frac{1}{2} & R & \mp1 \\
	p,\, \widetilde{p} & 1 &0 & \frac{1}{2} & R & 0
\end{array}
\end{equation}

\paragraph{Global Symmetries.}
The global symmetry of the \textit{electric} SQCD theory, depicted in the left panel of Equation \ref{eq: 3d_N=4_Unitary_Antisymmetric}, is\footnote{Here we consider symmetries that commute with our choice of $\NN=2$ subalgebra in the $\NN=2^*$ setup, therefore we list $U(1)_\tau$ as a flavor symmetry.}:
\begin{equation*}
    SU(F)\times U(1)_s\times U(1)_{\eta}\times U(1)_{\tau},
\end{equation*}
while the UV global symmetry of the \textit{mirror} flavored D-type quiver theory, depicted in the right panel of Equation \ref{eq: 3d_N=4_Unitary_Antisymmetric}, is:
\begin{equation*}
    \prod_{j=1}^{F-1}\,U(1)_{X_{j+1}-X_j} \times U(1)_{-X_{F}-X_{F-1}+s} \times U(1)_{\eta} \times U(1)_{\tau}.
\end{equation*}
Mirror symmetry predicts the enhancement of the topological symmetries $\prod_{j=1}^{F-1}\,U(1)_{X_{j+1}-X_j} \times U(1)_{-X_F-X_{F-1}+s}$ to $SU(F)\times U(1)_s$, which was verified by calculating the superconformal index of the two theories for small values of $N$ and $F$, providing a strong check in favor of the duality.

\paragraph{Mapping Gauge-Invariant Operators.}
Mirror symmetry provides a precise map between the order and disorder operators of the two theories:
\begin{itemize}
    \item The fundamental monopoles $\mathfrak M^{\pm}$ of the \textit{electric} SQCD theory are mapped to the mesonic operators in the \textit{mirror theory} constructed by contracting the fields connecting the two $U(1)$ flavor nodes.

    \item The composite operator $\lambda \tilde Q \tilde Q$ (and its conjugate $\tilde \lambda Q Q)$ are mapped to a collection of monopole operators of the mirror theory. More precisely, these monopoles carry GNO fluxes $\pm 1$ under consecutive gauge nodes starting from the rightmost gauge node with topological symmetry $U(1)_{-X_F-X_{F-1}+s}$.

\end{itemize}
There are other interesting operators, such as the mesons $Q \tilde{Q}$ and the operators $\tilde{Q} \lambda (\lambda \tilde{\lambda})^p \tilde{Q}$, which are mapped to monopole operators on the mirror side. These operators do not survive the $\NN=2$ deformation studied below, therefore we will not discuss them further.

\subsection{The Flow to the \texorpdfstring{$\mathcal N=2$}{N=2} Chiral-Planar Mirror Duals}
We now consider a supersymmetry-breaking mass deformation, implemented by turning on
a real mass for the $U(1)_\tau$ symmetry. 
This corresponds to turning on
a nonzero vacuum expectation value for the real scalar in the background vector multiplet associated with the $U(1)_\tau$ symmetry. Since $U(1)_\tau$ is a subgroup of the $\mathcal N=4$ R-symmetry, this deformation breaks
$SU(2)_C \times SU(2)_H \to U(1)_R$ and breaks supersymmetry from $\mathcal N=4$ to $\mathcal N=2$.

For generic values of the Coulomb branch parameters, all fields in the electric theory acquire a mass and the infrared theory is a topological quantum field theory. However, there exist distinguished loci on the Coulomb branch where
the effective mass of some fields vanish and
the flow instead leads to an interacting $\mathcal N=2$ theory. The identification of these vacua and the resulting effective descriptions can be carried out efficiently at the level of the $\mathbf{S}^3_b$ partition function, following the methods developed in \cite{Benvenuti:2024seb,Benvenuti:2025a}. 
The procedure is tedious and hardly illuminating, therefore
in this work we limit ourselves to a qualitative summary of the outcome.

We first consider the ``\textit{electric}" $\mathcal{N}=4$ SQCD theory. We move on the Coulomb branch in such a way as to reach an interacting vacuum where all fundamental chiral fields $Q$ and the anti-chiral component of the antisymmetric tensor $\tilde\lambda$ remain massless, while the adjoint $\Phi$, the $F$ anti-chiral fields $\tilde Q$, and the chiral component of the antisymmetric tensor $ \lambda$ acquire a mass and are integrated out\footnote{One can also consider a deformation where $Q$ and $\lambda$ remain massless. The resulting $\NN=2$ SQCD is qualitatively different, e.g. there are no gauge invariant chiral operators build out of two $Q$s and $\lambda$. We leave the study of this theory and its planar Abelian dual to future work.}. 
This process generates CS interactions, resulting in an $\mathcal N=2$ $U(2N)$ theory with $F$ fundamentals $ Q$ and an anti-chiral antisymmetric field $\tilde\lambda$ with CS level fixed by the matter content
\begin{equation*}
    \underbrace{(-\frac{F}{2}, -\frac{F}{2})}_{\text{F fundamentals}}+ \underbrace{(2N,0)}_\text{adjoint} + \underbrace{(1-N, 1-2N)}_{\text{antisymmetric}} = (1-\frac{F}{2}+N,\, 1-2N-\frac{F}{2});
\end{equation*}
notice, in particular, that the CS level for the diagonal $U(1) \subset U(2N)$ is $\ell=-\tfrac{3}{2}$ (refer to Appendix \ref{app: notation} for our conventions regarding CS levels).

\begin{figure}
    \centering
    \includegraphics[width=0.9\linewidth]{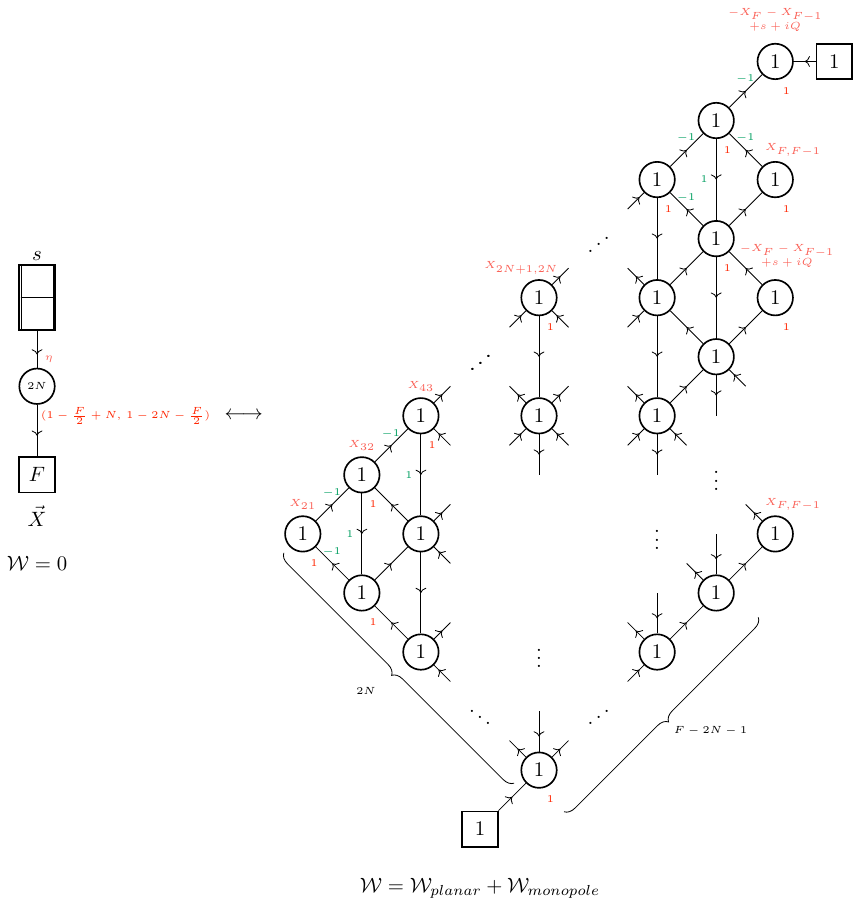}
    \caption{The $\mathcal N=2$ mirror dual of $U(2N)_{(1-\tfrac{F}{2}+N,\, 1-2N-\tfrac{F}{2})}$ SQCD with $F$ fundamental multiplets and an antisymmetric tensor is shown here. In our parametrization, the fields in the electric theory are assigned a trial R-charge of $1-R$, while the diagonal chiral fields have trial R-charge $R$ and the vertical ones $2-2R$ in the planar dual. The final column contains $2N$ $U(1)$ gauge nodes that are \textbf{not} connected by any matter or topological interactions to each other, and whose FI terms alternate between $-X_{F}-X_{F-1}+s+iQ(1-R)$ and $X_{F}-X_{F-1}$ from top to bottom. Unless otherwise stated, we set $R=0$ without loss of generality.}
    \label{fig: 3d_N=2_Unitary_Antisymmetric_Web}
\end{figure}

In direct analogy with the analysis of SQCD without tensor matter, the corresponding $\mathcal N=2$ mirror dual is obtained by Higgsing each $U(k)$ gauge node to its maximal torus $U(1)^k$. The resulting theory is a planar Abelian quiver, organized into columns of $k$ $U(1)$ gauge nodes. We show the resulting duality in Figure \ref{fig: 3d_N=2_Unitary_Antisymmetric_Web}.

The planar dual shown in Figure \ref{fig: 3d_N=2_Unitary_Antisymmetric_Web} shares many features with the planar theory dual to SQCD without tensor matter (refer to the discussion below Figure \ref{fig: 3d_N=4_N=2_Unit_Web} for details), and we only comment on the distinguishing features of this theory:
\begin{itemize}
    \item There is no gauge-invariant mesonic operator in the chiral ring of the planar mirror dual. Consequently, we conclude that there is no gauge-invariant monopole operator in the chiral ring of the electric theory.
    \item The monopole superpotential $\mathcal{W}_{monopole}$ contains additional monopole terms with generic GNO flux assignments shown in Equation \ref{eq: spicy monopole}. These monopoles also have an R-charge of $+2$, and are necessary to ensure the correct counting of global symmetry.  

    The additional $2N-2$ monopoles appearing in the superpotential are of the form:
      \begin{equation} \label{eq: spicy monopole}
      \mathcal{W}_{monopole} \supset
      \mathfrak{M}
        \begin{pmatrix}
            & &  & & + & \\ 
            & &  & + & & \\ 
            & & 0  &  & 0 \\ 
            & \iddots &  & - & & \\ 
            & & 0  &  & -  \\ 
            & &  & 0 & \\ 
            & &  &  & 0 \\
            & &  & \vdots & \vdots \\
        \end{pmatrix}
        \, + \mathfrak{M} 
        \begin{pmatrix}
            & &  & & 0 & \\ 
            & &  & 0 & & \\ 
            & & 0  &  & + \\ 
            & \iddots &  & + & & \\ 
            & & 0  &  & 0  \\ 
            & &  & - & \\ 
            & &  &  & - \\
            & &  & \vdots & \vdots \\
        \end{pmatrix} +\,\ldots
    \end{equation}

\noindent where all the fluxes that are not explicitly written vanish.
This set of monopole superpotential terms breaks the $U(1)^{2N}$ topological symmetries of the gauge nodes in the rightmost column to $U(1)^2$.
\end{itemize}

\paragraph{Matching Global Symmetries.}
The global symmetry of the \textit{electric} SQCD theory, shown in the left panel of Figure~\ref{fig: 3d_N=2_Unitary_Antisymmetric_Web}, is
\begin{equation*}
SU(F)\times U(1)_s \times U(1)_\eta \, .
\end{equation*}
The UV global symmetry of the \textit{mirror} theory, depicted in the right panel of Figure~\ref{fig: 3d_N=2_Unitary_Antisymmetric_Web}, is instead
\begin{equation}    \label{eq:symm_planar_U_antisymm_UV}
\prod_{j=1}^{F-1} U(1)_{X_{j+1}-X_j}
\times U(1)_{-X_F-X_{F-1}+s}
\times U(1)_\eta \, .
\end{equation}

In the absence of a monopole superpotential, each $U(1)$ gauge node would give rise to an independent topological $U(1)$ symmetry. The monopole superpotential breaks the topological symmetries associated with the nodes in each column of the planar quiver to a single diagonal $U(1)$. Consequently, the theory admits a $U(1)^{F-2}$ topological symmetry associated with the $F-2$ columns of the planar quiver, together with an additional $U(1)^2$ topological symmetry associated with the final column, due to the 
pattern of superpotential monopole operators with GNO flux under the last column~\eqref{eq: spicy monopole}. 
We performed superconformal index computations for low values of $N$ and $F$ which support the symmetry enhancement from \eqref{eq:symm_planar_U_antisymm_UV} to $SU(F)\times U(1)$ in the deep IR, see Table \ref{tab:SQCD_index_as}.

\begin{table}[ht]
\centering
\begin{tabular}{||c|c|c|c|c||} 
 \hline
 $N$ & $k$ & $F$ & $R$ & Superconformal Index \\
 \hline\hline

1 & 0 & 4 & $3/5$ & 
$\begin{array}{l}
1+6 x^{6/5}-17 x^2+20 x^{12/5}-64 x^{16/5} + \mathcal O(x^{18/5})
\end{array}$ \\
\hline

1 & $1/2$ & 5 & $3/5$ & 
$\begin{array}{l}
1+ 10x^{6/5} -26x^2+ 50x^{12/5}- 175x^{16/5} +\mathcal O(x^{18/5})
\end{array}$ \\
\hline

1 & 1 & 6 & $3/5$ & 
$\begin{array}{l}
1+15 x^{6/5}-37 x^2+105 x^{12/5}-384 x^{16/5} + \mathcal O(x^{18/5})
\end{array}$ \\
\hline

\end{tabular}
\caption{Values of the unrefined superconformal index of $\mathcal N=2$ $U(2N)_{(-k, -k-3N)}$ SQCD with $F$ anti-fundamental fields is reported here. The chiral multiplets are assigned a trial R-charge $1-R$. We verify that the index matches that of the corresponding Abelian mirror quiver (see Figure \ref{fig: 3d_N=2_Unitary_Antisymmetric_Web}).}
\label{tab:SQCD_index_as}
\end{table}

\paragraph{Mapping Gauge Invariant Operators.} The chiral ring of the \textit{electric} theory is generated by the $\binom{F}{2}$ composite mesonic operators $\mathbf M_{ij}:= \tilde\lambda_{\alpha\beta}  Q_i^{\alpha} Q_j^{\beta}$ (where $\alpha,\beta$ are gauge indices, and $i,j$ are flavor indices) with R-charge $3(1-R)$. 

We now provide details of the monopole operators dual to the composite mesons. One can check, using the monopole formula in Appendix \ref{app: monopole}, that all the monopole operators reported below have the same charges as the corresponding mesonic operator in the SQCD, including the R-charge.

\begin{itemize}
    \item The simplest operator is
    \[
    \mathbf M_{F, F-1} = \tilde\lambda_{\alpha\beta}\, Q_F^{\alpha} Q_{F-1}^{\beta} \, .
    \]
    In the mirror theory, this operator is mapped to a monopole carrying $-1$ GNO flux under the top gauge node of the final column: 
    
    \begin{equation*}
        \includegraphics[width=.5\linewidth]{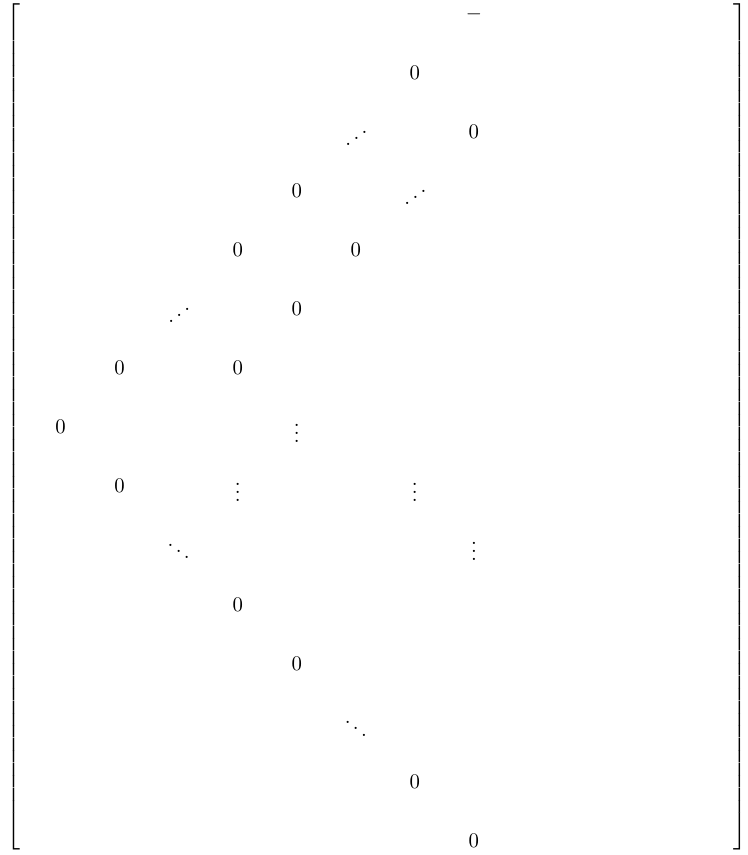}
    \end{equation*}

    \item More generally, the charge assignment propagates contiguously along the diagonal of the planar quiver. For example, the meson $\mathbf M_{F,F-1-I}$ is dual to a monopole with the flux configuration
    \begin{equation*}
        \includegraphics[width=.5\linewidth]{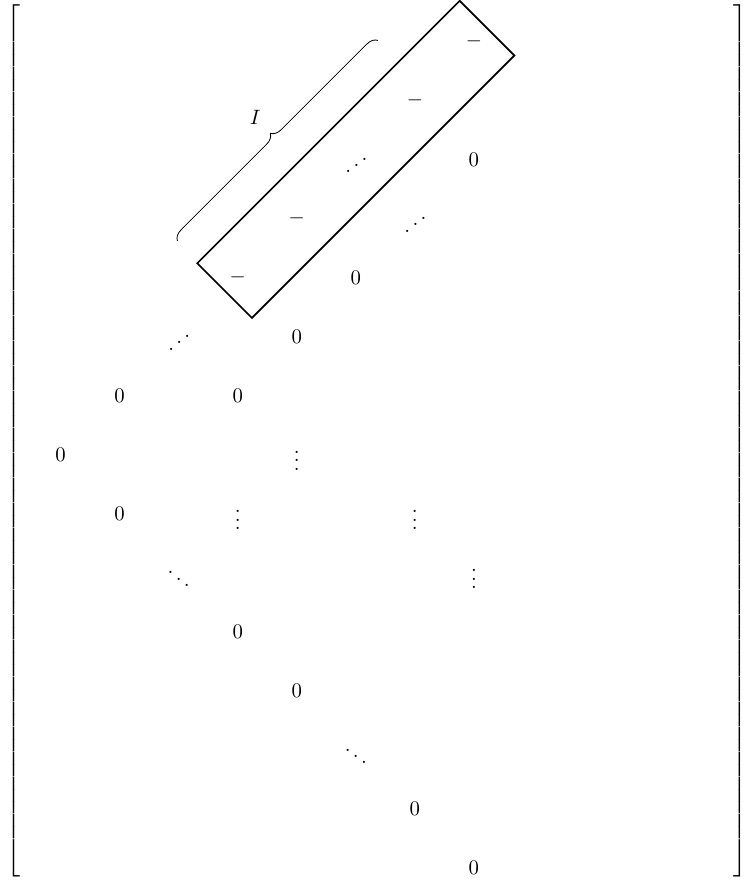}
    \end{equation*}

    \item Each column can host \textbf{at most} two nonzero GNO fluxes. This restriction follows from representation-theoretic constraints: the dual gauge-invariant operator transforms as a rank-two tensor, and therefore cannot accommodate more than two independent flux insertions in a given column. However, a $-1$ GNO flux may be reassigned to the $I^{\text{th}}$ column only if there is already a flux in the $(I-1)^{\text{th}}$ column. For example, the meson $\mathbf{M}_{F-1,F-1-I}$ is dual to a monopole with the flux configuration
     \begin{equation*}
        \includegraphics[width=.5\linewidth]{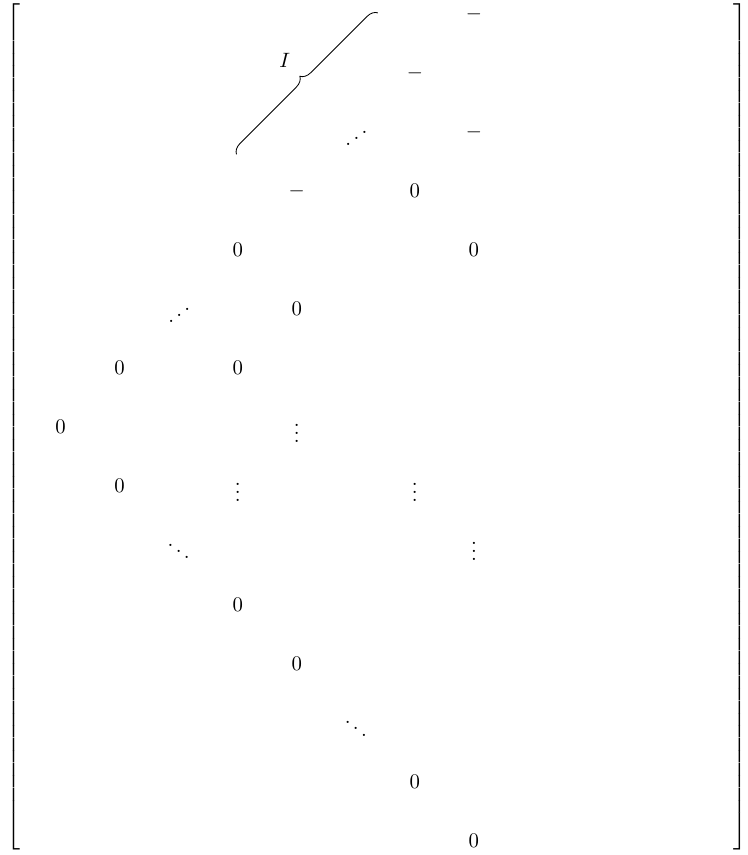}
    \end{equation*}

    Thus, a generic operator $\mathbf{M}_{F-J, F-1-I}$ is dual to a flux configuration with $I$ $+1$ GNO fluxes in the topmost diagonal, and $J$ ($<I)$ $+1$ GNO fluxes in the second diagonal

    \begin{equation*}
        \includegraphics[width=.5\linewidth]{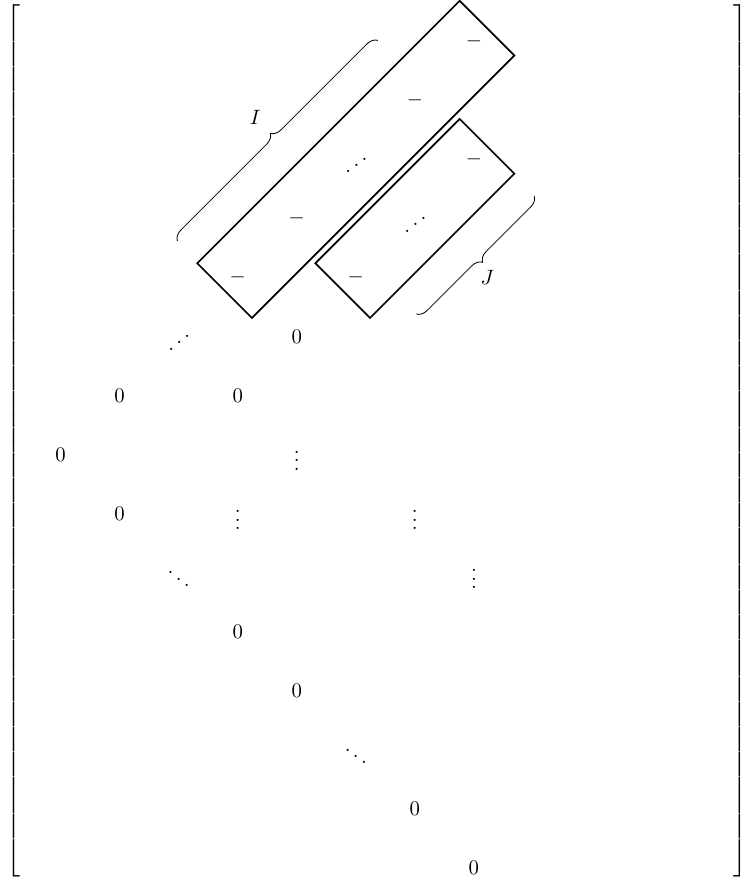}
    \end{equation*}
    
\end{itemize}

It would be interesting to further investigate the duality between
$SU(2N)_{1-\tfrac{F}{2}+N}$ SQCD with $F$ chiral multiplets and a rank-2
antisymmetric tensor, and its corresponding planar mirror dual. This duality can
be obtained in a straightforward manner by gauging the topological symmetry of
the $U(2N)$ theory \cite{witten2003sl2zactionthreedimensionalconformal}, which on
the mirror side corresponds to gauging one of the two $U(1)$ flavor symmetries.

This case is particularly rich, as the chiral ring is generated by a larger class
of operators, including composite mesons of the form $\tilde\lambda\,Q\, Q$,
baryonic operators $ Q^{2N}$, and the operator $\tilde{\lambda}^N$. 

The study of this duality is, however, complicated by an ambiguity in gauging the
$U(1)$ flavor symmetries on the mirror side: one may choose to gauge either of the
two $U(1)$ factors, leading to inequivalent presentations of the gauge-invariant
monopole operators. In particular, one choice yields a description in which the
composite operator $\tilde\lambda^N$ is mapped to a monopole with higher GNO
flux, whereas the alternative choice leads to a presentation in which the
composite mesons and baryonic operators are mapped to monopoles with higher GNO
flux. As such, we defer the resolution of this ambiguity and a more detailed study of these theories to future work. 

The paradigm of planar Abelianization admits a natural extension to theories with multiple rank-2 antisymmetric tensors. A particularly interesting case arises in the context of $\mathcal{N}=4$ $U(2N)$ SQCD with $F$ fundamentals and two rank-2 antisymmetric tensors: giving a VEV to the scalar component of one of the tensor multiplets Higgses the gauge group to $\mathcal{N}=4$ $USp(2N)$, yielding a theory with an adjoint multiplet and $F$ flavors. We anticipate an analogous mechanism in the $\NN=2$ setting. Concretely, starting from the chiral-planar mirror pair of $U(2N)$ SQCD with $F$ fundamental chiral multiplets and two chiral fields in the rank-2 antisymmetric representation, a suitable Higgsing of the gauge group should yield a planar Abelian dual of $USp(2N)$ CS-SQCD with $F$ fundamental multiplets and an adjoint chiral field — a family of dualities that lies beyond the reach of the single-tensor construction considered here in this paper, and is deferred to future work.
\subsection*{A Dual for \texorpdfstring{U(2N)$_{(0,-3N)}$ with $F\geq2N+2$ $\Box$ and $\overline{\scalebox{0.6}{$\ydiagram{1,1}$}}$
}{U(2N)0 with F}}
We conclude by proposing an Abelian quiver gauge theory that is infrared dual to
$U(2N)_{(0,-3N)}$ SQCD with $F$ fundamental chiral multiplets and a rank-2 antisymmetric tensor,
with $F\geq2N+2$ in Equation \ref{eq: 3d_N=2_Unitary_AS_k=0}. Consistency of Chern--Simons level quantization requires $F$ to be even.

This proposal is motivated by three complementary considerations. First, its structure
closely parallels the planar Abelian duals of $U(N)_0$ SQCD with $F$ fundamental chiral
multiplets proposed in \cite{Benvenuti:2026a}. Second, we have verified agreement of
superconformal indices for several low-rank and low-flavor examples, providing
nontrivial evidence in favor of the duality. Finally, under appropriate real mass
deformations, the proposed Abelian quiver flows to the known $\mathcal N=4$ descendant
theories discussed before. A more detailed \emph{a posteriori}
justification of this proposal will be given once we construct the analogous dual theory for $USp(2N)_0$ with $F$ fundamental multiplets, since both constructions share many qualitative features (see the discussion before Equation \ref{eq: u4_usp4_web}).
\newpage
\begin{equation}
\label{eq: 3d_N=2_Unitary_AS_k=0}
    \includegraphics[width=\linewidth]{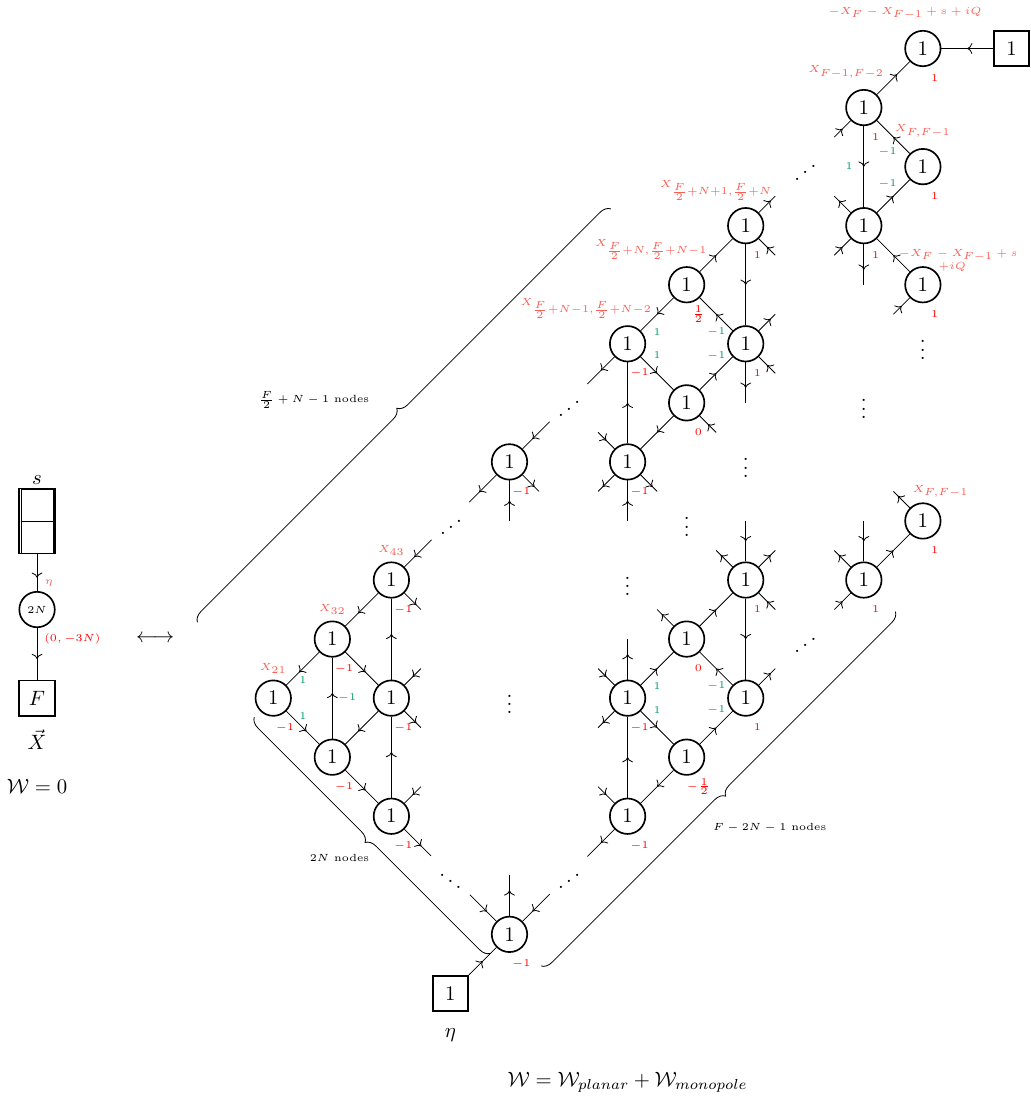}
\end{equation}
The planar dual shown in Equation \ref{eq: 3d_N=2_Unitary_AS_k=0} retains most of the features of the planar theory shown in Figure \ref{fig: 3d_N=2_Unitary_Antisymmetric_Web}; we only comment on the qualitative differences between the two theories here.
\begin{itemize}
     \item $\mathcal{W}_{\text{planar}}$ contains one cubic (quartic) term for each triangular (square) face of the quiver, with sign $-1$ ($+1$) if the arrows run clockwise (anticlockwise) around the face.
    \item $\mathcal{W}_{\text{monopole}}$ includes linear monopole superpotentials involving monopoles with GNO fluxes under two adjacent nodes within the same column. Specifically, there is a monopole $\mathfrak{M}^{\bigg(\begin{array}{c} +\\-\end{array}\bigg)}$ for each pair of vertically adjacent nodes. 

    In addition, $\mathcal W_{\text{monopole}}$ also contains additional monopole terms of the form described in Equation \ref{eq: spicy monopole}.
    
     The FI terms of the $U(1)$ gauge nodes are compatible with the monopole superpotential terms and are \textbf{not arbitrary}.
    \item There is a mixed CS term between each pair of nodes connected by an arrow. For downward-pointing arrows (\tikz{\draw[->-] (0,.5)--(.5,0);}, \tikz{\draw[->-] (.5,.5)--(0,0);}, and \tikz{\draw[->-] (0,.5)--(0,0);}), the mixed CS level is $+1$, while for upward-pointing arrows (\tikz{\draw[-<-] (0,.5)--(.5,0);}, \tikz{\draw[-<-] (.5,.5)--(0,0);}, and \tikz{\draw[-<-] (0,.5)--(0,0);}), it is $-1$.
    \item The CS level of a $U(1)_i$ gauge node is determined by its mixed CS couplings:
    \begin{equation}
        k_i = -\frac{1}{2} \sum_{j \neq i} k_{ij},
    \end{equation}
    where $k_{ij}$ denotes the level of the mixed CS term between $U(1)_i$ and $U(1)_j$. The sum runs over every gauge and flavor node of the quiver except for the node $U(1)_i$.
\end{itemize}

For concreteness, we consider the planar mirror dual of $U(4)_{(0,-6)}$ with $8$ anti-fundamental chiral multiplets and a rank-2 antisymmetric tensor in Equation \ref{eq: u4_8_squares}. This will serve as a prototype for the remainder of this paper.

\begin{equation}
\label{eq: u4_8_squares}
    \includegraphics[width=.6\linewidth]{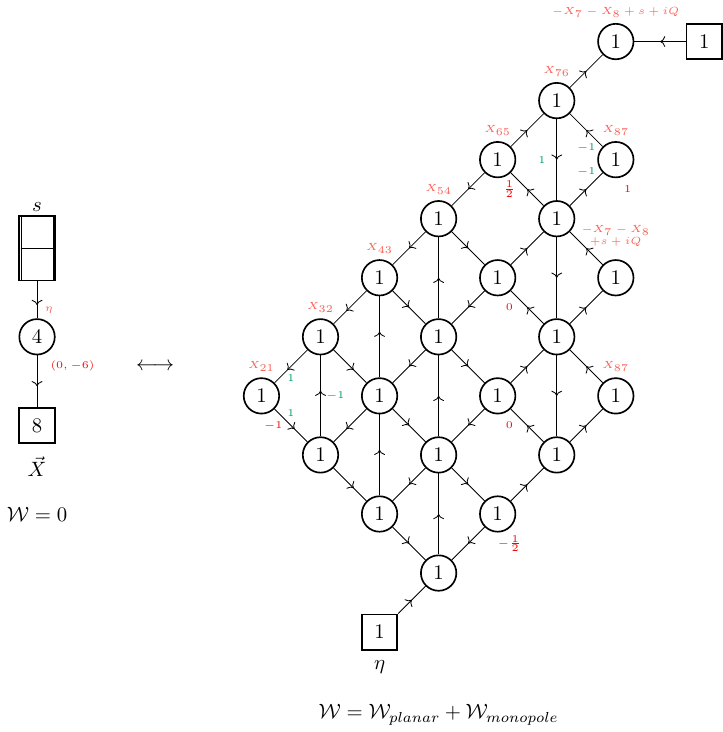}
\end{equation}

Once the rank-2 antisymmetric tensor acquires an expectation value, the theory is Higgsed to $USp(4)_0$ with $8$ fundamental chiral multiplets \footnote{Since the representations of $USp(2N)$ are pseudoreal, there is no intrinsic distinction between fundamental and antifundamental chiral multiplets. When we refer to a field as transforming in the fundamental representation of $USp(2N)$, we therefore mean that it transforms in the antifundamental representation of the $U(F)$ flavor symmetry, or vice versa.}, shown in Equation \ref{eq: usp4_8_squares}. Note that the $U(1)_s$ global symmetry is broken by the VEV of the antisymmetric field.

\begin{equation}
\label{eq: usp4_8_squares}
    \includegraphics[width=.6\linewidth]{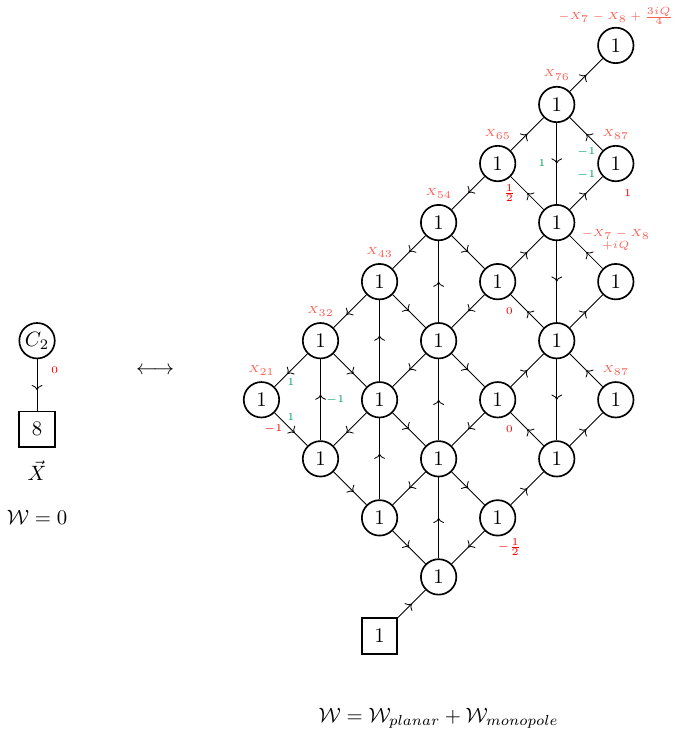}
\end{equation}

\section{Planar Abelian Duals of Symplectic CS-SQCD\texorpdfstring{$_3$}{3}} \label{sec: symplectic}

3d $\mathcal{N}=4$ gauge theories with a symplectic gauge group are known to admit two different mirror descriptions. The first is a unitary quiver with the shape of a $D$-type Dynkin diagram ($D$-type quiver, for brevity) \cite{Porrati:1996xi,Hori:1997zj,Kapustin:1998fa, Giveon:1998sr, Hanany:1999sj}. The second is a linear quiver with alternating orthogonal and symplectic gauge groups \cite{Feng:2000eq}, which may be obtained by means of Hanany-Witten brane setups in the presence of $O3$ planes, often referred to as Feng-Hanany constructions. 

\begin{figure}[H]
    \centering
    \includegraphics[width=1\linewidth]{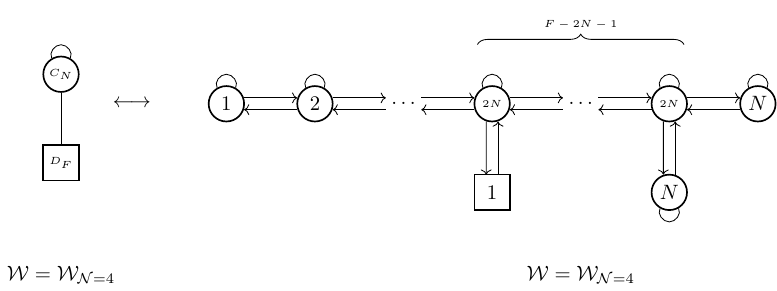}
    \caption{The $\mathcal{N}=4$ mirror duality between $USp(2N)$ SQCD and $F$ ($\geq 2N+2$) hypermultiplets and a $D$-type unitary quiver gauge theory is shown here.}
    \label{fig: 3d_N=4_USp_U_mirror}
\end{figure}

\begin{figure}[H]
    \centering
    \includegraphics[width=1\linewidth]{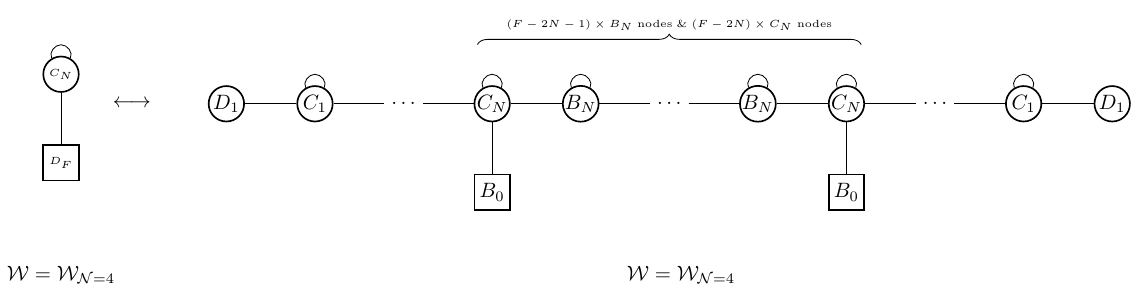}
    \caption{The $\mathcal{N}=4$ mirror duality between $USp(2N)$ SQCD and $F$ ($\geq 2N+2$) hypermultiplets and a linear, orthosymplectic quiver gauge theory is shown here.}
    \label{fig: 3d_N=4_USp_O_mirror}
\end{figure}

In this discussion, we will focus on the first mirror dual, leaving the analysis of orthosymplectic mirror duals to future work. In Figure \ref{fig: 3d_N=4_USp_U_mirror}, we show the $D$-type mirror dual of $USp(2N)$ SQCD, while in Figure \ref{fig: 3d_N=4_USp_O_mirror} we show the orthosymplectic mirror dual for completeness. 
We label gauge and flavor nodes associated with symplectic or orthogonal symmetries by their algebras $C_N$ and $D_N$, see Appendix \ref{app: notation} for details regarding our notation.
More details regarding these dualities may be also found in \cite{Nawata:2021nse,Nawata:2023rdx,Dey:2014tka}. 
\newpage

Both Figures \ref{fig: 3d_N=4_USp_U_mirror} and \ref{fig: 3d_N=4_USp_O_mirror} are presented in $\mathcal{N}=2$ language. In the orthosymplectic mirror of $USp(2N)$ (Figure \ref{fig: 3d_N=4_USp_O_mirror}), the flavor nodes labeled by $B_0$ represent half-hypermultiplets transforming in the fundamental representation of $C_N$.

\paragraph{\texorpdfstring{An Aside: The $\mathcal N=4$ Superpotential}{N=4 Superpotential}.}
The reason for denoting the flavor symmetry group on the \textit{electric} SQCD side as $D_F$ is that the $\mathcal{N}=4$ superpotential preserves precisely this amount of symmetry. A flavor node $D_F = SO(2F)$ therefore corresponds to $F$ hypermultiplets transforming in the fundamental of the gauge group $USp(2N)$. Equivalently, it is often said that there are $2F$ half-hypermultiplets transforming in the fundamental representation of the gauge group $USp(2N)$ and in the vector representation of the flavor group $SO(2F)$.

Since it is not immediately obvious why a $USp(2N)$ gauge theory should admit an $SO(2F)$ global symmetry, let us briefly justify this point (refer to \cite{Tachikawa:2013kta} for a detailed discussion). An $\mathcal{N}=4$ hypermultiplet can be written in $\mathcal{N}=2$ language as a pair of chiral multiplets $(Q^a_I,\tilde Q^I_a)$, where $a,b,\ldots$ denote gauge indices and $I,J,\ldots$ flavor indices. The fundamental representation of $USp(2N)$ is pseudoreal, which means that there exists an antisymmetric invariant tensor $J_{ab}$, with inverse denoted as $J^{ab}$. 

A crucial property of the representation theory of $USp(2N)$ is that the generators $\left(T^A\right)^a{}_b$ of the fundamental representation are symmetric once contracted with the antisymmetric tensor $J_{ab}$, namely:
\[
J_{ac} \left(T^A\right)^c{_b} = J_{bc} \left(T^A\right)^c{}_a, \qquad \textrm{compactly} \qquad (J\,T^A)_{ab} = (J\,T^A)_{ba},
\]
which follows from differentiating the defining relation $U^T J U = J$. These identities have important consequences for the structure of the theory -- in particular, if $Q^a_I$ transforms in the fundamental representation, then so does $J^{ab}\tilde Q_b^I$. It is therefore natural to introduce the purely fundamental vector
\[
\mathcal Q^a_i := \bigl(Q^a_1,\ldots,Q^a_F,\; J^{ab}\tilde Q_b^1,\ldots,J^{ab}\tilde Q_b^F\bigr),
\qquad i=1,\ldots,2F.
\]
With this redefinition, the superpotential can be cast in a manifestly $SO(2F)$-invariant form:

\begin{equation}\label{eq: symplectic superpotential}
\begin{split}
    \mathcal{W}_{\mathcal N=4} = \tilde{Q}^I_a \, \Phi^a_{\, b} Q^b_{I} &= \tilde{Q}_a^I J^{ab} J_{bc} \Phi^c_{\, d} \, Q^d_I = -\frac{1}{2} \, \mathcal{Q}^a_i (J\Phi)_{ab} \, \mathcal{Q}^b_j \, \eta^{ij},  \\ \text{where:}
   & \qquad\eta^{ij}=\begin{pmatrix}
0 & \mathbb{I}_F \\
\mathbb{I}_F & 0
\end{pmatrix},
\end{split}
\end{equation} where the symmetry of $(J\Phi)_{ab}$ has been employed. In this form, the symmetry group acting on $\mathcal{Q}$ is the set of matrices preserving $\eta$. To make it explicit, we perform the change of basis:

\begin{equation}\label{eq: change of basis}
\begin{split}
    iv_1^a = \frac{1}{\sqrt{2}} \left( Q_1^a + J^{ab} \tilde{Q}^1_b  \right), &\qquad v_2^a =\frac{1}{\sqrt{2}} \left( Q_1^a - J^{ab}\tilde{Q}^1_b \right),\\
   & \vdots \\
  iv_{2F-1}^a = \frac{1}{\sqrt{2}} \left( Q_F^a + J^{ab} \tilde{Q}^F_b  \right), &\qquad v_{2F}^a =\frac{1}{\sqrt{2}} \left( Q_F^a - J^{ab}\tilde{Q}^F_b \right).
   \end{split}
\end{equation} Notice that this change of basis respects the gauge structure, as $Q$ and $J\tilde{Q}$ transform in the same representation, and thus can be summed over. The superpotential now becomes:

\begin{equation}\label{eq: manifestly so superpotential}
    \mathcal{W} = \frac{1}{2} \,  v^a_i (J\Phi)_{ab} \, v^b_j \, \delta^{ij},
\end{equation} which is manifestly $SO(2F)$ invariant. One could repeat the same procedure for an $SO(N)$ gauge theory: now the fundamental representation is real and admits a symmetric invariant tensor, thus one will find a $USp(2F)$ flavor symmetry group.

\newpage
The reason we are interested in the $D$-type unitary dual of $USp(2N)$ SQCD is that it is closely related to the duality of $U(2N)$ with an antisymmetric tensor (see Section \ref{sec: antisymmetric}). In the remainder of this section, we will focus on the $USp(2N)$ dualities and analyze the $\mathcal{N}=2$ deformations of the duality shown in Figure \ref{fig: 3d_N=4_USp_U_mirror}, depending on whether we take a \textit{chiral} or \textit{planar} limit on the SQCD side, following the general guidelines of Section \ref{sec: introduction} and \ref{sec: antisymmetric}.

\begin{figure}[H]
    \centering
    \includegraphics[width=.8\linewidth]{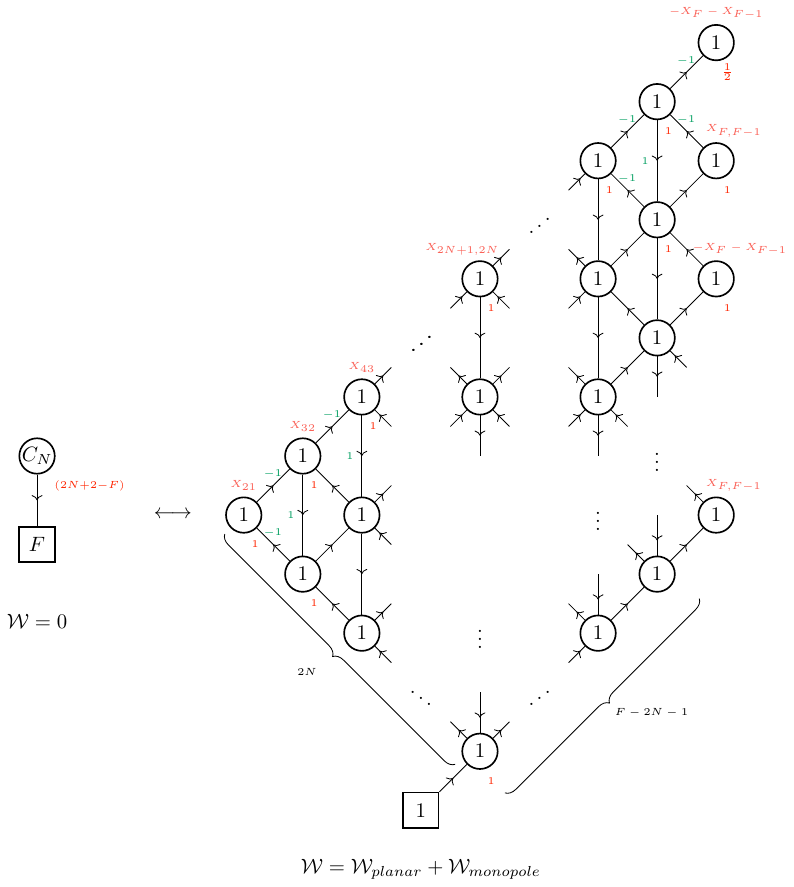}
    \caption{Schematic depiction of the $\mathcal{N}=2$ mirror-like duality between $USp(2N)$ CS-SQCD$_3$ with $F$ fundamental multiplets, and a planar Abelian quiver gauge theory. The FI terms of the mirror dual are shown explicitly (their mixing with the $U(1)_R$ symmetry has been suppressed for brevity), with the shorthand $X_{ij} := X_i - X_j$. The final column contains $2N$ $U(1)$ gauge nodes with no vertical lines or BF interactions and whose FI terms alternate between $-X_{F}-X_{F-1}$ and $X_{F}-X_{F-1}$ from top to bottom.}
    \label{fig: 3d_N=2_USp_Web}
\end{figure}

Let us start by analyzing the \textit{chiral limit} on the SQCD side. More precisely, we perform a supersymmetry breaking deformation from $\mathcal{N}=4$ to $\mathcal{N}=2$ in $USp(2N)$ SQCD that preserves the gauge group; at the level of the $\mathbf{S}^3_b$ partition function, one performs the same steps described in \cite{Benvenuti:2025a}. 
The deformation is such that the adjoint chiral and the $F$ antifundamental chiral multiplets acquire a mass and are integrated out, generating a Chern-Simons level equal to $2N + 2 -F$. We are left with an $\mathcal{N}=2$ theory with gauge group $USp(2N)_{2N+2-F}$ and $F$ chiral multiplets in the fundamental representation. 
The deformation also breaks the flavor symmetry from $SO(2F)$ to $U(F)$.
We can map the deformation on the mirror side of Figure \ref{fig: 3d_N=4_USp_U_mirror}, obtaining a planar Abelian quiver displayed in Figure \ref{fig: 3d_N=2_USp_Web}. This mapping admits an algorithmic generalization closely analogous to that of \cite{Benvenuti:2025a}, as further discussed in \cite{Comi:wip}.

\newpage

We conclude with a few remarks highlighting the main differences between the planar dual of $USp(2N)$ and the planar dual of $U(2N)$ shown in Figure \ref{fig: 3d_N=4_N=2_Unit_Web}: 
\begin{itemize}
    \item There is no flavor node attached to the topmost gauge node; hence, we cannot form any long meson on the planar side. 
    Therefore, in contrast with the unitary SQCD case, in general we do not expect a gauge-invariant chiral monopole in the SQCD side.
    However, in the \textit{s-confining} case $F=2N+2$, such gauge-invariant monopole exists on the SQCD side, which is mapped to a higher GNO flux monopole on the planar side (see Section \ref{subsec: example} for a worked example of this case). 
    \item In the last column, the full FI parameter of the topmost gauge node is $-X_F - X_{F-1} + \frac{3}{4}iQ(1-R)$. The second-highest node has FI parameter $X_F - X_{F-1}$, while the third-highest node has FI parameter $-X_F - X_{F-1} + iQ(1-R)$. The fourth-highest node again has FI parameter $X_F - X_{F-1}$, and this pattern continues down the column. Apart from the topmost node, which exhibits a different mixing, the structure is therefore determined by the second- and third-highest nodes.
    \item The monopole superpotential $\mathcal{W}_{monopole}$ now contains $2N-2$ additional terms:
    \begin{equation}
      \mathcal{W}_{monopole} \supset
      \mathfrak{M}
        \begin{pmatrix}
            & &  & & + & \\ 
            & &  & + & & \\ 
            & & 0  &  & 0 \\ 
            & \iddots &  & - & & \\ 
            & & 0  &  & -  \\ 
            & &  & 0 & \\ 
            & &  &  & 0 \\
            & &  & \vdots & \vdots \\
        \end{pmatrix}
        \, + \mathfrak{M} 
        \begin{pmatrix}
            & &  & & 0 & \\ 
            & &  & 0 & & \\ 
            & & 0  &  & + \\ 
            & \iddots &  & + & & \\ 
            & & 0  &  & 0  \\ 
            & &  & - & \\ 
            & &  &  & - \\
            & &  & \vdots & \vdots \\
        \end{pmatrix} +\,\ldots
    \end{equation}

\noindent where all the fluxes that are not explicitly written vanish.
Because of the R-charge mixing discussed above, these monopoles have R-charge equal to 2 and thus appear in the superpotential. 
This set of monopole superpotential terms breaks the $U(1)^{2N}$ topological symmetries of the gauge nodes in the rightmost column to $U(1)^2$, and is crucial to obtain the correct counting of global symmetries.

\end{itemize}

\paragraph{Matching Global Symmetries.}
The global symmetry of the \textit{electric} SQCD theory, shown in the left panel of Figure~\ref{fig: 3d_N=2_USp_Web}, is $U(F)$.

The UV global symmetry of the \textit{mirror} theory, depicted in the right panel of Figure~\ref{fig: 3d_N=2_USp_Web}, is instead
\begin{equation*}
\prod_{j=1}^{F-1} U(1)_{X_{j+1}-X_j}
\times U(1)_{-X_F-X_{F-1}}.
\end{equation*}

Analogously to the $\mathcal N=2$ chiral-planar mirror duals studied in Section \ref{sec: antisymmetric}, the theory admits a $U(1)^{F-2}$ topological symmetry associated with the $F-2$ columns of the planar quiver, together with an additional $U(1)^2$ topological symmetry associated with the final column, due to the presence of the extra monopole operators in the superpotential. As before, we consistently observe symmetry enhancement to $U(F)$ in the deep IR from index calculations. 

\paragraph{Mapping Gauge-Invariant Operators.} The chiral ring of the \textit{electric} theory is generated by the $\binom{F}{2}$ composite mesonic operators $\mathbf M_{ij}:=J_{ab}  Q_i^{a} Q_j^{b}$ (where $a,b$ are gauge indices, and $i,j$ are flavor indices, and $J$ is the $2N\times 2N$ symplectic matrix) with R-charge $2-2R$. The dual flux configurations are identical to those studied in Section \ref{sec: antisymmetric}. As such, a generic operator $\mathbf{M}_{F-J, F-1-I}$ is dual to a flux configuration with $I$ negative GNO fluxes on the topmost diagonal, and $J$ ($<I)$ negative GNO fluxes on the second diagonal

    \begin{equation*}
        \includegraphics[width=.5\linewidth]{Images/mon4.pdf}
    \end{equation*}

\subsection{\texorpdfstring{An $\NN=2$ Quotient Quiver Subtraction Proposal}{N=2 Quiver Subtraction}}\label{subsec: quiv_subtract}

Quiver subtraction \cite{Cabrera:2018ann} is a combinatorial operation on quivers which encodes the structure of Coulomb branches of $3d$ $\mathcal{N}=4$
quiver gauge theories. 

We briefly review the steps of the procedure: given a quiver $\mathcal{Q}$ and a subquiver $\mathcal{S}$, the subtraction $\mathcal{Q} - \mathcal{S}$ produces a remainder quiver whose gauge ranks are reduced node by node, with each step corresponding to a transverse slice between symplectic leaves. This has been extended to \textit{quotient quiver subtraction} \cite{Bennett:2024llh,Hanany:2023tvn,Bennett:2025edk}, which translates the question of gauging a Coulomb branch isometry subgroup into a straightforward procedure, bypassing complications from strongly coupled IR physics. In particular, gauging an $USp(2N)$ subgroup of the Coulomb branch symmetry of a unitary quiver (\textit{à la} \cite{Dey:2014tka}) proceeds by subtracting a linear quotient quiver from the long leg, followed by a splitting of the remaining node into two, producing a D-type quiver \cite{Bennett:2026gtm}. 
\newpage

We summarize this with an example for fixed values of $N$ and $F$:
\begin{enumerate}
    \item We begin with the $[4,7]$ asymmetric bifundamental block and its mirror dual (refer to \cite{Comi:2022aqo} for more details):
    \begin{equation}\label{eq: quiv_sub_n=4_1}
    \begin{tikzpicture}[baseline=(current bounding box.center)]
    \node at (0,0) (f1) [flavor] {$4$};
    \node at (1.5,0) (f2) [flavor] {$7$};
    \draw[->] (f1)++(.3,.1) -- (1.2,.1); 
    \draw[->] (f2)++(-.3,-.1) -- (.3,-.1);
 
    \node at (2.8,0) {$\longleftrightarrow$};

    \begin{scope}[xshift = -3.5cm, yshift = -1.5cm]
    \node at (4.2,0)  (n1)  [gauge] {$1$};
    \node at (5.7,0)  (n2)  [gauge] {$2$};
    \node at (7.2,0)  (n3)  [gauge] {$3$};
    \node at (8.7,0)  (n4a) [gauge] {$4$};
    \node at (10.2,0) (n4b) [gauge] {$4$};
    \node at (11.7,0) (n4c) [gauge] {$4$};
    \node at (13.2,0) (n4d) [gauge] {$4$};
    \node at (14.7,0) (n3b) [gauge] {$3$};
    \node at (16.2,0) (n2b) [gauge] {$2$};
    \node at (17.7,0) (n1b) [gauge] {$1$};
    \node at (8.7,-1.5) (fl1) [flavor] {$1$};
 
    \draw[-] (n1)  to[out=60,in=0]  (4.2,0.5)  to[out=180,in=120] (n1);
    \draw[-] (n2)  to[out=60,in=0]  (5.7,0.5)  to[out=180,in=120] (n2);
    \draw[-] (n3)  to[out=60,in=0]  (7.2,0.5)  to[out=180,in=120] (n3);
    \draw[-] (n4a) to[out=60,in=0]  (8.7,0.5)  to[out=180,in=120] (n4a);
    \draw[-] (n4b) to[out=60,in=0]  (10.2,0.5) to[out=180,in=120] (n4b);
    \draw[-] (n4c) to[out=60,in=0]  (11.7,0.5) to[out=180,in=120] (n4c);
    \draw[-] (n4d) to[out=60,in=0]  (13.2,0.5) to[out=180,in=120] (n4d);
    \draw[-] (n3b) to[out=60,in=0]  (14.7,0.5) to[out=180,in=120] (n3b);
    \draw[-] (n2b) to[out=60,in=0]  (16.2,0.5) to[out=180,in=120] (n2b);
    \draw[-] (n1b) to[out=60,in=0]  (17.7,0.5) to[out=180,in=120] (n1b);
 
    \draw[->] (n1)++(.3,.1) -- (5.4,.1); 
    \draw[->] (n2)++(-.3,-.1) -- (4.5,-.1);

    \draw[->] (n2)++(.3,.1) -- (6.9,.1); 
    \draw[->] (n3)++(-.3,-.1) -- (6,-.1);

    \draw[->] (n3)++(.3,.1) -- (8.4,.1); 
    \draw[->] (n4a)++(-.3,-.1) -- (7.5,-.1);

    \draw[->] (n4a)++(.3,.1) -- (9.9,.1); 
    \draw[->] (n4b)++(-.3,-.1) -- (9,-.1);

    \draw[->] (n4b)++(.3,.1) -- (11.4,.1); 
    \draw[->] (n4c)++(-.3,-.1) -- (10.5,-.1);

    \draw[->] (n4c)++(.3,.1) -- (12.9,.1); 
    \draw[->] (n4d)++(-.3,-.1) -- (12,-.1);

    \draw[->] (n4d)++(.3,.1) -- (14.4,.1); 
    \draw[->] (n3b)++(-.3,-.1) -- (13.5,-.1);

    \draw[->] (n3b)++(.3,.1) -- (15.9,.1); 
    \draw[->] (n2b)++(-.3,-.1) -- (15,-.1);

    \draw[->] (n2b)++(.3,.1) -- (17.4,.1); 
    \draw[->] (n1b)++(-.3,-.1) -- (16.5,-.1);
    
    \draw[->] (n4a)++(-.1,-.3) -- (8.6,-1.2);
    \draw[->] (fl1) ++(.1,.3)--(8.8,-.3);
    \end{scope}
\end{tikzpicture}
    \end{equation}

    \item We gauge a $USp(4)$ subgroup of the \textit{electric} $[4,7]$ block; the resulting \textit{electric} theory is $USp(4)$ with $7$ hypermultiplets. 
    
    This corresponds to subtracting a linear quiver from the mirror dual:
    \begin{equation} \label{eq: quiv_sub_n=4_2}
        \begin{tikzpicture}[baseline=(current bounding box.center)]
     \begin{scope}[xshift = -4.2cm]
    \node at (4.2,0)  (n1)  [gauge] {$1$};
    \node at (5.7,0)  (n2)  [gauge] {$2$};
    \node at (7.2,0)  (n3)  [gauge] {$3$};
    \node at (8.7,0)  (n4a) [gauge] {$4$};
    \node at (10.2,0) (n4b) [gauge] {$4$};
    \node at (11.7,0) (n4c) [gauge] {$4$};
    \node at (13.2,0) (n4d) [gauge] {$4$};
    \node at (14.7,0) (n3b) [gauge] {$3$};
    \node at (16.2,0) (n2b) [gauge] {$2$};
    \node at (17.7,0) (n1b) [gauge] {$1$};
    \node at (8.7,-1.5) (fl1) [flavor] {$1$};
 
    \draw[-] (n1)  to[out=60,in=0]  (4.2,0.5)  to[out=180,in=120] (n1);
    \draw[-] (n2)  to[out=60,in=0]  (5.7,0.5)  to[out=180,in=120] (n2);
    \draw[-] (n3)  to[out=60,in=0]  (7.2,0.5)  to[out=180,in=120] (n3);
    \draw[-] (n4a) to[out=60,in=0]  (8.7,0.5)  to[out=180,in=120] (n4a);
    \draw[-] (n4b) to[out=60,in=0]  (10.2,0.5) to[out=180,in=120] (n4b);
    \draw[-] (n4c) to[out=60,in=0]  (11.7,0.5) to[out=180,in=120] (n4c);
    \draw[-] (n4d) to[out=60,in=0]  (13.2,0.5) to[out=180,in=120] (n4d);
    \draw[-] (n3b) to[out=60,in=0]  (14.7,0.5) to[out=180,in=120] (n3b);
    \draw[-] (n2b) to[out=60,in=0]  (16.2,0.5) to[out=180,in=120] (n2b);
    \draw[-] (n1b) to[out=60,in=0]  (17.7,0.5) to[out=180,in=120] (n1b);
 
    \draw[->] (n1)++(.3,.1) -- (5.4,.1); 
    \draw[->] (n2)++(-.3,-.1) -- (4.5,-.1);

    \draw[->] (n2)++(.3,.1) -- (6.9,.1); 
    \draw[->] (n3)++(-.3,-.1) -- (6,-.1);

    \draw[->] (n3)++(.3,.1) -- (8.4,.1); 
    \draw[->] (n4a)++(-.3,-.1) -- (7.5,-.1);

    \draw[->] (n4a)++(.3,.1) -- (9.9,.1); 
    \draw[->] (n4b)++(-.3,-.1) -- (9,-.1);

    \draw[->] (n4b)++(.3,.1) -- (11.4,.1); 
    \draw[->] (n4c)++(-.3,-.1) -- (10.5,-.1);

    \draw[->] (n4c)++(.3,.1) -- (12.9,.1); 
    \draw[->] (n4d)++(-.3,-.1) -- (12,-.1);

    \draw[->] (n4d)++(.3,.1) -- (14.4,.1); 
    \draw[->] (n3b)++(-.3,-.1) -- (13.5,-.1);

    \draw[->] (n3b)++(.3,.1) -- (15.9,.1); 
    \draw[->] (n2b)++(-.3,-.1) -- (15,-.1);

    \draw[->] (n2b)++(.3,.1) -- (17.4,.1); 
    \draw[->] (n1b)++(-.3,-.1) -- (16.5,-.1);
    
    \draw[->] (n4a)++(-.1,-.3) -- (8.6,-1.2);
    \draw[->] (fl1) ++(.1,.3)--(8.8,-.3);
    \end{scope}
 
    \node at (6.75,-3.5) {$-$};
 
    \begin{scope}[xshift = -5.2cm, yshift = -3.5cm]
    
    \node at (13.2,0) (n4d) [gauge] {$4$};
    \node at (14.7,0) (n3b) [gauge] {$3$};
    \node at (16.2,0) (n2b) [gauge] {$2$};
    \node at (17.7,0) (n1b) [gauge] {$1$};


    \draw[-] (n4d) to[out=60,in=0]  (13.2,0.5) to[out=180,in=120] (n4d);
    \draw[-] (n3b) to[out=60,in=0]  (14.7,0.5) to[out=180,in=120] (n3b);
    \draw[-] (n2b) to[out=60,in=0]  (16.2,0.5) to[out=180,in=120] (n2b);
    \draw[-] (n1b) to[out=60,in=0]  (17.7,0.5) to[out=180,in=120] (n1b);
 
    
    \draw[->] (n4d)++(.3,.1) -- (14.4,.1); 
    \draw[->] (n3b)++(-.3,-.1) -- (13.5,-.1);

    \draw[->] (n3b)++(.3,.1) -- (15.9,.1); 
    \draw[->] (n2b)++(-.3,-.1) -- (15,-.1);

    \draw[->] (n2b)++(.3,.1) -- (17.4,.1); 
    \draw[->] (n1b)++(-.3,-.1) -- (16.5,-.1);

    \end{scope}

\end{tikzpicture}
    \end{equation}
    The resulting quiver is shown below:
    \begin{equation} \label{eq: quiv_sub_n=4_3}
        \begin{tikzpicture}[baseline=(current bounding box.center)]
     \begin{scope}[xshift = -8.2cm]
    \node at (4.2,0)  (n1)  [gauge] {$1$};
    \node at (5.7,0)  (n2)  [gauge] {$2$};
    \node at (7.2,0)  (n3)  [gauge] {$3$};
    \node at (8.7,0)  (n4a) [gauge] {$4$};
    \node at (10.2,0) (n4b) [gauge] {$4$};
    \node at (11.7,0) (n4c) [gauge, red,thick] {$4$};
    \node at (8.7,-1.5) (fl1) [flavor] {$1$};
 
    \draw[-] (n1)  to[out=60,in=0]  (4.2,0.5)  to[out=180,in=120] (n1);
    \draw[-] (n2)  to[out=60,in=0]  (5.7,0.5)  to[out=180,in=120] (n2);
    \draw[-] (n3)  to[out=60,in=0]  (7.2,0.5)  to[out=180,in=120] (n3);
    \draw[-] (n4a) to[out=60,in=0]  (8.7,0.5)  to[out=180,in=120] (n4a);
    \draw[-] (n4b) to[out=60,in=0]  (10.2,0.5) to[out=180,in=120] (n4b);
    \draw[-,red, thick] (n4c) to[out=60,in=0]  (11.7,0.5) to[out=180,in=120] (n4c);

    \draw[->] (n1)++(.3,.1) -- (5.4,.1); 
    \draw[->] (n2)++(-.3,-.1) -- (4.5,-.1);

    \draw[->] (n2)++(.3,.1) -- (6.9,.1); 
    \draw[->] (n3)++(-.3,-.1) -- (6,-.1);

    \draw[->] (n3)++(.3,.1) -- (8.4,.1); 
    \draw[->] (n4a)++(-.3,-.1) -- (7.5,-.1);

    \draw[->] (n4a)++(.3,.1) -- (9.9,.1); 
    \draw[->] (n4b)++(-.3,-.1) -- (9,-.1);

    \draw[->,red] (n4b)++(.3,.1) -- (11.4,.1); 
    \draw[->, red] (n4c)++(-.3,-.1) -- (10.5,-.1);

    \draw[->] (n4a)++(-.1,-.3) -- (8.6,-1.2);
    \draw[->] (fl1) ++(.1,.3)--(8.8,-.3);
    \end{scope}
    \end{tikzpicture}
    \end{equation}

    \item While rebalancing is not required, the $U(4)$ node highlighted in {\color{red}{red}} must be split into $U(2)\;\&\; U(2)$ (which may be regarded as a consequence of orientifolding the Hanany-Witten setup corresponding to the mirror pairs shown in Equation \ref{eq: quiv_sub_n=4_1}), both of which inherit the bifundamental matter content of the $U(4)$ gauge node. The resulting quiver is the familiar D-type quiver, which we recognize as the mirror dual of $USp(4)$ with $7$ hypermultiplets:
     \begin{equation} \label{eq: quiv_sub_n=4_4}
        \begin{tikzpicture}[baseline=(current bounding box.center)]
     \begin{scope}[xshift = -8.2cm]
    \node at (4.2,0)  (n1)  [gauge] {$1$};
    \node at (5.7,0)  (n2)  [gauge] {$2$};
    \node at (7.2,0)  (n3)  [gauge] {$3$};
    \node at (8.7,0)  (n4a) [gauge] {$4$};
    \node at (10.2,0) (n4b) [gauge] {$4$};
    \node at (11.7,0)  (n2a) [gauge] {$2$};
    \node at (10.2,-1.5) (n2b) [gauge] {$2$};
    \node at (8.7,-1.5) (fl1) [flavor] {$1$};
 
    \draw[-] (n1)  to[out=60,in=0]  (4.2,0.5)  to[out=180,in=120] (n1);
    \draw[-] (n2)  to[out=60,in=0]  (5.7,0.5)  to[out=180,in=120] (n2);
    \draw[-] (n3)  to[out=60,in=0]  (7.2,0.5)  to[out=180,in=120] (n3);
    \draw[-] (n4a) to[out=60,in=0]  (8.7,0.5)  to[out=180,in=120] (n4a);
    \draw[-] (n4b) to[out=60,in=0]  (10.2,0.5) to[out=180,in=120] (n4b);
     \draw[-] (n2a) to[out=60,in=0]  (11.7,.5) to[out=180,in=120] (n2a);
    \draw[-] (n2b) to[out=-60,in=0] (10.2,-2) to[out=180,in=-120] (n2b);

    \draw[->] (n1)++(.3,.1) -- (5.4,.1); 
    \draw[->] (n2)++(-.3,-.1) -- (4.5,-.1);

    \draw[->] (n2)++(.3,.1) -- (6.9,.1); 
    \draw[->] (n3)++(-.3,-.1) -- (6,-.1);

    \draw[->] (n3)++(.3,.1) -- (8.4,.1); 
    \draw[->] (n4a)++(-.3,-.1) -- (7.5,-.1);

    \draw[->] (n4a)++(.3,.1) -- (9.9,.1); 
    \draw[->] (n4b)++(-.3,-.1) -- (9,-.1);

    \draw[->] (n4b)++(.3,.1) -- (11.4,.1); 
    \draw[->] (n2a)++(-.3,-.1) -- (10.5,-.1);

     \draw[->] (n4b)++(-.1,-.3) -- (10.1,-1.2);
    \draw[->] (n2b) ++(.1,.3)--(10.3,-.3);

    \draw[->] (n4a)++(-.1,-.3) -- (8.6,-1.2);
    \draw[->] (fl1) ++(.1,.3)--(8.8,-.3);
    \end{scope}
    \end{tikzpicture}
    \end{equation}

\end{enumerate}

Here we provide some evidence that a similar mechanism should hold for $\mathcal N=2$ \textit{chiral-planar} mirror duals as well\footnote{Note that we only display the global topology of the quiver diagrams here and suppress the CS, FI and superpotential terms for brevity. These can be recovered following our usual conventions, elaborated in Appendix \ref{app: notation} .}. This expectation can be argued for given the results derived in this paper, as we explain below by showing how such a procedure should work in an example. We consider the $\mathcal N=2$ analog of the example studied before:
\begin{enumerate}
    \item We begin with the $[4,7]$ asymmetric chiral bifundamental block and its mirror dual (which can be obtained via a suitable supersymmetry breaking real mass deformation of Equation \ref{eq: quiv_sub_n=4_1}):

    \begin{equation} \label{eq: quiv_sub_n=2_1}
    \begin{tikzpicture}[baseline=(current bounding box.center)]
    \node at (0,0) (f1) [flavor] {$4$};
    \node at (1.5,0) (f2) [flavor] {$7$};
    \draw[->-] (f1)--(f2);
 
    \node at (2.8,0) {$\longleftrightarrow$};

    \begin{scope}[xshift = 2.5cm, yshift = -3.5cm]
        \node at (0,0) (g11) [gauge,black] {$1$};
       
       \node at (1,1) (g21) [gauge,black] {$1$};
       \node at (1,-1) (g22) [gauge,black] {$1$};
       
       \node at (2,2) (g31) [gauge,black]{$1$};
       \node at (2,0) (g32) [gauge,black] {$1 $};
       \node at (2,-2) (g33) [gauge,black] {$1$};

       \node at (3,3) (g41) [gauge,black] {$1$};
       \node at (3,1) (g42) [gauge,black] {$1$};
       \node at (3,-1) (g43) [gauge,black] {$1$};
       \node at (3,-3) (g44) [gauge,black] {$1$};
       \node at (2,-4) (f41) [flavor,black] {$1$};

       \node at (4,4) (g51) [gauge,black] {$1$};
       \node at (4,2) (g52) [gauge,black] {$1$}; 
       \node at (4,0) (g53) [gauge,black] {$1$};
       \node at (4,-2) (g54) [gauge,black] {$1$};

       \node at (5,5) (g61) [gauge,black] {$1$};
       \node at (5,3) (g62) [gauge,black] {$1$};
       \node at (5,1) (g63) [gauge,black]{$1$};
       \node at (5,-1) (g64) [gauge,black] {$1$};
       
       \node at (6,6) (g71) [gauge,black] {$1$};
       \node at (6,4) (g72) [gauge,black] {$1$};
       \node at (6,2) (g73) [gauge,black]{$1$};
       \node at (6,0) (g74) [gauge,black] {$1$};

       \node at (7,5) (g81) [gauge,black] {$1$};
       \node at (7,3) (g82) [gauge,black] {$1$};
       \node at (7,1) (g83) [gauge,black] {$1$};

       \node at (8,4) (g91) [gauge,black] {$1$};
       \node at (8,2) (g92) [gauge,black] {$1$};

       \node at (9,3) (g101) [gauge,black] {$1$};


       \draw[->-] (g22)--(g11);
       \draw[->-] (g11)--(g21);
       \draw[->-] (g21)--(g22);
       \draw[->-] (g33)--(g22);
       \draw[->-] (g22)--(g32);
       \draw[->-] (g32)--(g21);
       \draw[->-] (g21)--(g31);
       \draw[->-] (g31)--(g41);
       \draw[->-] (g31)--(g32);
       \draw[->-] (g32)--(g33);
       \draw[->-] (g42)--(g31);
       \draw[->-] (g42)--(g43);
       \draw[->-] (g32)--(g42);
       \draw[->-] (g33)--(g43);
       \draw[->-] (g43)--(g32);
       \draw[->-] (g41)--(g42);
       \draw[->-] (g43)--(g44);
       \draw[->-] (g44)--(g33);
       \draw[->-] (f41)--(g44);

       \draw[->-] (g51)--(g52);
       \draw[->-] (g52)--(g53);
       \draw[->-] (g53)--(g54);

       \draw[->-] (g41)--(g51);
       \draw[->-] (g42)--(g52);
       \draw[->-] (g43)--(g53);
       \draw[->-] (g44)--(g54);

       \draw[->-] (g51)--(g61);
       \draw[->-] (g52)--(g62);
       \draw[->-] (g53)--(g63);
       \draw[->-] (g54)--(g64);

       \draw[->-] (g52)--(g41);
       \draw[->-] (g53)--(g42);
       \draw[->-] (g54)--(g43);
       
       \draw[->-] (g62)--(g51);
       \draw[->-] (g63)--(g52);
       \draw[->-] (g64)--(g53);

       \draw[->-] (g61)--(g62);
       \draw[->-] (g62)--(g63);
       \draw[->-] (g63)--(g64);

       \draw[->-] (g71)--(g72);
       \draw[->-] (g72)--(g73);
       \draw[->-] (g73)--(g74);

       \draw[->-] (g61)--(g71);
       \draw[->-] (g62)--(g72);
       \draw[->-] (g63)--(g73);
       \draw[->-] (g64)--(g74);

       \draw[->-] (g72)--(g61);
       \draw[->-] (g73)--(g62);
       \draw[->-] (g74)--(g63);

       \draw[->-] (g81)--(g71);
       \draw[->-] (g82)--(g72);
       \draw[->-] (g83)--(g73);
       \draw[->-] (g74)--(g83);

       \draw[->-] (g81)--(g82);
       \draw[->-] (g82)--(g83);

       \draw[->-] (g72)--(g81);
       \draw[->-] (g73)--(g82);

       \draw[->-] (g91)--(g92);
       \draw[->-] (g92)--(g101);
       \draw[->-] (g101)--(g91);
       \draw[->-] (g91)--(g81);
       \draw[->-] (g92)--(g82);
       \draw[->-] (g83)--(g92);
       \draw[->-] (g82)--(g91);
    \end{scope}
\end{tikzpicture}
    \end{equation}
    The $U(4)$ flavor node of the bifundamental has CS level $-\tfrac{7}{2}$, which corresponds to a CS level $-7$ for its $USp(4)$ subgroup in our conventions, see Appendix \ref{app: notation}.

    \item We gauge a $USp(4)$ subgroup of the \textit{electric} theory with CS level $6$, which is a natural $\NN=2$ deformation of the $\NN=4$ $USp(4)$ gauging. The CS term at level $6$ is obtained when one integrates out the adjoint chiral inside the $\NN=4$ vector multiplet.
    The resulting \textit{electric} theory is $USp(4)_{-1}$ with $7$ chiral multiplets. 

    \newpage
    Analogous to Equation \ref{eq: quiv_sub_n=4_2}, this corresponds to subtracting a planar subquiver (which can be obtained via a suitable supersymmetry breaking real mass deformation of Equation \ref{eq: quiv_sub_n=4_2}) from the mirror dual:

    \begin{equation}
    \begin{tikzpicture}[baseline=(current bounding box.center)]

    \begin{scope}[xshift = 0cm, yshift = 2cm]
        \node at (0,0) (g11) [gauge,black] {$1$};
       
       \node at (1,1) (g21) [gauge,black] {$1$};
       \node at (1,-1) (g22) [gauge,black] {$1$};
       
       \node at (2,2) (g31) [gauge,black]{$1$};
       \node at (2,0) (g32) [gauge,black] {$1 $};
       \node at (2,-2) (g33) [gauge,black] {$1$};

       \node at (3,3) (g41) [gauge,black] {$1$};
       \node at (3,1) (g42) [gauge,black] {$1$};
       \node at (3,-1) (g43) [gauge,black] {$1$};
       \node at (3,-3) (g44) [gauge,black] {$1$};
       \node at (2,-4) (f41) [flavor,black] {$1$};

       \node at (4,4) (g51) [gauge,black] {$1$};
       \node at (4,2) (g52) [gauge,black] {$1$}; 
       \node at (4,0) (g53) [gauge,black] {$1$};
       \node at (4,-2) (g54) [gauge,black] {$1$};

       \node at (5,5) (g61) [gauge,black] {$1$};
       \node at (5,3) (g62) [gauge,black] {$1$};
       \node at (5,1) (g63) [gauge,black]{$1$};
       \node at (5,-1) (g64) [gauge,black] {$1$};
       
       \node at (6,6) (g71) [gauge,black] {$1$};
       \node at (6,4) (g72) [gauge,black] {$1$};
       \node at (6,2) (g73) [gauge,black]{$1$};
       \node at (6,0) (g74) [gauge,black] {$1$};

       \node at (7,5) (g81) [gauge,black] {$1$};
       \node at (7,3) (g82) [gauge,black] {$1$};
       \node at (7,1) (g83) [gauge,black] {$1$};

       \node at (8,4) (g91) [gauge,black] {$1$};
       \node at (8,2) (g92) [gauge,black] {$1$};

       \node at (9,3) (g101) [gauge,black] {$1$};


       \draw[->-] (g22)--(g11);
       \draw[->-] (g11)--(g21);
       \draw[->-] (g21)--(g22);
       \draw[->-] (g33)--(g22);
       \draw[->-] (g22)--(g32);
       \draw[->-] (g32)--(g21);
       \draw[->-] (g21)--(g31);
       \draw[->-] (g31)--(g41);
       \draw[->-] (g31)--(g32);
       \draw[->-] (g32)--(g33);
       \draw[->-] (g42)--(g31);
       \draw[->-] (g42)--(g43);
       \draw[->-] (g32)--(g42);
       \draw[->-] (g33)--(g43);
       \draw[->-] (g43)--(g32);
       \draw[->-] (g41)--(g42);
       \draw[->-] (g43)--(g44);
       \draw[->-] (g44)--(g33);
       \draw[->-] (f41)--(g44);

       \draw[->-] (g51)--(g52);
       \draw[->-] (g52)--(g53);
       \draw[->-] (g53)--(g54);

       \draw[->-] (g41)--(g51);
       \draw[->-] (g42)--(g52);
       \draw[->-] (g43)--(g53);
       \draw[->-] (g44)--(g54);

       \draw[->-] (g51)--(g61);
       \draw[->-] (g52)--(g62);
       \draw[->-] (g53)--(g63);
       \draw[->-] (g54)--(g64);

       \draw[->-] (g52)--(g41);
       \draw[->-] (g53)--(g42);
       \draw[->-] (g54)--(g43);
       
       \draw[->-] (g62)--(g51);
       \draw[->-] (g63)--(g52);
       \draw[->-] (g64)--(g53);

       \draw[->-] (g61)--(g62);
       \draw[->-] (g62)--(g63);
       \draw[->-] (g63)--(g64);

       \draw[->-] (g71)--(g72);
       \draw[->-] (g72)--(g73);
       \draw[->-] (g73)--(g74);

       \draw[->-] (g61)--(g71);
       \draw[->-] (g62)--(g72);
       \draw[->-] (g63)--(g73);
       \draw[->-] (g64)--(g74);

       \draw[->-] (g72)--(g61);
       \draw[->-] (g73)--(g62);
       \draw[->-] (g74)--(g63);

       \draw[->-] (g81)--(g71);
       \draw[->-] (g82)--(g72);
       \draw[->-] (g83)--(g73);
       \draw[->-] (g74)--(g83);

       \draw[->-] (g81)--(g82);
       \draw[->-] (g82)--(g83);

       \draw[->-] (g72)--(g81);
       \draw[->-] (g73)--(g82);

       \draw[->-] (g91)--(g92);
       \draw[->-] (g92)--(g101);
       \draw[->-] (g101)--(g91);
       \draw[->-] (g91)--(g81);
       \draw[->-] (g92)--(g82);
       \draw[->-] (g83)--(g92);
       \draw[->-] (g82)--(g91);
    \end{scope}

    \begin{scope}[xshift = 2cm, yshift = -6cm]

       \node at (5,3) {$-$};
       
       \node at (6,6) (g71) [gauge,black] {$1$};
       \node at (6,4) (g72) [gauge,black] {$1$};
       \node at (6,2) (g73) [gauge,black]{$1$};
       \node at (6,0) (g74) [gauge,black] {$1$};

       \node at (7,5) (g81) [gauge,black] {$1$};
       \node at (7,3) (g82) [gauge,black] {$1$};
       \node at (7,1) (g83) [gauge,black] {$1$};

       \node at (8,4) (g91) [gauge,black] {$1$};
       \node at (8,2) (g92) [gauge,black] {$1$};

       \node at (9,3) (g101) [gauge,black] {$1$};



       \draw[->-] (g71)--(g72);
       \draw[->-] (g72)--(g73);
       \draw[->-] (g73)--(g74);

       \draw[->-] (g81)--(g71);
       \draw[->-] (g82)--(g72);
       \draw[->-] (g83)--(g73);
       \draw[->-] (g74)--(g83);

       \draw[->-] (g81)--(g82);
       \draw[->-] (g82)--(g83);

       \draw[->-] (g72)--(g81);
       \draw[->-] (g73)--(g82);

       \draw[->-] (g91)--(g92);
       \draw[->-] (g92)--(g101);
       \draw[->-] (g101)--(g91);
       \draw[->-] (g91)--(g81);
       \draw[->-] (g92)--(g82);
       \draw[->-] (g83)--(g92);
       \draw[->-] (g82)--(g91);
    \end{scope}
\end{tikzpicture}
\end{equation}
\newpage
    The resulting quiver is shown in Equation \ref{eq: quiv_sub_n=2_3}:
\begin{equation}
\label{eq: quiv_sub_n=2_3}
    \begin{tikzpicture}[baseline=(current bounding box.center)]

    \begin{scope}[xshift = 0cm, yshift = 2cm]
        \node at (0,0) (g11) [gauge,black] {$1$};
       
       \node at (1,1) (g21) [gauge,black] {$1$};
       \node at (1,-1) (g22) [gauge,black] {$1$};
       
       \node at (2,2) (g31) [gauge,black]{$1$};
       \node at (2,0) (g32) [gauge,black] {$1 $};
       \node at (2,-2) (g33) [gauge,black] {$1$};

       \node at (3,3) (g41) [gauge,black] {$1$};
       \node at (3,1) (g42) [gauge,black] {$1$};
       \node at (3,-1) (g43) [gauge,black] {$1$};
       \node at (3,-3) (g44) [gauge,black] {$1$};
       \node at (2,-4) (f41) [flavor,black] {$1$};

       \node at (4,4) (g51) [gauge,black] {$1$};
       \node at (4,2) (g52) [gauge,black] {$1$}; 
       \node at (4,0) (g53) [gauge,black] {$1$};
       \node at (4,-2) (g54) [gauge,black] {$1$};

       \node at (5,5) (g61) [gauge,red] {$1$};
       \node at (5,3) (g62) [gauge,red] {$1$};
       \node at (5,1) (g63) [gauge,red]{$1$};
       \node at (5,-1) (g64) [gauge,red] {$1$};
       
      \begin{scope}[on background layer]
            \draw[line width=5pt, red!50, line cap=round](g61)--(g62)--(g63)--(g64) ;
      \end{scope}


       \draw[->-] (g22)--(g11);
       \draw[->-] (g11)--(g21);
       \draw[->-] (g21)--(g22);
       \draw[->-] (g33)--(g22);
       \draw[->-] (g22)--(g32);
       \draw[->-] (g32)--(g21);
       \draw[->-] (g21)--(g31);
       \draw[->-] (g31)--(g41);
       \draw[->-] (g31)--(g32);
       \draw[->-] (g32)--(g33);
       \draw[->-] (g42)--(g31);
       \draw[->-] (g42)--(g43);
       \draw[->-] (g32)--(g42);
       \draw[->-] (g33)--(g43);
       \draw[->-] (g43)--(g32);
       \draw[->-] (g41)--(g42);
       \draw[->-] (g43)--(g44);
       \draw[->-] (g44)--(g33);
       \draw[->-] (f41)--(g44);

       \draw[->-] (g51)--(g52);
       \draw[->-] (g52)--(g53);
       \draw[->-] (g53)--(g54);

       \draw[->-] (g41)--(g51);
       \draw[->-] (g42)--(g52);
       \draw[->-] (g43)--(g53);
       \draw[->-] (g44)--(g54);

       \draw[->-,red,thick] (g51)--(g61);
       \draw[->-,red,thick] (g52)--(g62);
       \draw[->-,red,thick] (g53)--(g63);
       \draw[->-,red,thick] (g54)--(g64);

       \draw[->-] (g52)--(g41);
       \draw[->-] (g53)--(g42);
       \draw[->-] (g54)--(g43);
       
       \draw[->-,red,thick] (g62)--(g51);
       \draw[->-,red,thick] (g63)--(g52);
       \draw[->-,red,thick] (g64)--(g53);

       \draw[->-] (g61)--(g62);
       \draw[->-] (g62)--(g63);
       \draw[->-] (g63)--(g64);

    \end{scope}

\end{tikzpicture}
\end{equation}

    \item 
    Comparing with our results described in the previous section, we find it consistent to split the column of $U(1)$ nodes (highlighted in {\color{red}{red}}) into two groups, alternating from top to bottom.
    We recognize the resulting quiver as the planar mirror dual of $USp(4)_{-1}$ with $7$ chiral multiplets:
    \begin{equation}
        \begin{tikzpicture}[baseline=(current bounding box.center)] 
    
       \node at (0,0) (g11) [gauge,black] {$1$};
       
       \node at (1,1) (g21) [gauge,black] {$1$};
       \node at (1,-1) (g22) [gauge,black] {$1$};
       
       \node at (2,2) (g31) [gauge,black]{$1$};
       \node at (2,0) (g32) [gauge,black] {$1 $};
       \node at (2,-2) (g33) [gauge,black] {$1$};

       \node at (3,3) (g41) [gauge,black] {$1$};
       \node at (3,1) (g42) [gauge,black] {$1$};
       \node at (3,-1) (g43) [gauge,black] {$1$};
       \node at (3,-3) (g44) [gauge,black] {$1$};
       \node at (2,-4) (f41) [flavor,black] {$1$};

       \node at (4,4) (g51) [gauge,black] {$1$};
       \node at (4,2) (g52) [gauge,black] {$1$}; 
       \node at (4,0) (g53) [gauge,black] {$1$};
       \node at (4,-2) (g54) [gauge,black] {$1$};

       \node at (5,5) (g61) [gauge,black] {$1$};
       \node at (5,3) (g62) [gauge,black] {$1$};
       \node at (5,1) (g63) [gauge,black]{$1$};
       \node at (5,-1) (g64) [gauge,black] {$1$};


       \draw[->-] (g22)--(g11);
       \draw[->-] (g11)--(g21);
       \draw[->-] (g21)--(g22);
       \draw[->-] (g33)--(g22);
       \draw[->-] (g22)--(g32);
       \draw[->-] (g32)--(g21);
       \draw[->-] (g21)--(g31);
       \draw[->-] (g31)--(g41);
       \draw[->-] (g31)--(g32);
       \draw[->-] (g32)--(g33);
       \draw[->-] (g42)--(g31);
       \draw[->-] (g42)--(g43);
       \draw[->-] (g32)--(g42);
       \draw[->-] (g33)--(g43);
       \draw[->-] (g43)--(g32);
       \draw[->-] (g41)--(g42);
       \draw[->-] (g43)--(g44);
       \draw[->-] (g44)--(g33);
       \draw[->-] (f41)--(g44);

       \draw[->-] (g51)--(g52);
       \draw[->-] (g52)--(g53);
       \draw[->-] (g53)--(g54);

       \draw[->-] (g41)--(g51);
       \draw[->-] (g42)--(g52);
       \draw[->-] (g43)--(g53);
       \draw[->-] (g44)--(g54);

       \draw[->-] (g51)--(g61);
       \draw[->-] (g52)--(g62);
       \draw[->-] (g53)--(g63);
       \draw[->-] (g54)--(g64);

       \draw[->-] (g52)--(g41);
       \draw[->-] (g53)--(g42);
       \draw[->-] (g54)--(g43);
       
       \draw[->-] (g62)--(g51);
       \draw[->-] (g63)--(g52);
       \draw[->-] (g64)--(g53);
    \end{tikzpicture}
    \end{equation}
\end{enumerate}
As such, we expect a generalization of $\mathcal N=4$ quiver subtraction to $\mathcal N=2$ \textit{chiral-planar} mirror duals, but we defer a systematic exploration of this paradigm to the future. 
\section{Planar Limits of Symplectic Gauge Symmetries}\label{sec: d quiver}

In this Section we consider a different $\NN=2$ deformation of the mirror duality in Figure \ref{fig: 3d_N=4_USp_U_mirror}, where we take a planar limit on the SQCD side corresponding to a chiral limit on the $D$-type mirror that preserves the non-Abelian structure of all gauge nodes.
We obtain an $\mathcal{N}=2$ unitary $D$-type quiver on the right-hand side, and a planar Abelian quiver on the left-hand side. 
Since the resulting $\mathcal{N}=2$ electric theory has some peculiar features, it is worth discussing an explicit example before presenting the general proposal.

\paragraph{An Example: The Planar Limit of \texorpdfstring{$USp(2)$ SQCD$_3$ with $F$ fundamental multiplets}{USp(2) SQCD3 with F}}

\begin{figure}[H]
    \centering
    \includegraphics[width=0.9\linewidth]{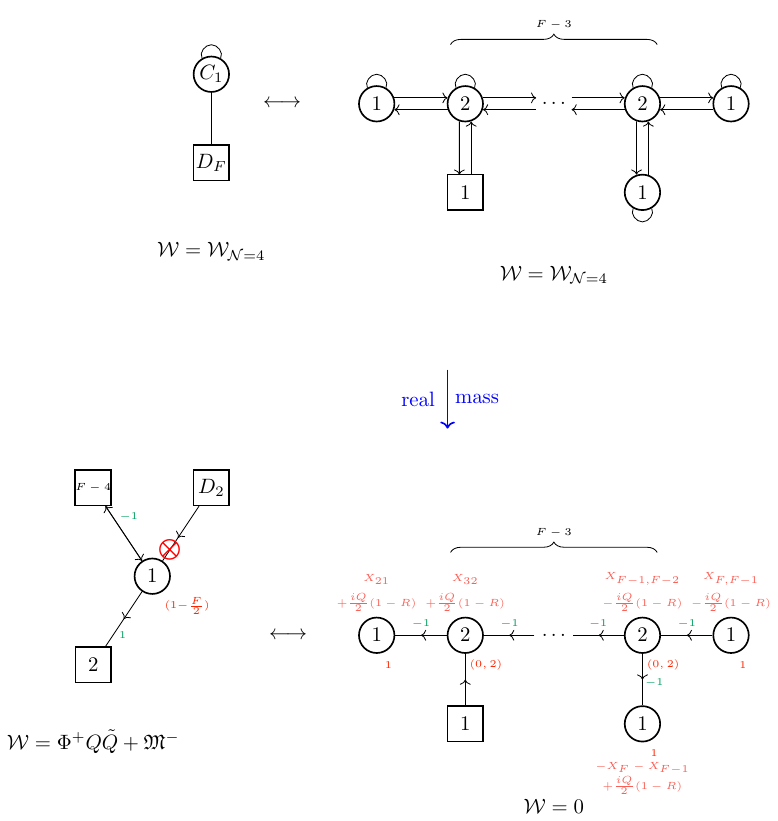}
    \caption{Schematic depiction of the $\mathcal{N}=2$ mirror-like duality obtained by introducing a real mass deformation that Higgses the $USp(2)$ gauge group to $U(1)$, while leaving the non-Abelian gauge symmetries of the mirror theory intact (bottom row). The spectrum of the \textit{electric} $U(1)$ theory is given in Equation~\ref{eq: spectrum_USp2_plan}; accordingly, the field $\Phi^+$ is represented by $\color{red}{\bigotimes}$, the field $p$ is represented by a double arrow because it transform in the fundamental of $U(F-4)$ and in the fundamental (charge $+1$) of the gauge node, while the remaining fields follow our standard conventions.}
    \label{fig: 3d_N=4_N=2_USp_U_Web}
\end{figure}

We start from the duality shown in Figure \ref{fig: 3d_N=4_USp_U_mirror}, focusing on $USp(2)$ with $F$ hypermultiplets on the \textit{electric} side. We then perform a real mass deformation breaking supersymmetry to $\mathcal{N}=2$ \textit{and} Higgsing the gauge group to $U(1)$. The resulting massless spectrum can be read off from Table \ref{eq: spectrum_USp2_plan}, while the full $\mathcal{N}=2$ duality is displayed in Figure \ref{fig: 3d_N=4_N=2_USp_U_Web}.

\begin{equation}\label{eq: spectrum_USp2_plan}
\begin{array}{c|c|c|c|c|c}
	 & U(1)_{gauge} & U(F-4)& U(2) & SO(4) & U(1)_R \\
	\hline
	\Phi^+ & +2 & \mathbf{1} & \mathbf{1}& \mathbf{1} & 2R \\
	q & +1 & \mathbf{1} & \overline{\Box}& \mathbf{1} & 1-R \\
	Q,\, \tilde{Q} & -1 &  \mathbf{1} & \mathbf{1} & \Box & 1-R  \\
    p & +1 & \Box & \mathbf{1} & \mathbf{1} & 1-R
\end{array}
\end{equation}
\newpage

\noindent Some comments on the duality shown in Figure \ref{fig: 3d_N=4_N=2_USp_U_Web} are in order: 

\begin{itemize}
 \item The coefficient of the mixing between the $U(1)_R$ symmetry and the topological $U(1)_{\text{top}}$ symmetry equals $\frac{iQ}{2}(1+\frac{2-F}{2}{R})$. Then, using Equation \ref{eq: monopole r contribution}, we can verify that the monopole $\mathfrak{M}^-$ has R-charge 2: \newpage

    \begin{equation}\label{eq: R_charge_usp2_planar}
    \begin{split}
        R[\mathfrak{M}^-] =&  1 + \frac{2-F}{2}R -\frac{1}{2} \cdot 4(1-R-1) - \frac{1}{2}\cdot 2(1-R-1) \\
        &-\frac{1}{2} \cdot (F-4)(1-R-1) - \frac{1}{2}\cdot 2 \cdot(2R-1) = 2,
    \end{split}
    \end{equation} and we can also verify that it is gauge invariant by means of Equations \ref{eq: monopole gauge contribution} and \ref{eq: monopole gauge charge}:

    \begin{equation}\label{eq: gauge_charge_usp2_planar}
        Q_{gauge}[\mathfrak{M}^-] = \left( 1 - \frac{F}{2} \right)(-1) + \frac{1}{2} \cdot 4 - \frac{1}{2} \cdot 2 - \frac{1}{2} \cdot (F-4) -\frac{1}{2} \cdot 2 \cdot 2=0. 
    \end{equation}

    \noindent Thus, the monopole $\mathfrak{M}^-$ must be present in the superpotential. This is expected, as a monopole potential is generated by the Polyakov mechanism \cite{Polyakov1977} when $USp(2)$ is Higgsed to $U(1)$. The monopole superpotential ensures that no independent topological symmetry survives in the electric theory.
    
    \item As also shown in Equation \ref{eq: spectrum_USp2_plan}, there is a field with gauge charge $+2$ in the quiver. This is a remnant of the adjoint field of $USp(2)$ that survives the Higgsing to $U(1)$.
    \item The $U(F-4)$ fields $p$ have gauge charge +1 but transform in the fundamental of $U(F-4)$; therefore, in our conventions, they are not bifundamental fields, which is why they are displayed with an arrow with arrowheads pointing in both directions. To make them ``true" bifundamental fields, one simply needs to change the signs of the corresponding fugacities in the $\mathbf{S}^3_b$ partition function, being careful to apply the same changes to the FI terms on the mirror dual. Upon this redefinition, the $U(F-4)$ and the $U(2)$ fields recombine to give a $U(F-2)$ symmetry group. 
    \item In Equation~\ref{eq: spectrum_USp2_plan}, the fields $Q$ and $\tilde{Q}$, corresponding to the $D_2$ node in Figure~\ref{fig: 3d_N=4_N=2_USp_U_Web}, have been written as transforming in the fundamental representation of $SO(4)$. It is not \emph{a priori} obvious that an $SO(4)$ global symmetry is present by inspecting the Higgsed $\mathbf{S}^3_b$ partition function alone, since the relevant term only encodes information about the weights of the representation. Consequently, there is a potential ambiguity in distinguishing, for instance, an $SO(4)$ flavor contribution from an $SO(2) \times SO(2)$ one in the matter sector. This ambiguity is resolved by the presence of a residual superpotential, inherited from the original $\mathcal{N}=4$ theory. For the present case, the gauge group is sufficiently small that the details of the Higgsing procedure can be explicitly worked out, allowing one to compute the surviving superpotential. Using the isomorphism $USp(2) \cong SU(2)$, we can represent the fields explicitly via the Pauli matrices. Up to an overall normalization, we then find:

    \begin{equation}\label{eq: USP2 superpotential}
        \mathcal{\mathcal{W}} = \left( \tilde{Q}^I_1 , \, \tilde{Q}^I_2 \right) \begin{pmatrix}
            \Phi_3 & \Phi^+ \\
            \Phi^- & -\Phi_3
        \end{pmatrix} 
        \begin{pmatrix}
            Q^1_I \\
            Q^2_I
        \end{pmatrix} = \tilde{Q}^I_1 \Phi_3 Q^1_I  - \tilde{Q}^I_2 \Phi_3 Q^2_I + \tilde{Q}^I_2 \Phi^- Q^1_I + \tilde{Q}^I_1 \Phi^+ Q^2_I.
    \end{equation} 
    Matching the partition function with the mirror theory constrains the choice of Higgsing: one must retain a field with charge $\pm 2$ under the unbroken $U(1)$ subgroup. Choosing the positive sign for definiteness, the only surviving component of $\Phi$ is $\Phi^+$. Consequently, in Equation~\ref{eq: USP2 superpotential} only the last term remains, while $\Phi_3$ and $\Phi^-$ acquire masses and are integrated out. Similarly, $\tilde{Q}^I_2$ and $Q^1_I$ are lifted. The fields $Q$ and $\tilde{Q}$ of Table~\ref{eq: spectrum_USp2_plan} are then identified with $Q^2_I$ and $\tilde{Q}^I_1$, respectively.

    To make the leftover symmetry group manifest, one can perform a change of basis analogous to Equation~\ref{eq: change of basis}, with the explicit choice $J^{12}= +1$ and $J^{21}=-1$:

    \begin{equation}\label{eq: change of basis example}
       \begin{split}
    iv^1_{2I-1} = \frac{1}{\sqrt{2}} \left( Q^1_I +  \tilde{Q}^I_2  \right), &\qquad v^1_{2I} =\frac{1}{\sqrt{2}} \left( Q_I^1 - \tilde{Q}^I_2 \right)\\
   iv_{2I-1}^2 = \frac{1}{\sqrt{2}} \left( Q_I^2 -  \tilde{Q}^I_1  \right), &\qquad v_{2I}^2 =\frac{1}{\sqrt{2}} \left( Q_I^2 + \tilde{Q}^I_1 \right).
   \end{split}
    \end{equation} 
    After Higgsing, only the $v^2$ component of each field survives. In this basis, the residual superpotential becomes
\begin{equation}\label{eq: higgsed superpotential}
\mathcal{W}_{\rm orth} = \frac{1}{2} v^2_I \Phi^+ v^2_I,
\end{equation}
which is manifestly $SO$ invariant. Since only two pairs of fields have the correct charge to couple to $\Phi^+$ after Higgsing, it is natural that an $SO(4)$ symmetry remains in the low-energy theory.

    \item From the previous analysis, we can state that the global symmetry breaking pattern upon Higgsing is $SO(2F) \rightarrow U(F-2) \times SO(4)$, up to possible discrete factors that we do not take into account. 
    Note that one could integrate out $\Phi^+$ and keep $\Phi^-$ massless. In this case, the surviving component would be $v^1$, but the $SO(4)$ symmetry would remain intact. 
    In general, several different chiral limits are possible on the mirror side depending on the relative shift in fugacities. Although the vacua are different, as can be inferred from the differences in divergent phases of the $\mathbf{S^3_b}$ partition function corresponding to such shifts, the resulting chiral quiver remains a D-type quiver. On the electric side, all these different vacua reflect the possible embeddings of the residual flavor symmetry group $U(F-2) \times SO(4)$ in the original flavor symmetry group $SO(2F)$. 
\end{itemize}
\newpage
 To illustrate all the features discussed above, we consider a simple example of the proposed duality, obtained for $F=5$. 
\begin{equation}
\label{eq: 3d_N=4_N=2_USp_U_F=5}
    \includegraphics[width=\textwidth]{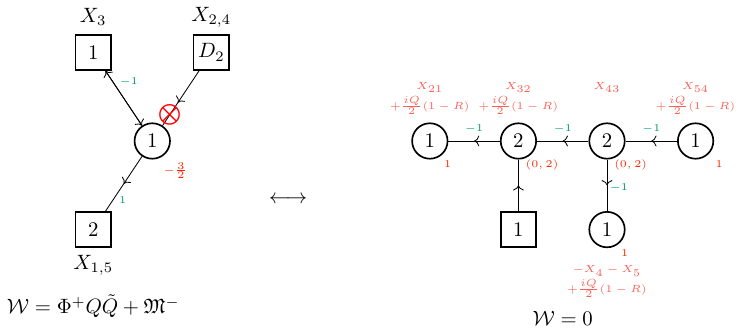}
\end{equation}

\noindent In equation \ref{eq: 3d_N=4_N=2_USp_U_F=5} we have maintained the roles the fields had in Figure \ref{fig: 3d_N=4_N=2_USp_U_Web}. To gain further insight into the symmetries of the theory, one can also compute the superconformal index for both sides of the duality. If we turn off all fugacities except the topological $\eta$, we find:

\begin{equation}\label{eq: index_usp2_planar}
  \mathcal{I}=  1+12 x^{8/7}+\left(\eta-16\right) x^2+54 x^{16/7} + O(x^{22/7}),
\end{equation}

\noindent where we fix $R=3/7$. It can be explicitly checked that this index is the same for both sides of the duality. Notice that the monopole with flux $-1$ correctly appears at order $x^2$. Due to the constraint imposed by the monopole superpotential $\mathcal{W}= \mathfrak{M}^-$, we are then forced to set $\eta = 1$, effectively eliminating one of the $16$ conserved currents corresponding to the topological symmetry.
If one wishes to keep track of all symmetries, it is possible to define fugacities for all flavors by setting $\phi_j := e^{iX_j}$. The coefficient of the term $-x^2$, which keeps track of the conserved currents, is then:

\begin{equation}\label{eq: index_conserved currents}
    \mathcal{I} \rvert_{-x^2} = \left(\phi _2 \phi _4+\frac{\phi _4}{\phi _2}+\frac{\phi _2}{\phi _4}+\frac{1}{\phi _2 \phi _4} + 2\right)+\left(\phi _3 \phi _1+\phi _3 \phi _5+\frac{\phi _5}{\phi _1}+\frac{1}{\phi _3 \phi _1}+\frac{\phi _1}{\phi _5}+\frac{1}{\phi _3 \phi _5} + 3 \right).
\end{equation}

\noindent The first summand can be traced back to the character of the adjoint representation of the Lie algebra $\mathfrak{so}(4) \cong \mathfrak{su}(2) \oplus \mathfrak{su}(2)$ by the change of fugacities $X_2 \rightarrow X_2 + X_4, \, \, X_4 \rightarrow X_4 - X_2$, which corresponds to the redefinition of exponentiated fugacities $\phi_2 \rightarrow \phi_2 \phi_4, \, \, \phi_4 \rightarrow \phi_4/\phi_2$. This change of variables disentangles and makes manifest the two $\mathfrak{su}(2)$ contributions. As for the second summand, it is not hard to see that upon sending $X_3 \rightarrow -X_3$ (i.e. $\phi_3 \rightarrow 1/\phi_3$) the character of the adjoint of $\mathfrak{u}(3)$ is reproduced. 

\begin{figure}[H]
    \centering
    \includegraphics[width=0.9\linewidth]{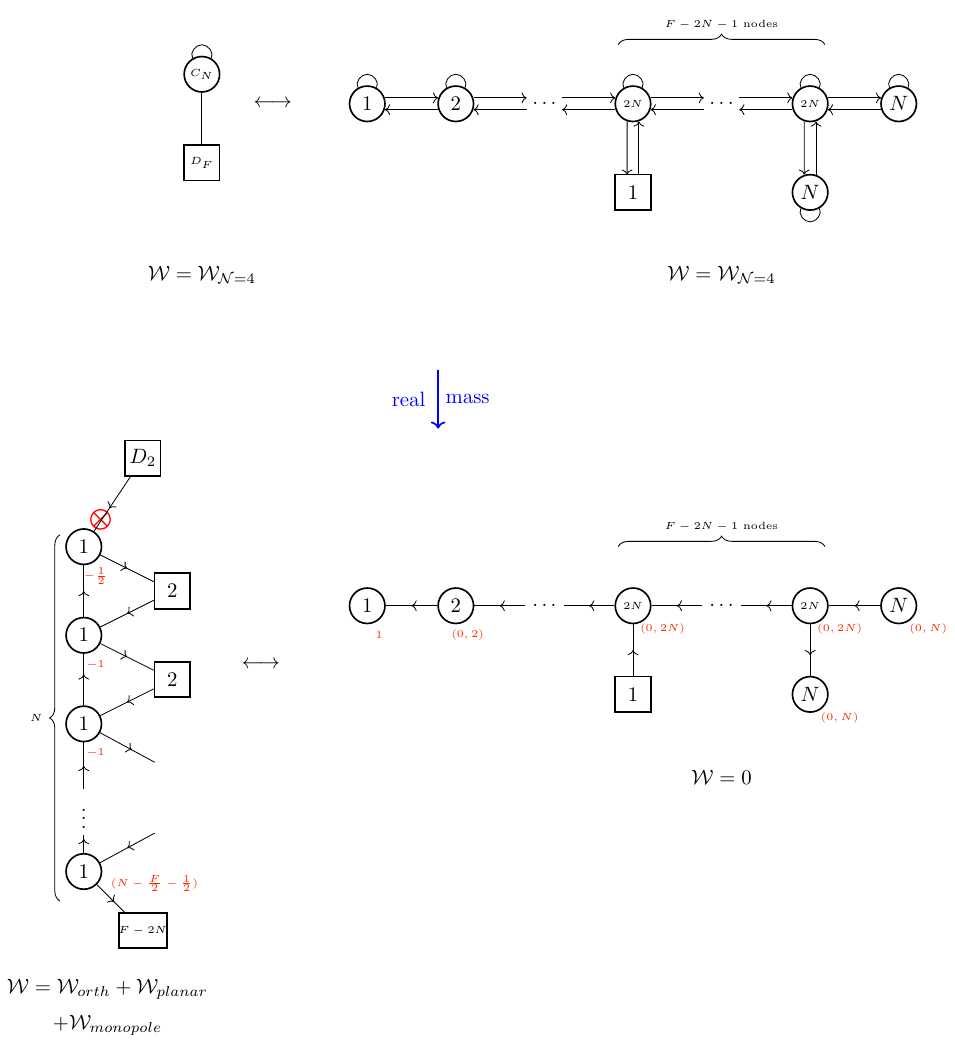}
    \caption{Schematic depiction of the $\mathcal{N}=2$ mirror-like duality obtained by introducing a real mass deformation that Higgses the $USp(2N)$ gauge symmetry to $U(1)^{N}$, while leaving the non-Abelian gauge symmetries of the mirror theory intact (bottom row). For clarity, FI terms and mixed CS interactions have been suppressed.}
    \label{fig: 3d_N=4_GenN_USp_U}
\end{figure}

\paragraph{The Planar Limit of \texorpdfstring{$USp(2N)$ SQCD$_3$ with $F$ fundamental multiplets}{USp(2N) SQCD3 with F}} The results of the previous discussion can easily be generalized to the case of $USp(2N)$. After considering a mass deformation analogous to the one described above, we reach the $\mathcal{N}=2$ duality displayed in figure \ref{fig: 3d_N=4_GenN_USp_U}.

\noindent Similarly to the $USp(2)$ case, there is an ``orthogonal" superpotential $\mathcal{W}_{orth}$ preserving the $SO(4)$ symmetry (analogous to Equation \ref{eq: higgsed superpotential}). In addition, there is a planar superpotential $\mathcal{W}_{planar}$ that contains a term for each closed triangle in the quiver. Both superpotentials are remnants of the $\mathcal{N}=4$ one. The monopole superpotential $\mathcal{W}_{monopole}$ is also modified accordingly, so that \emph{all} topological symmetries in the electric theory are broken.

We emphasize again that multiple chiral limits of the mirror are possible, depending on which fundamental and anti-fundamental fields become massive and are integrated out; these correspond to analogous deformations on the electric side that Higgs the gauge group. As before, the symmetry breaking pattern is independent of the particular fields that survive the deformation.

\section{Mapping Mass Flows to the Mirror Dual}\label{sec: mass flow}
The planar Abelian duals of symplectic SQCD constructed in Section \ref{sec: symplectic} arise from specific axial mass deformations of known $\mathcal N=4$ mirror pairs. Since these $\mathcal N=2$ duals descend from $\mathcal N=4$ theories, they exist only for a restricted range of parameters:
\begin{equation*}
    F = 2N+2+|k|,
\end{equation*}
where the CS level $k$ is fixed in terms of the rank and flavor content of the $\mathcal N=4$ parent. For values of $(F,k,N)$ that do not satisfy this relation, it is not \emph{a priori} guaranteed that the dual description will remain a planar Abelian quiver. Supersymmetry is broken for 
\begin{equation*}
    F < 2N+2-|k|;
\end{equation*}
this bound coincides with the regime in which the Aharony-like dual gauge group has a negative rank, signaling the absence of a supersymmetric vacuum \cite{Benini_2011a}.

The key observation of \cite{Benvenuti:2026a} is that \textbf{real mass deformations provide a systematic bridge} between $\mathcal N=4$ descendants and more general $\mathcal N=2$ theories. Real masses shift Chern--Simons levels, integrate out matter fields, and move the theory along controlled RG-flow trajectories in the $(F,k,N)$ parameter space. By tracking these deformations simultaneously in the non-Abelian theory and its planar mirror, we can follow how the quiver structure, CS interactions, and monopole superpotentials evolve, thereby extending planar Abelian duality beyond the original $\mathcal N=4$ descendants.\footnote{We remind the reader that a superfield transforming in a representation $\rho$ of a group $G$ has a positive (negative) mass if it shifts the CS level of $G$ by a positive (negative) amount upon being integrated out. Our main computational tool in this paper is the $\mathbf S^3_b$ partition function, which is built with double-sine functions $s_b(x)$; in terms of the asymptotics of these functions, a field is said to acquire a positive (negative) mass if the argument of the corresponding $s_b$ function is large and negative (positive).} 
The deformation of the planar Abelian dual corresponding to a given real mass deformation of the SQCD can be quite involved. For all such deformations studied in this paper we followed the real mass deformation on the $\mathbf{S}_b^3$ partition function and checked that the divergent prefactors cancel out between the electric and magnetic side. 
We do not present this analysis in detail and only report the resulting deformations.

\begin{figure}[H]
    \centering
\resizebox{.95\hsize}{!}{ 

\begin{tikzpicture}[scale=1.5,>=Stealth]

\draw[->,thick] (0,0) -- (6.5,0) node[right] {$F$};
\draw[->,thick] (0,0) -- (0,6.5) node[above] {$|k|$};

\node[left] at (0,3.5) {$2N+2$};
\node[below] at (3.5,0) {$2N+2$};


\draw[dashed,thick] (3.5,0) -- (0,3.5);

\draw[thick,blue] (.3,3.2) -- (3,5.9);  
\draw[thick,orange] (3.5,0) -- (6,2.5); 
\draw[thick,green] (3.5,0) -- (6,0);
\draw[thick,red] (0,3.5) -- (0,5.5);


\node at (5.3,0.6) {zone $1$};
\node at (3,2.5) { zone $2$};
\node at (.7,4.5) { zone $3$};

\node[rotate=0] at (0.9,1.2) { $\cancel{SUSY}$};

\begin{scope}[shift={(7,5)}]
    \node[anchor=west] at (0,0) {
    $
    \begin{array}{r|l}
     
        \multicolumn{2}{c}{\textbf{Mirror-like duality}}\\
        \hline
       \text{Zone 1}  & \eqref{eq: 3d_USp_gen_k}  \\
       \text{Zone 2}  & \eqref{eq: Planar_N=4D_kl2NNegM_Fin}  \\
       \text{Zone 3}  & \eqref{eq: Planar_N=4D_k_BF_Col_NegM_Fin}   \\
      
       \tikz{\path[draw,thick,blue](0,0)--(1,0);} & \eqref{eq: Planar_N=4D_k_2N+1NegM_Fin} \\
       \tikz{\path[draw,thick,orange](0,0)--(1,0);} & \eqref{eq: 3d_USp_gen_k_N=4}, \eqref{eq: 3d_USp_N=4_D} \\
       \tikz{\path[draw,thick,green](0,0)--(1,0);} & \eqref{eq: 3d_USp_k=0} \\
       \tikz{\path[draw,thick,red](0,0)--(1,0);} & \eqref{eq: Planar_N=4D_TQFT_BF_Col_NegM_Fin} \\
       \tikz{\path[draw,thick,dashed](0,0)--(1,0);} & \text{confinement; see Section \ref{sec: confinement}} 
    \end{array}
    $
    };

\end{scope}

\end{tikzpicture}
}
    \caption{Phase diagram of $USp(2N)_k$ SQCD with $F$ fundamental fields in the $(F,\,|k|)$-plane for fixed $N$. Each point in the plane with integer $F$ and $k$ corresponds to a different SQCD theory. The diagram is partitioned into various zones, within which the planar Abelian duals take qualitatively distinct forms. 
    Theories sitting on the orange line can be obtained from $\NN=4$ SQCD by turning on an axial mass. Theories sitting on the red line are TQFTs, while theories sitting on the dashed line s-confine.
    The blue line intercepts the vertical axis at $|k|=2N$.
    }
    \label{fig:KF_plane}
\end{figure}
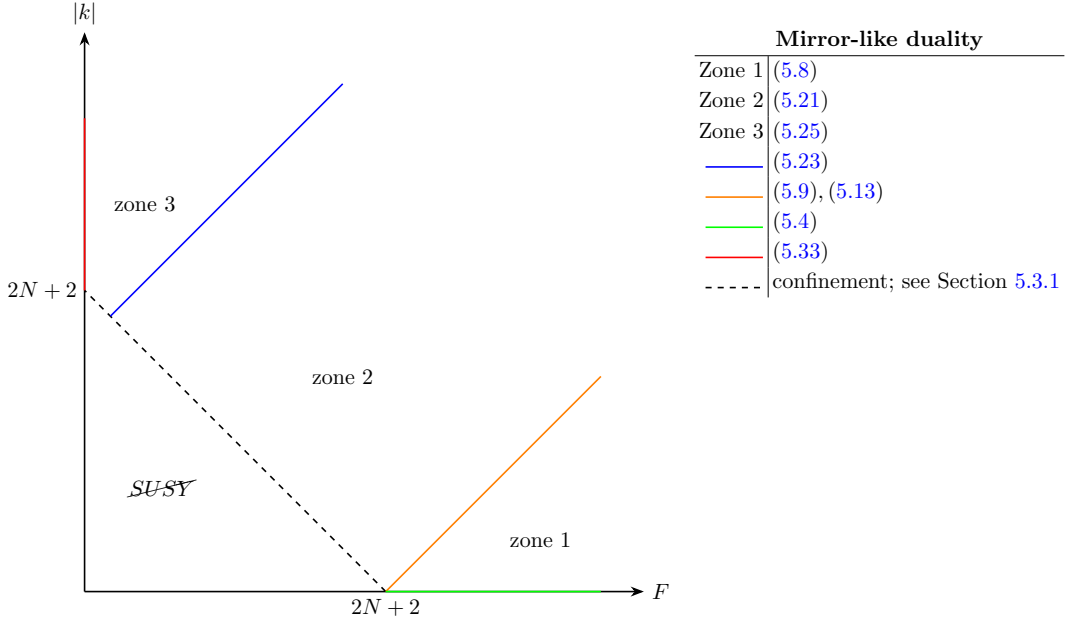

Analogous to \cite{Benvenuti:2026a}, we organize $\NN=2$ $USp(2N)_k$ SQCD$_3$ with $F$ chiral multiplets in a two-dimensional parameter space spanned by $(|k|,F)$ for fixed $N$. Each point in this plane with integer $k$ and $F$ corresponds to a distinct theory. In this paper, we focus on SQCD with fundamentals, with CS level $ k\leq0$. The case $k>0$ is analogous. We show this organization in Figure \ref{fig:KF_plane}.
\newpage

We extend this approach here to study $\mathcal N=2$ symplectic SQCD for generic ranks, flavors, and Chern--Simons levels, and then focus on a representative example to highlight the features and subtleties of this construction. Our main tool in this exercise is the $\mathbf S^3_b$ partition function - at each step, we are able to consistently match the asymptotic phases generated by the integrating out procedure, which provides strong evidence in favor of the dualities generated along the RG flow trajectories.

The planar Abelian dualities discussed in the previous section, denoted as $\NN=4$ descendants, apply to SQCD theories sitting on the orange line in Figure \ref{fig:KF_plane}.
By introducing additional real mass deformation we obtain planar Abelian duals for theories sitting above this line, namely in Zones 2 and 3. 
Furthermore, we propose planar Abelian duals for theories in Zone 1 and on the green line $k=0$. 
We do not provide a direct derivation of these latter dualities via real mass deformation, but we provide extensive checks of their validity. 
Overall, we present a planar Abelian dual for all $USp(2N)$ CS-SQCD theories with fundamental matter.
\\

For the remainder of this section, we adopt a compact notation for the Abelian planar quivers studied here, since most of the information required to define the $\NN=2$ planar Abelian quivers is encoded in the shape of the quiver itself. The compact quiver notation used here was first introduced in \cite{Benvenuti:2026a}, we summarize it here for completeness:

\begin{equation}    \label{eq:quiv:notation_compact}
    \begin{array}{|c|c|}
        \hline
        \text{\textbf{Quiver notation}} & \text{\textbf{Field theory}} 
        \\ 
        \hline
        \begin{tikzpicture}[baseline=(current bounding box).center]
        \node[gaugefill,black] at (0,0) (g1) {}; 
        \draw[gray] (g1) ++(-0.5,0) node {$u$};
        \end{tikzpicture}
        &  \begin{array}{c} U(1) \; \NN=2 \text{ vector multiplet}\end{array}
        \\ 
        \hline
        \begin{tikzpicture}[baseline=(current bounding box).center]
        \node[flavorfill,black] at (0,0) (g1) {}; 
        \draw[gray] (g1) ++(-0.5,0) node {$X$};
        \end{tikzpicture}
        &  \begin{array}{c} \text{$U(1)$ Flavor node}\end{array}
        \\ 
        \hline
        \begin{tikzpicture}[baseline=(current bounding box).center,scale=0.7]
        \node[gaugefill,gray] at (0,1) (g1) {}; 
        \node[gaugefill,gray] at (1,0) (g2) {}; 
        \draw[->-,black] (g1) --  (g2);
        \draw[gray] (g1) ++(-0.5,0) node {$u$};
        \draw[gray] (g2) ++(0.5,0) node {$v$};
        \end{tikzpicture},
        \begin{tikzpicture}[baseline=(current bounding box).center,scale=0.7]
        \node[gaugefill,gray] at (1,1) (g1) {}; 
        \node[gaugefill,gray] at (0,0) (g2) {}; 
        \draw[->-,black] (g1) --  (g2);
        \draw[gray] (g1) ++(-0.5,0) node {$u$};
        \draw[gray] (g2) ++(0.5,0) node {$v$};
        \end{tikzpicture}
        &  \begin{array}{c} \text{bifundamental chiral with R-charge $1-R$}  \text{ and mixed CS level $+1$}  \end{array}
        \\ 
        \hline
        \begin{tikzpicture}[baseline=(current bounding box).center,scale=0.7]
        \node[gaugefill,gray] at (0,1) (g1) {}; 
        \node[gaugefill,gray] at (0,0) (g2) {}; 
        \draw[->-,black] (g1) --  (g2);
        \draw[gray] (g1) ++(-0.5,0) node {$u$};
        \draw[gray] (g2) ++(0.5,0) node {$v$};
        \end{tikzpicture}
        &  \begin{array}{c} \text{bifundamental chiral with R-charge $2-2R$}  \text{ and mixed CS level $+1$}  \end{array}
        \\ 
        \hline
        \begin{tikzpicture}[baseline=(current bounding box).center,scale=0.7]
        \node[gaugefill,gray] at (0,1) (g1) {}; 
        \node[gaugefill,gray] at (1,0) (g2) {}; 
        \draw[-<-,black] (g1) --  (g2);
        \draw[gray] (g1) ++(-0.5,0) node {$u$};
        \draw[gray] (g2) ++(0.5,0) node {$v$};
        \end{tikzpicture},
        \begin{tikzpicture}[baseline=(current bounding box).center,scale=0.7]
        \node[gaugefill,gray] at (1,1) (g1) {}; 
        \node[gaugefill,gray] at (0,0) (g2) {}; 
        \draw[-<-,black] (g1) --  (g2);
        \draw[gray] (g1) ++(-0.5,0) node {$u$};
        \draw[gray] (g2) ++(0.5,0) node {$v$};
        \end{tikzpicture}
        &  \begin{array}{c} \text{bifundamental chiral with R-charge $R$}  \text{ and mixed CS level $-1$}  \end{array}
    
        \\ 
        \hline
        \begin{tikzpicture}[baseline=(current bounding box).center,scale=0.7]
        \node[gaugefill,gray] at (0,1) (g1) {}; 
        \node[gaugefill,gray] at (0,0) (g2) {}; 
        \draw[-<-,black] (g1) --  (g2);
        \draw[gray] (g1) ++(-0.5,0) node {$u$};
        \draw[gray] (g2) ++(0.5,0) node {$v$};
        \end{tikzpicture}
        &  \begin{array}{c} \text{bifundamental chiral with R-charge $2R$}  \text{ and mixed CS level $-1$}  \end{array}
        
        \\ 
        \hline
        \begin{tikzpicture}[baseline=(current bounding box).center]
        \node[gaugefill,gray] at (0,0) (g1) {$1$}; 
        \node[gaugefill,gray] at (1.5,0) (g2) {$1$}; 
        \draw[BFline,black] (g1) --  (g2);
        \draw[gray] (g1) ++(-0.5,0) node {$u$};
        \draw[gray] (g2) ++(0.5,0) node {$v$};
        \end{tikzpicture}
        &  \text{Mixed CS term at level $-2$}
      
        \\ \hline
    \end{array}
\end{equation}
Additionally, each gauge or flavor node $U(1)_i$ carries a Chern--Simons term at level $k_i$, determined by the mixed CS couplings through
\begin{equation}
    k_i \;=\; -\frac{1}{2}\sum_{j\neq i} k_{ij},
\end{equation}
which contributes to the localized partition function as
\[
    e^{- \pi i\, k_i\, u_i^2}.
\]

For symplectic theories, assigning a positive or negative real mass $\mposneg$ removes one chiral field from the spectrum and shifts the CS level by $\pm1$ without affecting the rank of the gauge group. On the $(k,F)-$plane these correspond to diagonal moves toward the left:
    \begin{equation}
\begin{tikzpicture}[baseline=(current bounding box).center]
\node[label=right:{SQCD}] at (0,0) (SQCD) {$\bullet$};
\node[] at (-1,1) (mm) {$\mneg$};
\node[] at (-1,-1) (pm) {$\mpos$};
\path[draw,->] (SQCD) -- (mm);
\path[draw,->] (SQCD) -- (pm);
\end{tikzpicture}
\end{equation} 
    which can be stated more succinctly: 
    \begin{equation*}
        (F,k)\overset{\mposneg}{\longrightarrow} (F-1, k\pm1).
    \end{equation*}

As such, for the remainder of this section, we will study the effect of these $\mposneg$ deformations on the electric theory and map their effect to the planar dual. 

\subsection{A Dual for \texorpdfstring{USp(2N)$_0$ SQCD with $F\geq2N+2$ $\Box$ and its Deformations}{USp(2N)0 with F}}
We begin by proposing an Abelian quiver gauge theory that is infrared dual to $USp(2N)_0$ SQCD with $F$ fundamental multiplets, where $F\geq2N+2$. Consistency of Chern--Simons level quantization requires $F$ to be even. 

This proposal is motivated by three complementary considerations. First, its structure closely parallels the planar Abelian duals of $U(N)_0$ SQCD with $F$ fundamental chiral multiplets proposed in \cite{Benvenuti:2026a}. Second, we have verified agreement of superconformal indices for several low-rank and low-flavor examples, providing nontrivial evidence for the duality. Finally, upon performing appropriate real mass deformations, the proposed Abelian quiver correctly flows to the known $\mathcal N=4$ descendant theories discussed in earlier sections.

We propose the dual shown in Equation \ref{eq: 3d_USp_k=0}.

\begin{equation}
\label{eq: 3d_USp_k=0}
    \includegraphics[width=\textwidth]{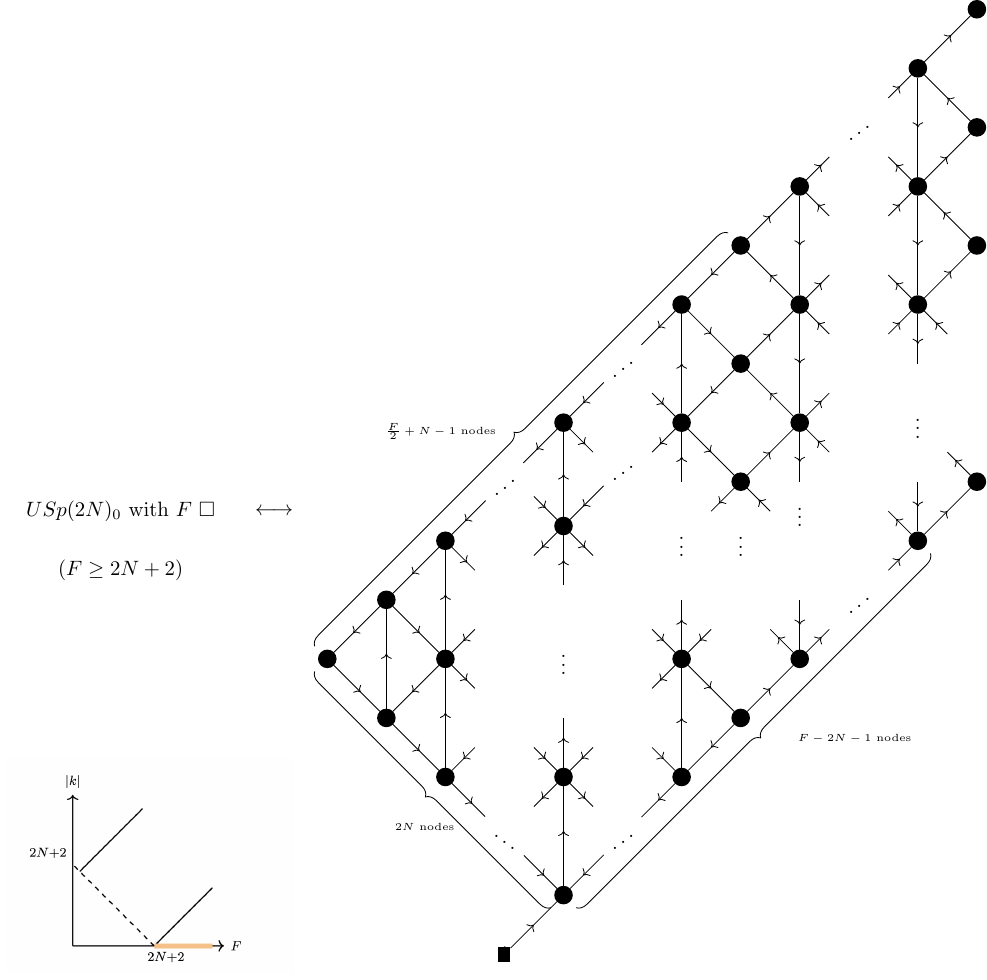}
\end{equation}

Here and in the rest of the paper we draw on the bottom left the region of the plane parametrized by $F$ and $k$ where the duality holds.
Across the duality, flavor and topological symmetries are exchanged, which is a characteristic feature of mirror-like dualities. We comment on some additional properties of these duals:
\begin{itemize}
    \item $\mathcal{W}_{\text{planar}}$ contains one cubic (quartic) term for each triangular (square) face of the quiver, with sign $-1$ ($+1$) if the arrows run clockwise (anticlockwise) around the face.
    \item For every column except for the right-most one, $\mathcal{W}_{\text{monopole}}$ includes linear monopole superpotentials involving monopoles with GNO fluxes under two adjacent nodes within the same column. Specifically, there is a monopole $\mathfrak{M}^{\bigg(\begin{array}{c} +\\-\end{array}\bigg)}$ for each pair of adjacent nodes.
    There are $2N-2$ additional monopole interactions of the form considered in Equation \ref{eq: spicy monopole}.
    
     The FI terms of the $U(1)$ gauge nodes are compatible with the monopole superpotential terms and are \textbf{not arbitrary}.
     \item The assignment of R-charges and CS levels follows the general conventions established in Equation \ref{eq:quiv:notation_compact}.
\end{itemize}
Overall, the information required to fully specify the planar Abelian theory is completely encoded in the quiver diagram. 

\subsubsection*{Mapping Multiple Negative Mass Deformations}
Starting from the duality between $USp(2N)_0$ SQCD with $F$ fundamental multiplets and its planar Abelian dual (see Equation \ref{eq: 3d_USp_k=0}), we now introduce negative real mass deformations and study the corresponding RG flows in both descriptions.

On the “electric” side, this deformation drives the theory to $USp(2N)_{-1}$ SQCD with $F-1$ fundamental multiplets. The analysis of the RG trajectory in the planar Abelian dual is more subtle. We propose that the corresponding deformation is implemented by Higgsing all the chiral fields along the bottom-left diagonal, while the chiral field connecting the bottom-most gauge and flavor nodes acquires a positive mass. This Higgsing identifies all gauge nodes along the diagonal with the final node, and the planar superpotential correspondingly reduces to mass terms.

\newpage
In quiver notation, we propose the duality shown in Equation \ref{eq: 3d_USp_Negm}.
\begin{equation}
\label{eq: 3d_USp_Negm}
    \includegraphics[width=\textwidth]{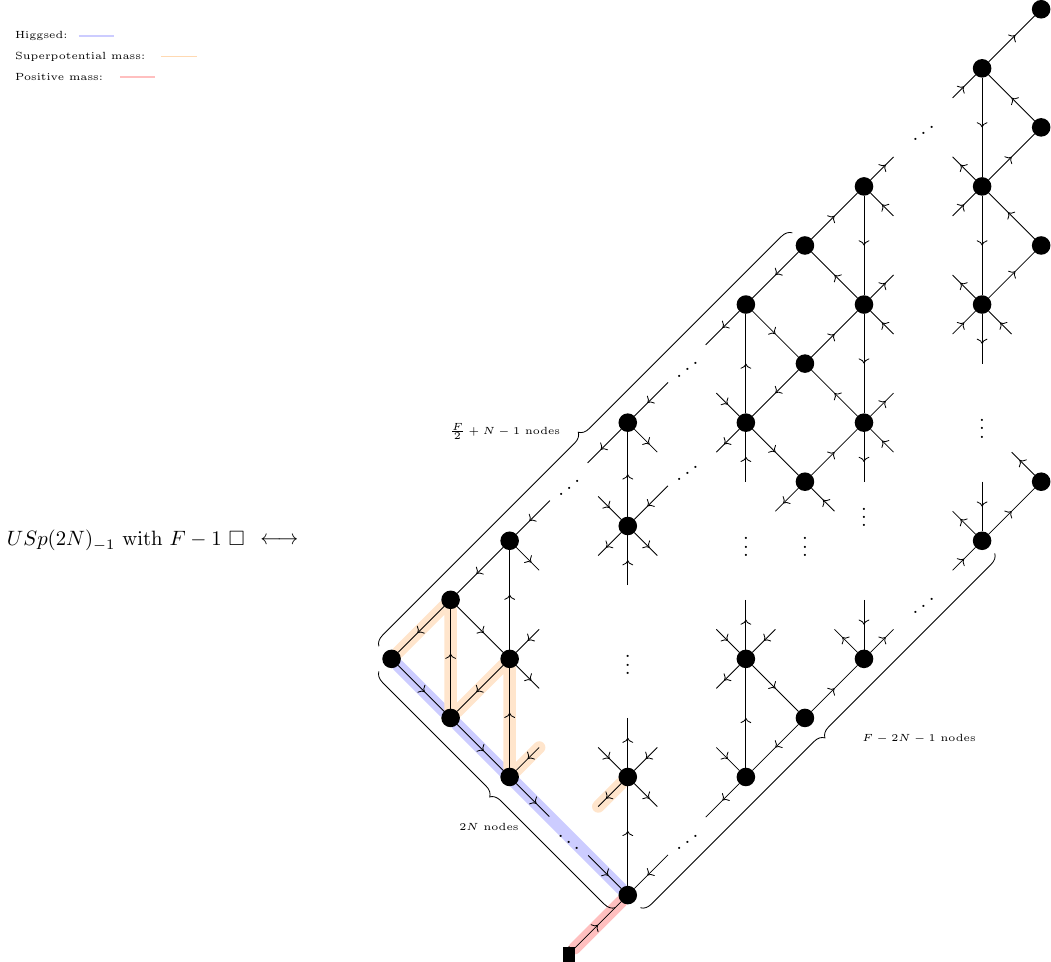}
\end{equation}

Here, the chiral fields shown in blue are those that acquire VEVs, while those in orange become  massive due to the planar superpotential. The chiral field highlighted in pink acquires a positive mass.

After this Higgsing, the planar superpotential reduces to mass terms for the fields, as 
\newpage
illustrated in Equation \ref{eq: Higgsing_diag_separate}:
\begin{equation}\label{eq: Higgsing_diag_separate}
\includegraphics[width=1\linewidth]{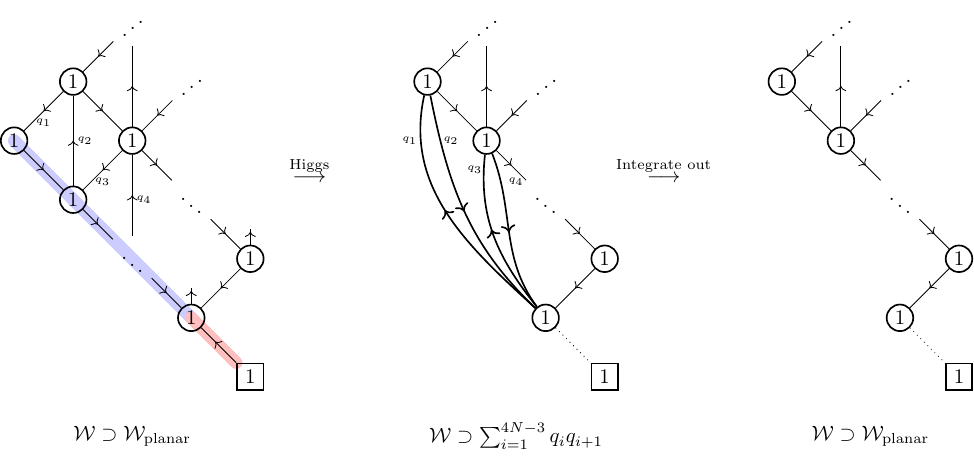}
\end{equation}

Upon integrating out the massive fields, the $U(1)$ gauge node at the bottom of the quiver becomes $s$-confining. We can therefore apply a local duality to confine this node, yielding an equivalent description.

As a result, we obtain the planar Abelian dual of $USp(2N)_{-1}$ SQCD with $F-1$ fundamental multiplets, shown in Equation~\ref{eq: 3d_USp_k=-1}.

\newpage
\begin{equation}
\label{eq: 3d_USp_k=-1}
    \includegraphics[width=\textwidth]{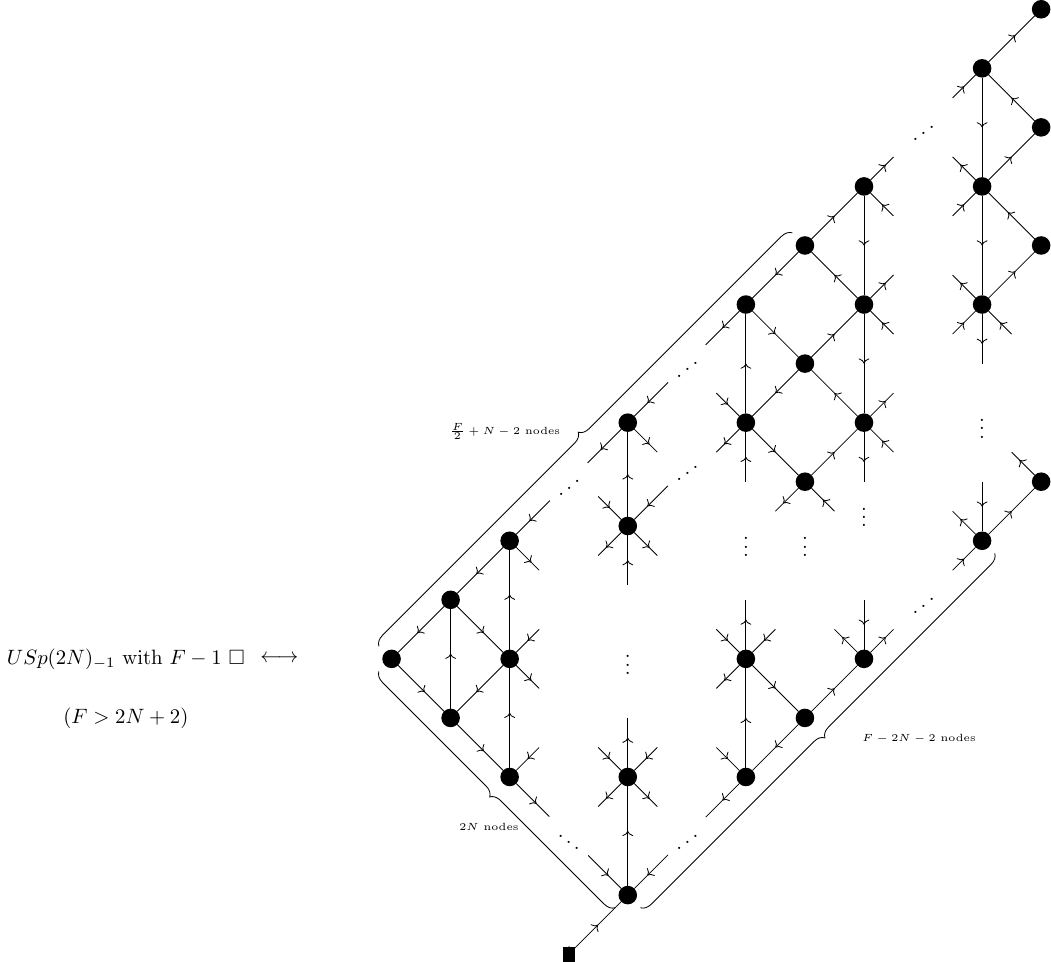}
\end{equation}

The resulting dual quiver closely resembles the Abelian dual before the mass deformation. 
Indeed, locally in the quiver the two theories present similar type of matter content and interactions. 
The difference lies in the overall dimensions of the quiver itself: the long side of the quiver is one node shorter than before and the relative position of the “column of squares” changed as well.
A similar trend was observed in the case of Unitary SQCD in \cite{Benvenuti:2026a}: the set of SQCD theories can be divided into "Zones", as in Figure \ref{fig:KF_plane}, and within each Zone the planar Abelian duals have the same qualitative structure and are differentiated by their sizes (or position of relevant features such as the “column of squares”).
Here we find that the case of Symplectic SQCD is analogous, and the partition of the space of SQCD theories into Zones in Figure \ref{fig:KF_plane} is useful in this case as well.

\newpage
We can introduce
further negative mass deformations. These are mapped analogously to the planar Abelian dual, leading to the duality for $USp(2N)_k$ SQCD with $F$ fundamental multiplets shown in Equation \ref{eq: 3d_USp_gen_k}.

\begin{equation}
\label{eq: 3d_USp_gen_k}
    \includegraphics[width=\textwidth]{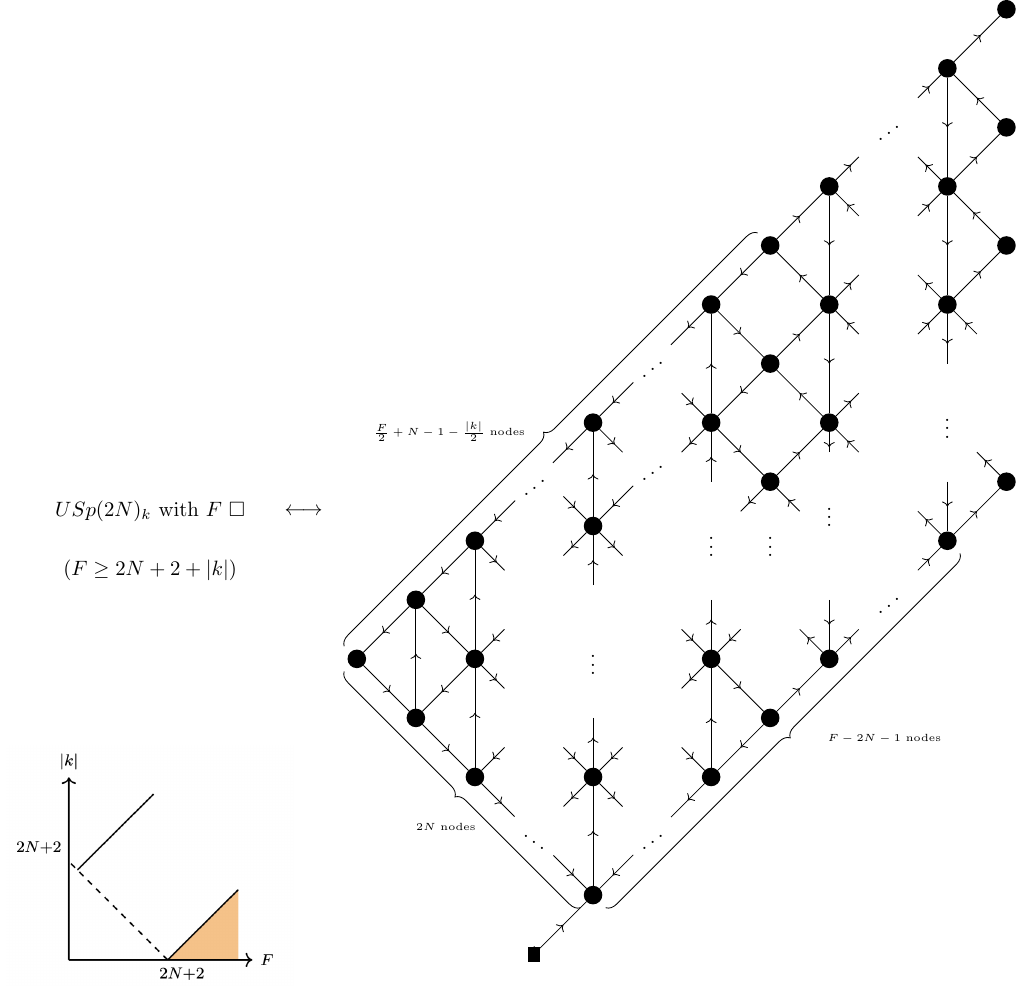}
\end{equation}
\newpage
In this way, we can follow the RG flow trajectories induced by multiple negative mass deformations, wherein the position of the column of squares shifts towards the left each time. The limiting case sitting in Zone 1, namely $USp(2N)_{2N+2-F}$ SQCD with $F$ fundamental multiplets, is shown in Equation \ref{eq: 3d_USp_gen_k_N=4}.

\begin{equation}
\label{eq: 3d_USp_gen_k_N=4}
    \includegraphics[width=\textwidth]{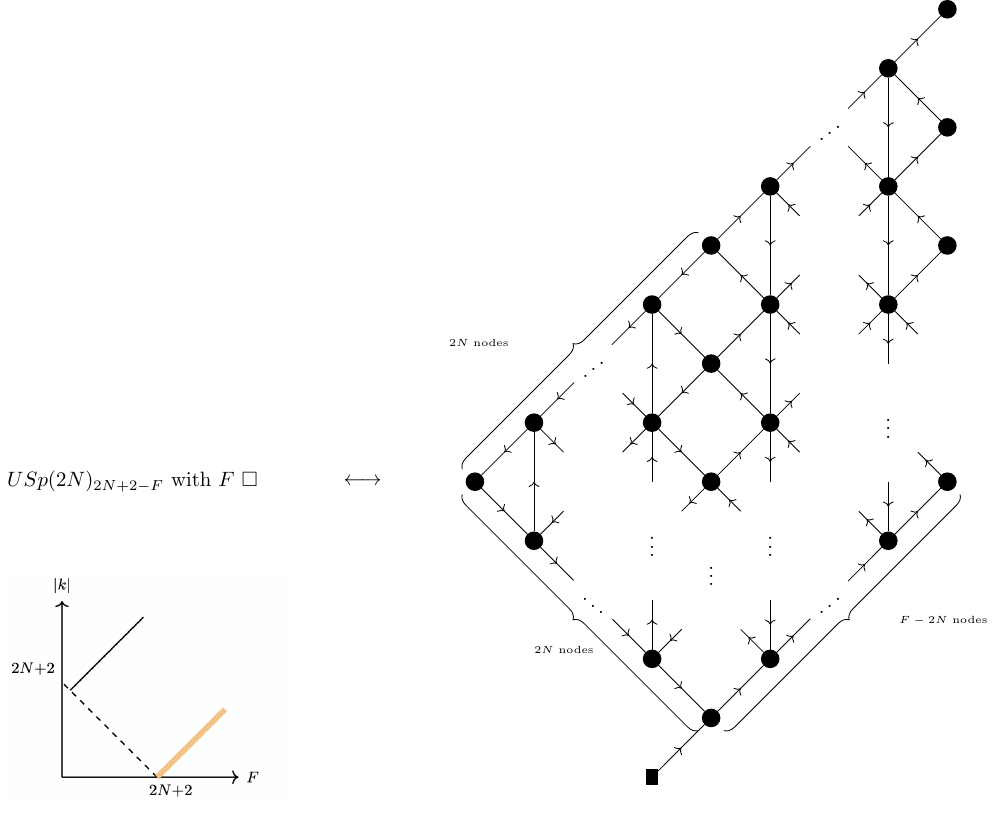}
\end{equation}

We remind the reader that this theory is an $\NN=4$ descendant. The equivalence of the presentation of this quiver with the standard form of $\NN=4$ descendant theories (see Figure \ref{fig: 3d_N=2_USp_Web}) can be shown via a sequence of local dualizations following the procedure outlined in Appendix~B of \cite{Benvenuti:2026a}. Specifically, starting from the quiver in Equation \ref{eq: 3d_USp_gen_k_N=4}, one dualizes each column from left to right, stopping at the $(2N-1)$-th column, then repeats the process up to the $(2N-2)$-th column, and so on. After performing this sequence $2N-1$ times, the resulting quiver matches the planar Abelian dual of the $\NN=4$ descendant SQCD. 

For concreteness, we demonstrate the effect of a negative mass deformation on the duality shown in Equation \ref{eq: usp4_8_squares}. The resulting duality between $USp(4)_{-1}$ SQCD with $7$ fundamental fields is shown in Equation \ref{eq: usp4_7_squares}.

\begin{equation}
\label{eq: usp4_7_squares}
    \includegraphics[width=.7\linewidth]{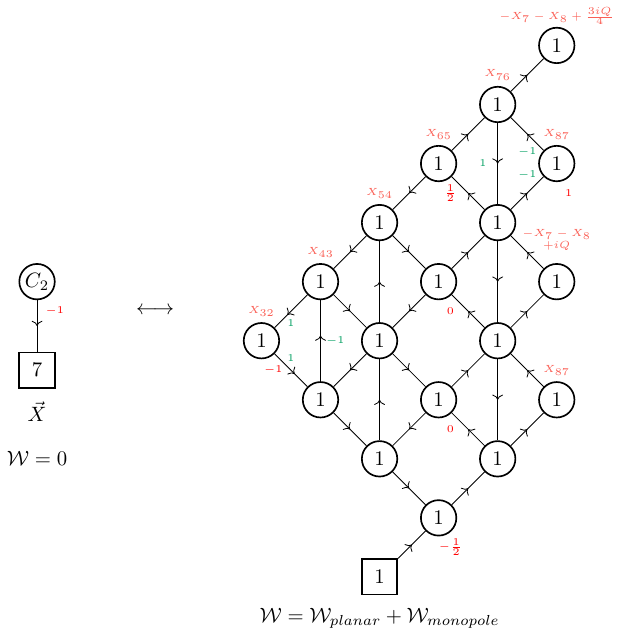}
\end{equation}

We recognize the dual pair shown in Equation \ref{eq: usp4_7_squares} as an $\NN=4$ descendant pair; as such, we can recast it in a more familiar presentation via a sequence of local dualizations schematically summarized in Equation \ref{eq: usp4_7_squares_dualization}.

\begin{equation}
\label{eq: usp4_7_squares_dualization}
    \includegraphics[width=\linewidth]{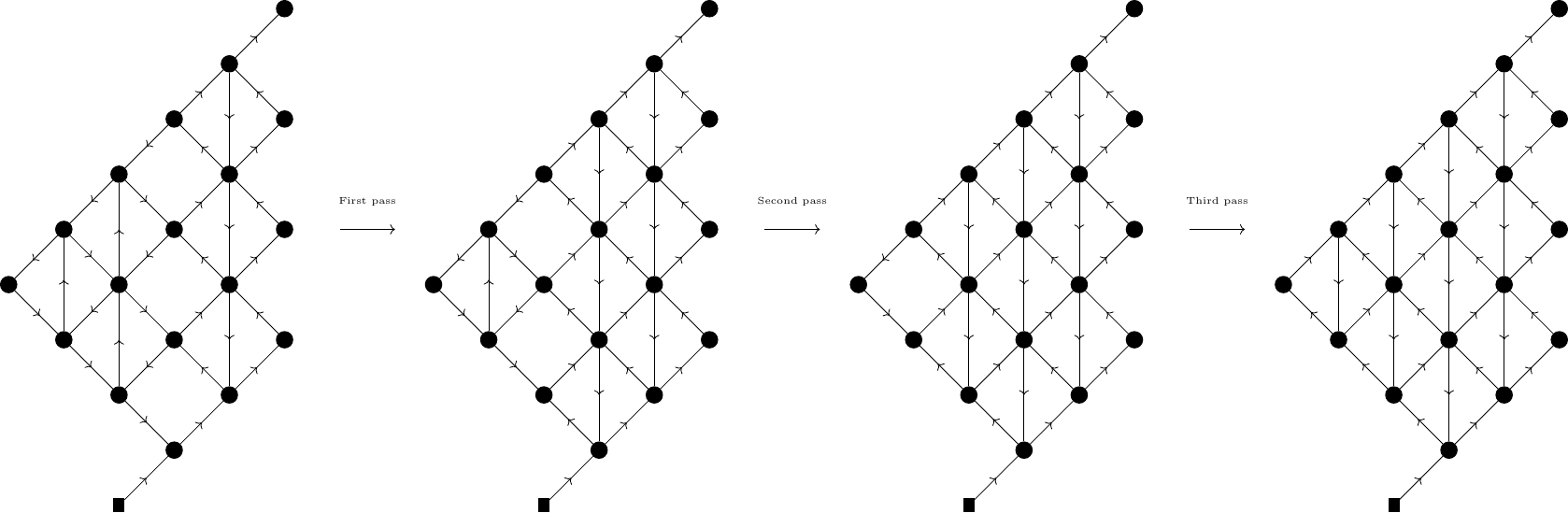}
\end{equation}

Before moving on to the study of further mass deformations let us discuss briefly the analogous problem in the case of $U(2N)$ SQCD with an antisymmetric. 
Similar to the case of $USp(2N)$ SQCD discussed here and the case of $U(2N)$ SQCD analyzed in \cite{Benvenuti:2026a} one can study mass deformations for fundamental matter in the presence of an antisymmetric tensor.
We note that the deformations discussed up to now only affect the left region of the planar Abelian dual, while the planar duals of $USp(2N)$ and $U(2N)$ with antisymmetric only differ by a fundamental field in the right region of the quiver. 
Therefore one can analyze such mass deformations in the case of $U(2N)$ with antisymmetric in a completely analogous way, and the process of giving a VEV to the antisymmetric commutes naturally with these mass deformations, even when mapped to the planar Abelian dual side. 
We summarize this connection between the planar mirror dual of $U(4)_{(0,-6)}$ SQCD with a rank-2 antisymmetric tensor and $8$ fundamental chiral multiplets, and $USp(4)_0$ SQCD with $8$ fundamental multiplets in Equation \ref{eq: u4_usp4_web}.

One can study the planar Abelian dual of $U(2N)_k$ with $F$ fundamentals and a rank-2 antisymmetric for generic values of $N,F$ and $k$ with the techniques used here. We expect that the planar Abelian dual of such theories will be identical to the corresponding dual for $USp(2N)$ SQCD, with an additional fundamental chiral on the right side of the quiver. There may be some exceptions to this expectation when considering edge cases, such as the case $F=0$, but we do not discuss them here in detail.
\newpage

\begin{equation}
\label{eq: u4_usp4_web}
    \includegraphics[width=.9\linewidth]{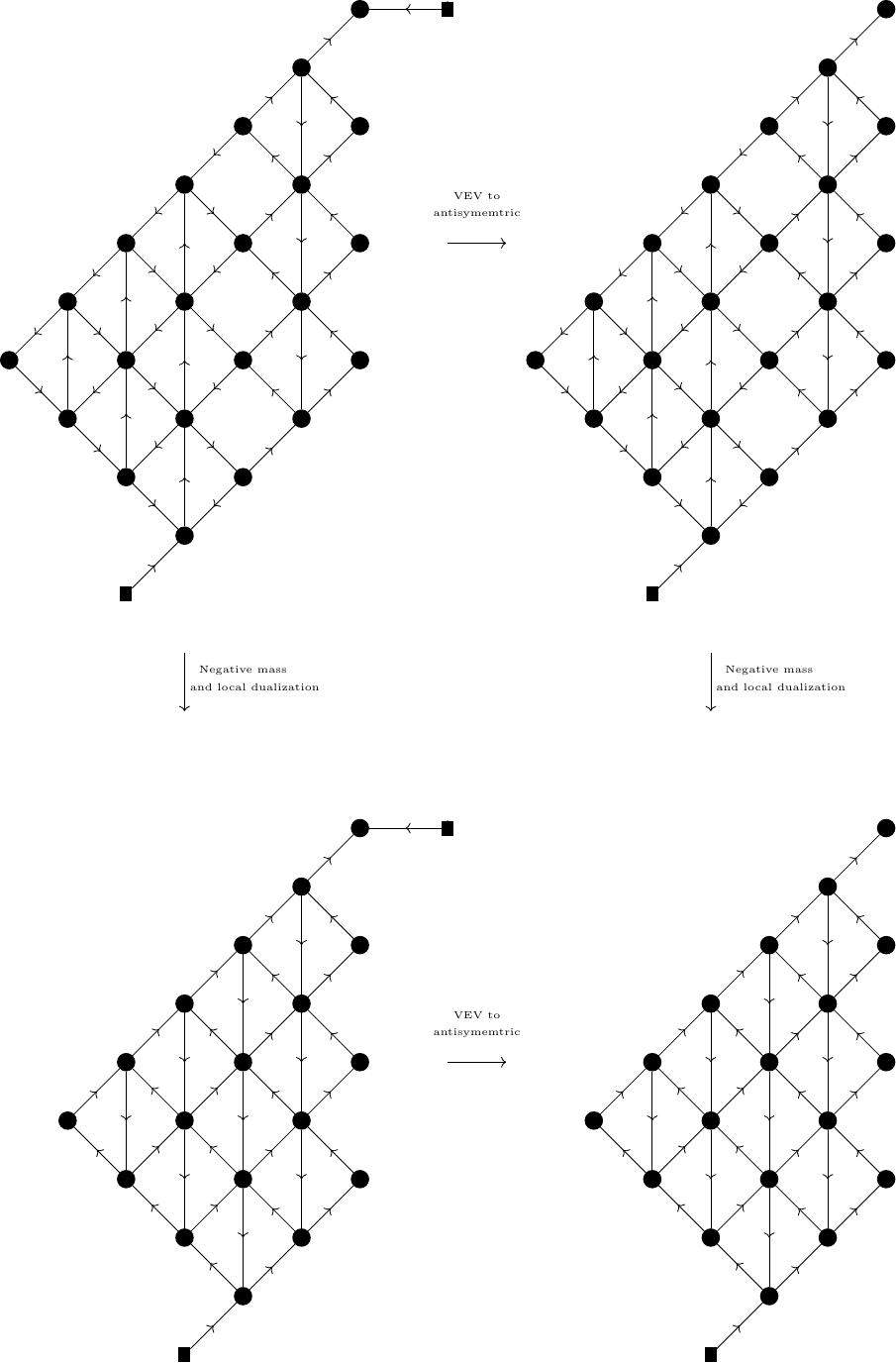}
\end{equation}

Thus, our analysis of $USp(2N)$ theories naturally sheds light on
$U(2N)$ gauge theories with a rank-2 antisymmetric tensor. Understanding how to construct planar mirror duals for $USp(2N)_k$ SQCD with
arbitrary $F$ naturally informs the corresponding construction for
$U(2N)_{(\tilde k,\tilde k-3N)}$ SQCD with $F$ fundamentals and a rank-2 antisymmetric tensor.
This provides an \emph{a posteriori} justification for our choice to begin from
the $U(2N)$ theory in Section \ref{sec: antisymmetric}.

\subsection{Mass Deformations of the \texorpdfstring{$\NN=4$ Descendant}{N=4 Descendant}}\label{subsec: n=4_d}
We now consider massive deformations of the planar dual of $USp(2N)_{(2+2N-F)}$ SQCD with $F$ fundamental fields, shown in Equation \ref{eq: 3d_USp_N=4_D}. 
We remind the reader that this is an $\NN=4$ descendant theory.
This allows us to derive a planar Abelian dual for SQCD theories sitting in Zones 2 and 3 in Figure \ref{fig:KF_plane}, which is the main goal of this Section.

\begin{equation}
\label{eq: 3d_USp_N=4_D}
    \includegraphics[width=.9\linewidth]{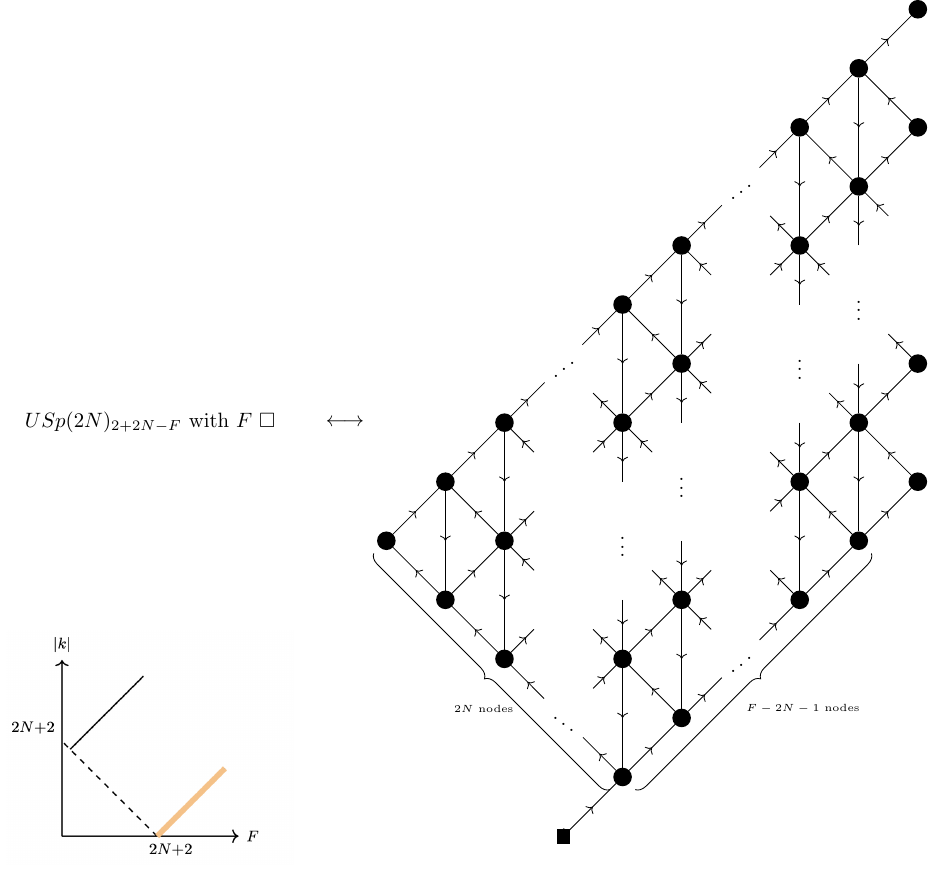}
\end{equation}

We have the possibility of studying the RG flow trajectories induced by both positive and negative mass deformations $\mposneg$ for these theories. Across the duality, a real mass for a quark in SQCD maps to an FI term for the gauge nodes contained in a single column of the planar Abelian quiver.

Consequently, giving a real mass to the $i$-th quark corresponds to turning on a large FI parameter for one (or two) adjacent columns of the Abelian quiver. Although different choices of $i$ are equivalent at the level of the IR-SCFT, being related by Weyl transformations of the $U(F)$ flavor symmetry, this symmetry is not manifest in the planar quiver Lagrangian. As a result, the UV analysis of the deformation depends on the specific choice of $i$, even though the corresponding IR deformations describe the same flow.

\begin{figure}
    \centering
    \includegraphics[width=.8\linewidth]{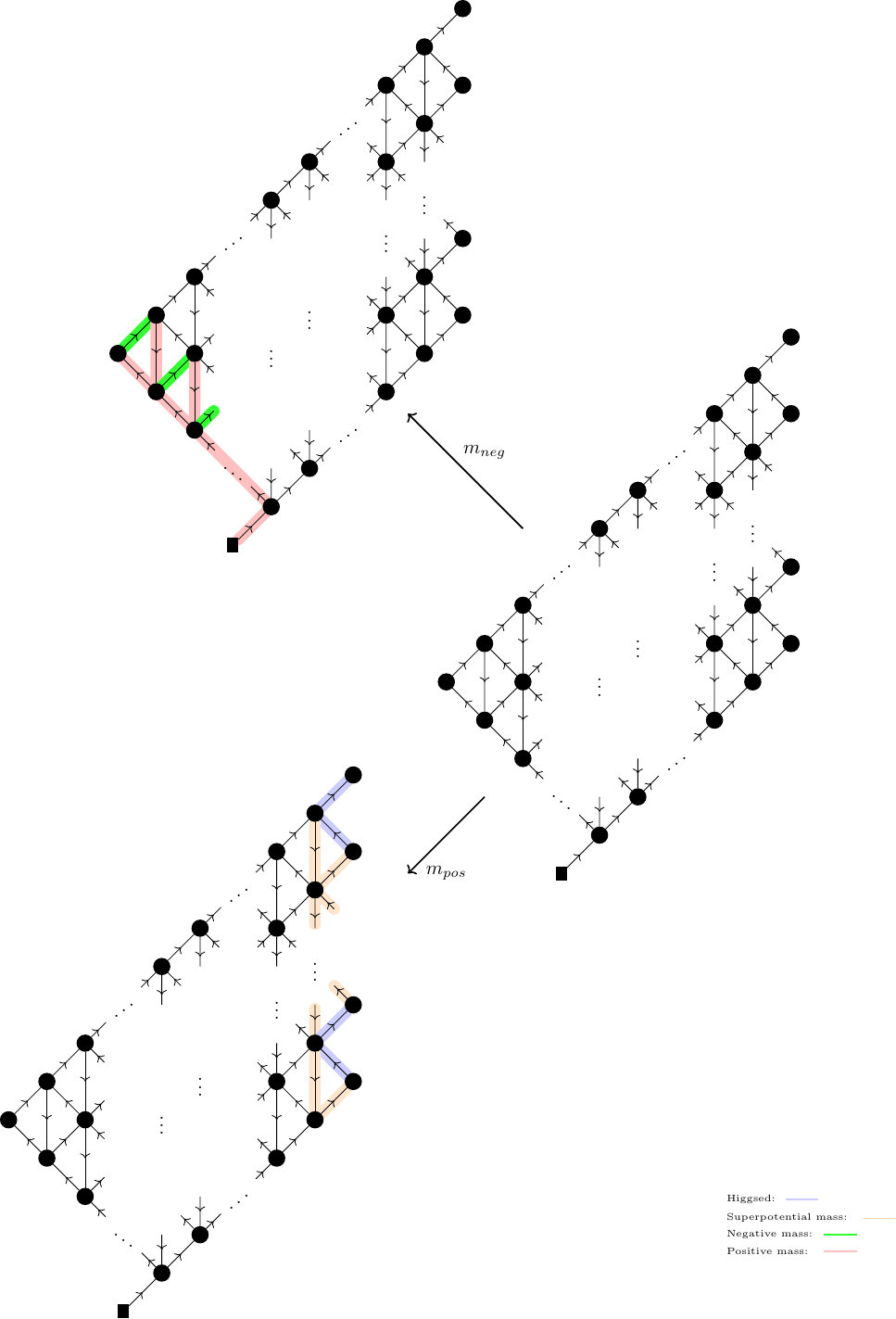}
    \caption{Deformations of the planar Abelian quiver dual to $USp(2N)_{(2+2N-F)}$ SQCD with $F$ fundamentals.
    Chiral fields highlighted in pink acquire a large positive mass, fields highlighted in green acquire a large negative mass, and fields highlighted in blue acquire a VEV, thereby Higgsing the nodes that they connect to a diagonal $U(1)$.
    }
    \label{fig: Def_Planar_Start}
\end{figure}

For concreteness, we adopt the following conventions for the two possible deformations:
\begin{equation}
\begin{array}{ll}
\mpos: & \text{ large mass for the last quark } Q^F \leftrightarrow X_F \to -\infty
\\
 \mneg: & \text{ large mass for the first quark } Q^1 \leftrightarrow  X_1 \to +\infty
\end{array}
\end{equation}
Therefore, the deformation $\mpos$ corresponds to evaluating large FIs limits for the nodes in the rightmost column in the quiver, while the deformation $\mneg$ corresponds to a large FI limit for the leftmost node. We show these flows in Figure \ref{fig: Def_Planar_Start}.

\newpage
We identify the following vacua corresponding to these deformations:
\begin{itemize}
    \item[$\mpos$: ] The $\mpos$ deformation has the effect of Higgsing the ``saw"-like structure at the end of the quiver, as shown in Figure \ref{fig: Def_Planar_Start}. There, the chiral fields in blue acquire Higgs VEVs, while the planar superpotential reduces to mass terms for the other fields (refer to Equation \ref{eq: Higgsing_diag_separate}).

    \item[$\mneg$: ] In this deformation, the real scalars in the vector multiplets associated with the gauge groups along the bottom-left diagonal acquire large VEVs. As a consequence, the chiral multiplets connecting these nodes acquire large positive real masses, while those connecting them to the rest of the quiver acquire negative real masses. These fields are highlighted in green in Figure \ref{fig: Def_Planar_Start}. 

    Integrating out the massive fields yields the quiver shown in Equation \ref{eq: Planar_N=4D_NegM}.
    \begin{equation}
    \label{eq: Planar_N=4D_NegM}
        \includegraphics[width=.9\linewidth]{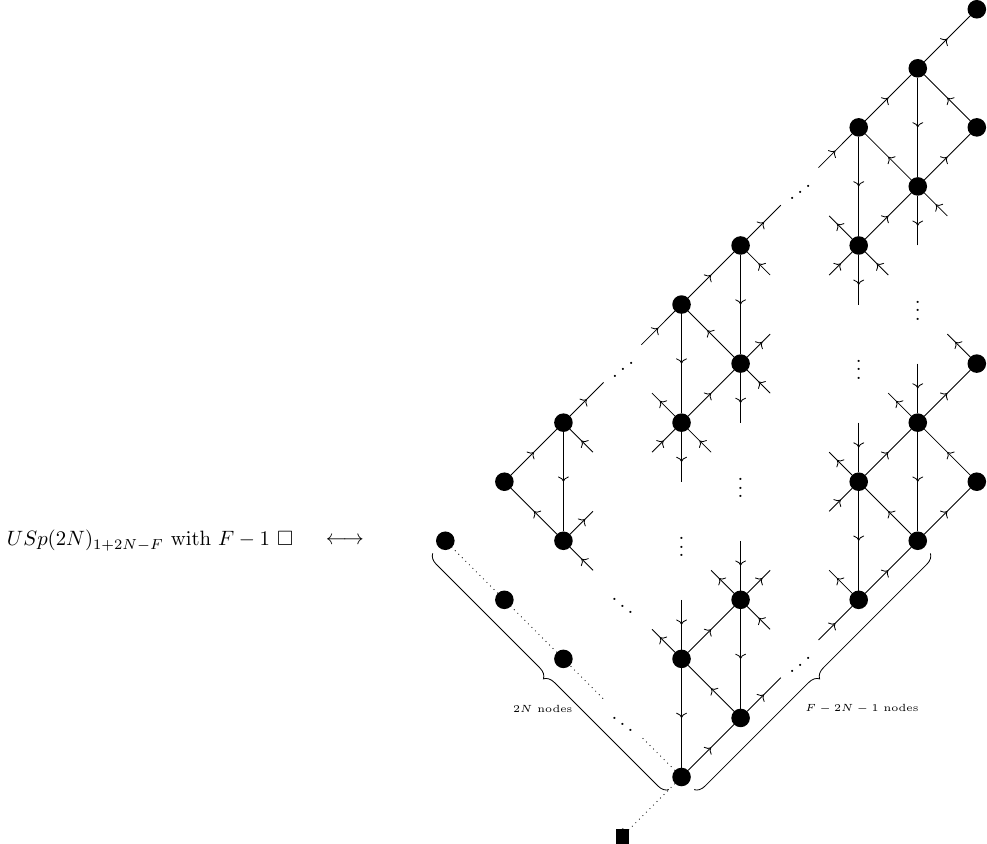}
    \end{equation}

     We remind the reader that pure $U(1)_{\pm1}$ Yang-Mills theory is almost trivial and its path integral can be performed exactly \cite{Kapustin:1999ha, witten2003sl2zactionthreedimensionalconformal}:

    \begin{equation}\label{eq:tftintegration}
    \includegraphics[]{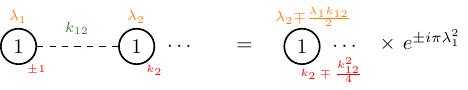}
    \end{equation}
    
    In the context of the quiver in Equation \eqref{eq: Planar_N=4D_NegM}, the sequential contraction proceeds as follows \cite{witten2003sl2zactionthreedimensionalconformal}:
    \begin{equation}
    \label{eq: witten sl2z}
    \begin{split}
    &
    \begin{tikzpicture}[baseline=(current bounding box).center]
        \node at (0,0) (g1) [gauge,black] {$1$};
        \draw[CScolor] (g1)++(.45,-.2) node {\tiny$2$};
        \nodeCS(1.5,0)(g2,1,2);
        \nodeCS(5,0)(g3,1,2);
        \nodeCS(6.5,0)(g4,1,1);
        \path (g2) edge[draw=none] node[midway] {$\dots$} (g3);
        \draw[BFline] (g1) -- node[midway,above,BFcolor] {\tiny -2} (g2);
        \draw[BFline] (g2) -- node[midway,above,BFcolor] {\tiny -2} ++(1,0);
        \draw[BFline] (g3)++(-1,0) -- node[midway,above,BFcolor] {\tiny -2} (g3);
        \draw[BFline] (g3) -- node[midway,above,BFcolor] {\tiny -2} (g4);
        \draw[->-] (g1) -- ++(150:1);
        \draw (g1)++(190:1) node {\tiny$\vdots$};
        \draw[-<-] (g1) -- ++(210:1);
    \end{tikzpicture}
    \\
    =&
    \begin{tikzpicture}[baseline=(current bounding box).center]
         \node at (0,0) (g1) [gauge,black] {$1$};
        \draw[CScolor] (g1)++(.45,-.2) node {\tiny$2$};
        \nodeCS(1.5,0)(g2,1,2);
        \nodeCS(5,0)(g3,1,1);
        \path (g2) edge[draw=none] node[midway] {$\dots$} (g3);
        \draw[BFline] (g1) -- node[midway,above,BFcolor] {\tiny -2} (g2);
        \draw[BFline] (g2) -- node[midway,above,BFcolor] {\tiny -2} ++(1,0);
        \draw[BFline] (g3)++(-1,0) -- node[midway,above,BFcolor] {\tiny -2} (g3);
        \draw[->-] (g1) -- ++(150:1);
        \draw (g1)++(190:1) node {\tiny$\vdots$};
        \draw[-<-] (g1) -- ++(210:1);
    \end{tikzpicture}
    \\
    & \vdots
    \\
    =&
    \begin{tikzpicture}[baseline=(current bounding box).center]
         \node at (0,0) (g1) [gauge,black] {$1$};
        \draw[CScolor] (g1)++(.45,-.2) node {\tiny$1$};
        \draw[->-] (g1) -- ++(150:1);
        \draw (g1)++(190:1) node {\tiny$\vdots$};
        \draw[-<-] (g1) -- ++(210:1);
    \end{tikzpicture}
    \end{split}
    \end{equation}

    The resulting quiver diagram is shown in Equation \ref{eq: Planar_N=4D_NegM_Contracted}.
    \begin{equation}
    \label{eq: Planar_N=4D_NegM_Contracted}
        \includegraphics[width=.9\linewidth]{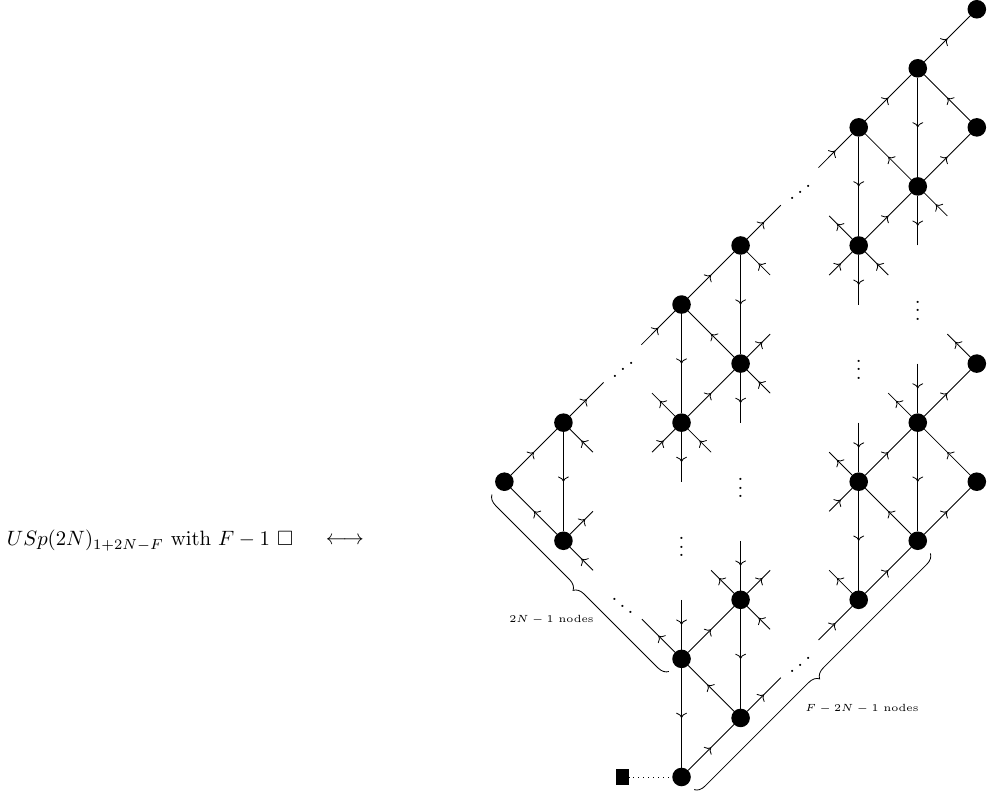}
    \end{equation}
\end{itemize}

\begin{figure}
    \centering
    \includegraphics[width=0.7\linewidth]{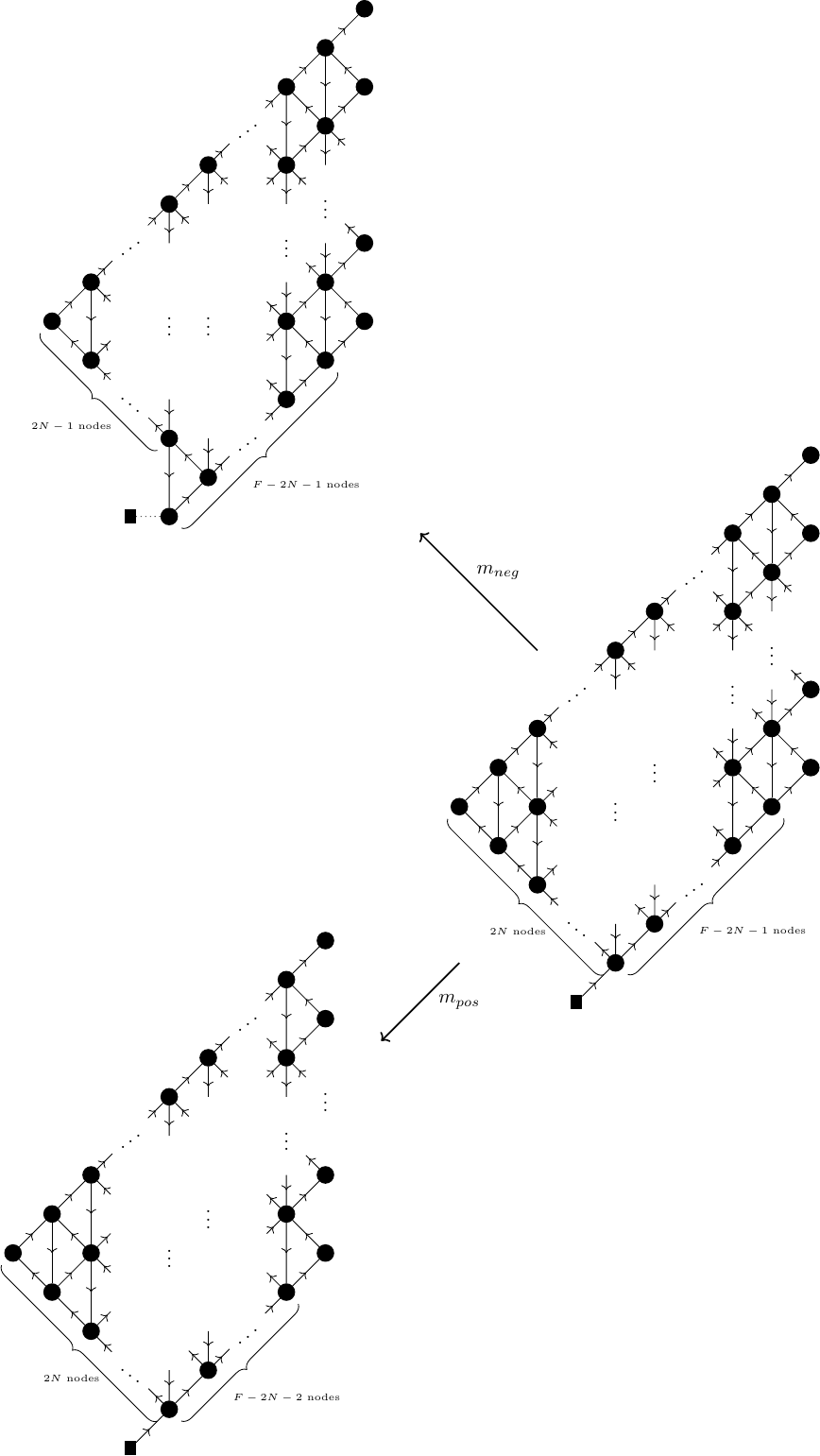}
    \caption{Deformations of the planar quivers dual to the SQCD mass deformations $\mposneg$ are shown here.}
    \label{fig: Def_Planar_Finish}
\end{figure}

We summarize these mass flows in Figure \ref{fig: Def_Planar_Finish}.
Notice that a $\mpos$ deformation produces another $\mathcal N=4$ descendant since the resulting quiver has the same global topology as the original dual. Hence, this deformation provides a \textbf{nontrivial self-consistency check} of the duality originally proposed in Section \ref{sec: symplectic}. Thus, if we start from an $\NN=4$ descendant, successive $\mpos$ deformations will produce the corresponding $\NN=4$ descendant. 

In contrast, the deformation $\mneg$ yields quivers with a modified geometry, leading to different planar structures. 
Indeed, the resulting SQCD is no longer an $\NN=4$ descendant, and instead lies in Zone 2 of Figure \ref{fig:KF_plane}.
We will now systematically study the effect of multiple $\mneg$ deformations in order to obtaine a planar Abelian dual of all SQCD theories sitting above the $\NN=4$-descendant line in Figure \ref{fig:KF_plane}.
The dual deformations on the planar Abelian side are analogous to those studied in \cite{Benvenuti:2026a} for the case of Unitary SQCD, therefore we gloss over the details and present the resulting planar Abelian quivers.

Starting from the duality \ref{eq: Planar_N=4D_NegM_Contracted},
we consider the RG trajectory induced by another $\mneg$ deformation. The effect of this deformation is shown in Equation \ref{eq: Planar_N=4D_2NegM_Contracted}.

\begin{equation}
\label{eq: Planar_N=4D_2NegM_Contracted}
    \includegraphics[width=.9\linewidth]{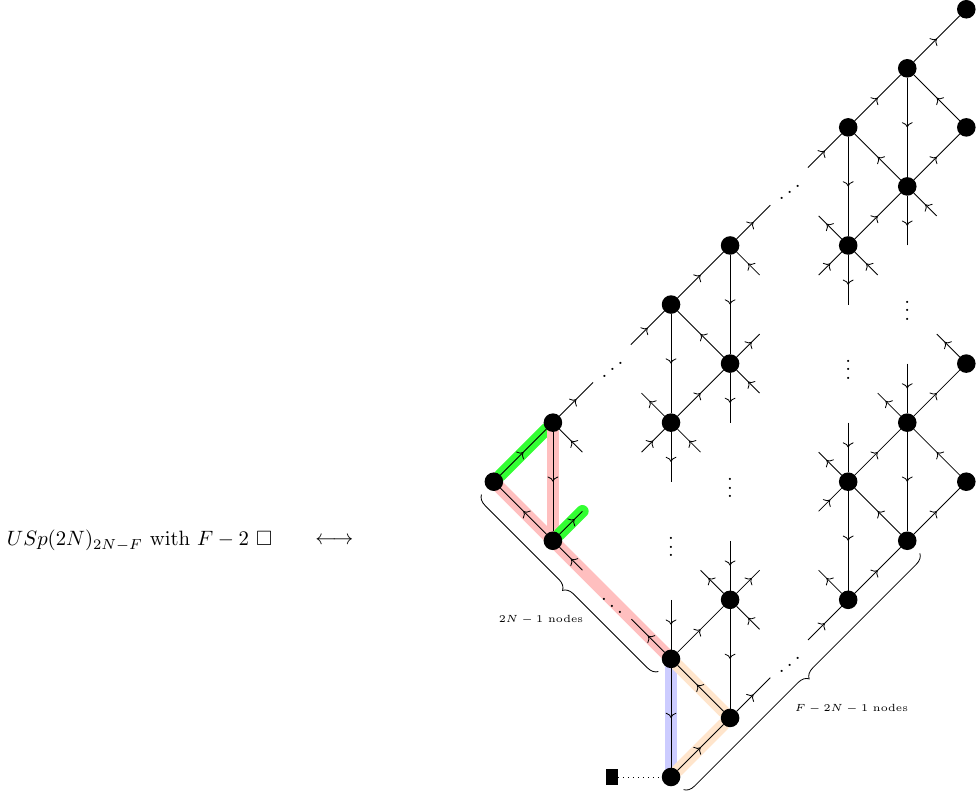}
\end{equation}
\newpage

Integrating out the massive fields yields the quiver shown in Equation \ref{eq: Planar_N=4D_2NegM_Contracted_Fin}.
\begin{equation}
\label{eq: Planar_N=4D_2NegM_Contracted_Fin}
    \includegraphics[width=.9\linewidth]{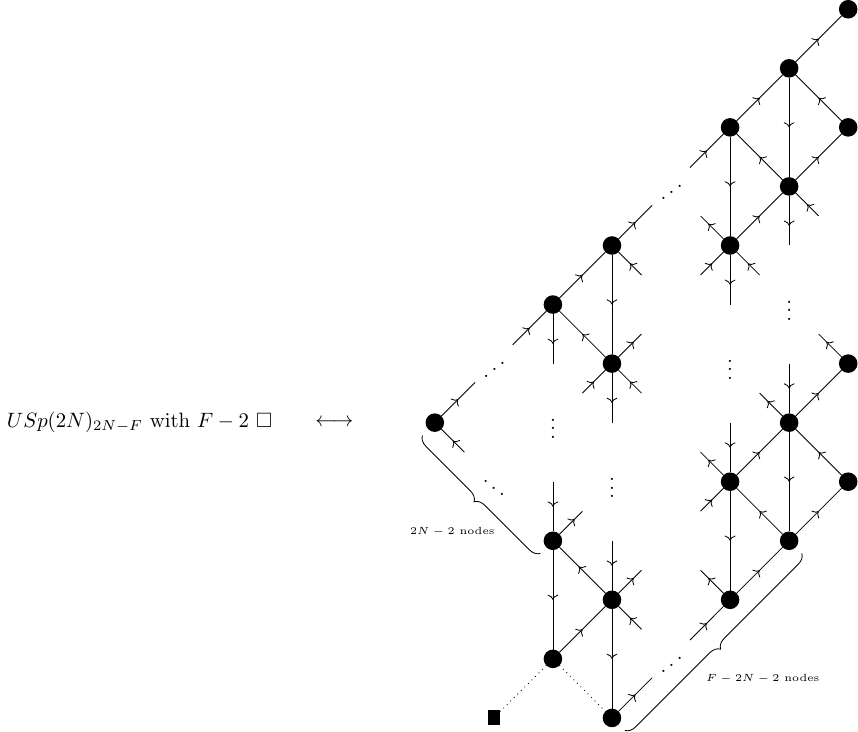}
\end{equation}

\newpage
Further negative mass deformations are mapped analogously to the planar Abelian dual.
Therefore we can follow the RG flow trajectory induced by the first $k\,(<\, 2N)$ $\mneg$ deformations. The resulting quiver is shown in Equation \ref{eq: Planar_N=4D_kl2NNegM_Fin}. 

\begin{equation}
\label{eq: Planar_N=4D_kl2NNegM_Fin}
    \includegraphics[width=.9\linewidth]{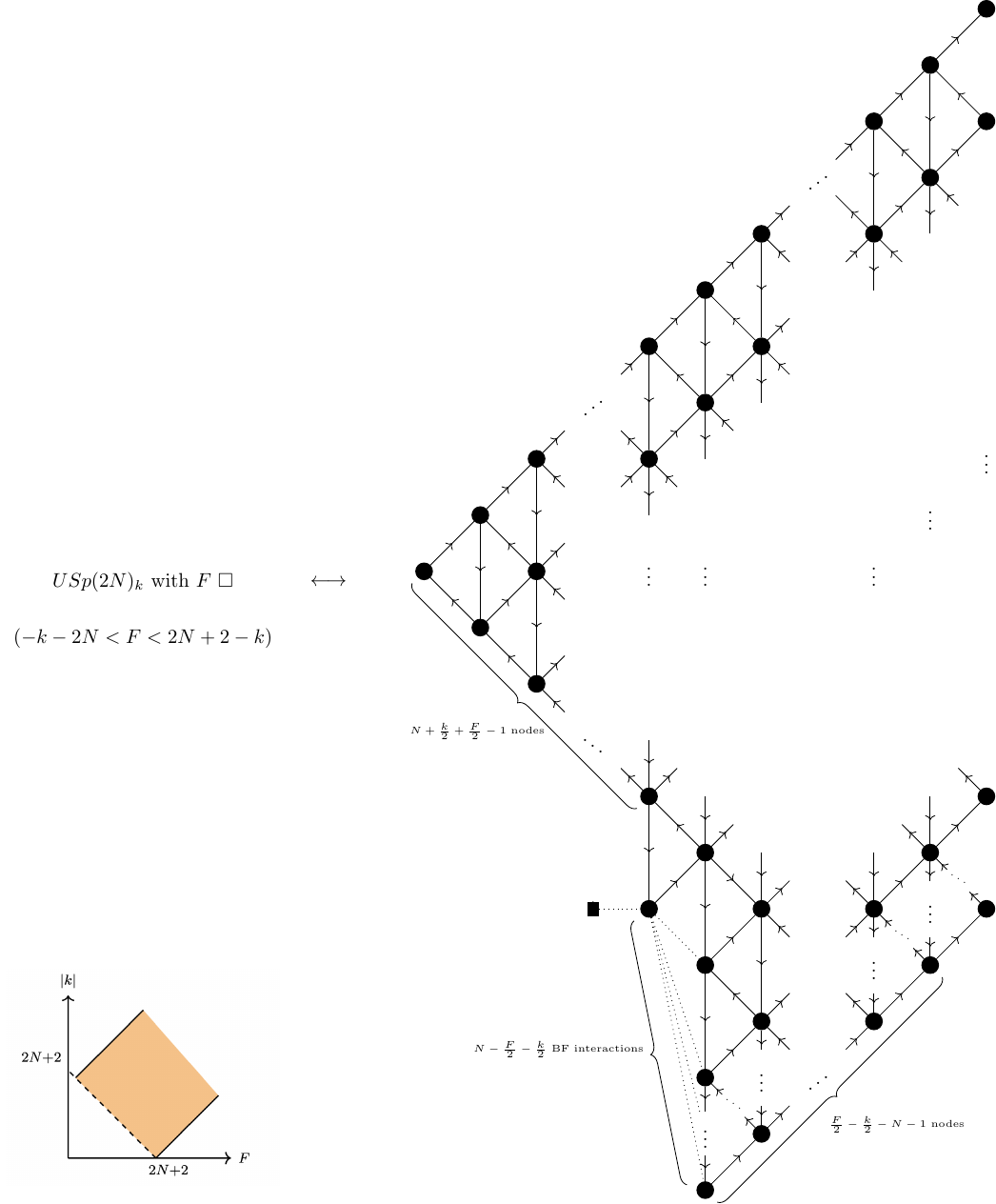}
\end{equation}

\newpage
The limiting case, where we perform $2N$ negative mass deformations $\mneg$, results in the quiver shown in Equation \ref{eq: Planar_N=4D_k_2NNegM_Fin}.

\begin{equation}
\label{eq: Planar_N=4D_k_2NNegM_Fin}
    \includegraphics[width=.9\linewidth]{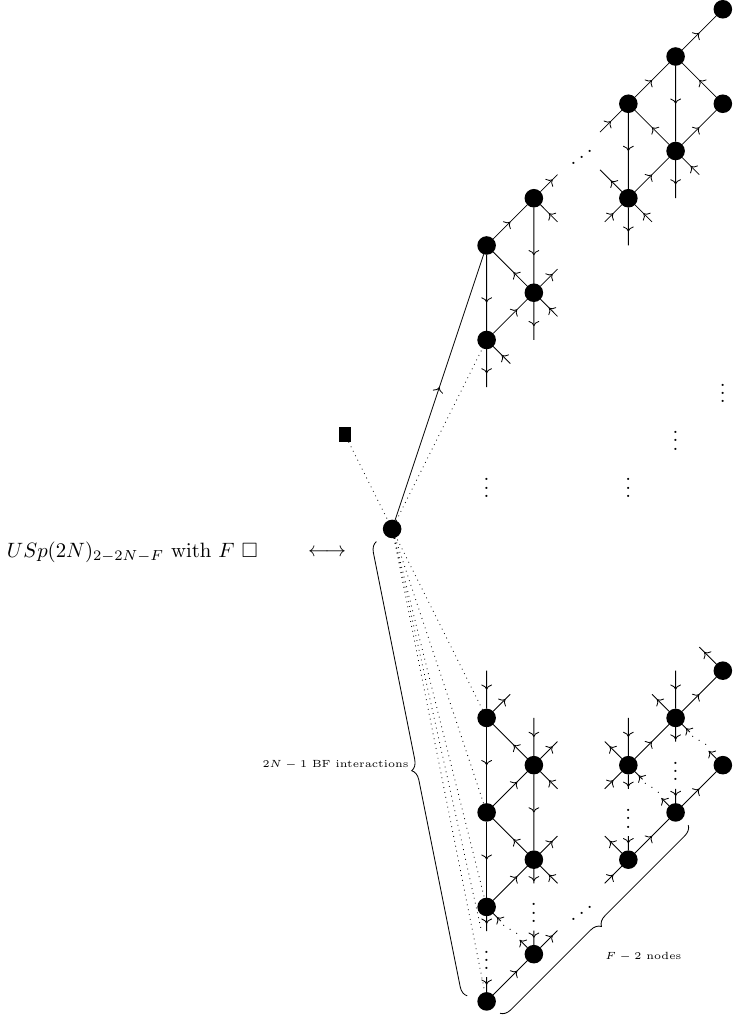}
\end{equation}

\newpage
The RG flow induced by the $(2N+1)^{\text{th}}$ $\mneg$ deformation results in the quiver shown in Equation \ref{eq: Planar_N=4D_k_2N+1NegM_Fin}.

\begin{equation}
\label{eq: Planar_N=4D_k_2N+1NegM_Fin}
    \includegraphics[width=.9\linewidth]{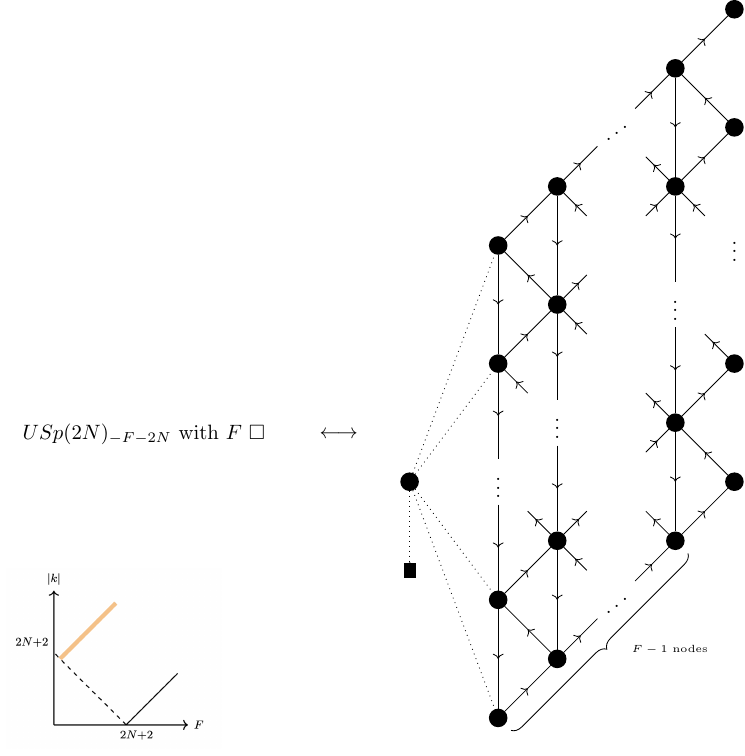}
\end{equation}

\newpage
Subsequent $\mneg$ deformations act on the bulk of the quiver and produce qualitatively different planar Abelian dual quivers. 
The qualitative difference between the quivers obtained up to now and the quivers obtained by further mass deformations is the reason for our choice to differentiate between Zones 2 and 3 in Figure \ref{fig:KF_plane}.
Starting from the duality in \eqref{eq: Planar_N=4D_k_2N+1NegM_Fin},
the action of a single $\mneg$ deformation on the bulk of the quiver is shown in Equation \ref{eq: columns}.

\begin{equation}
\label{eq: columns}
    \includegraphics[width=.9\linewidth]{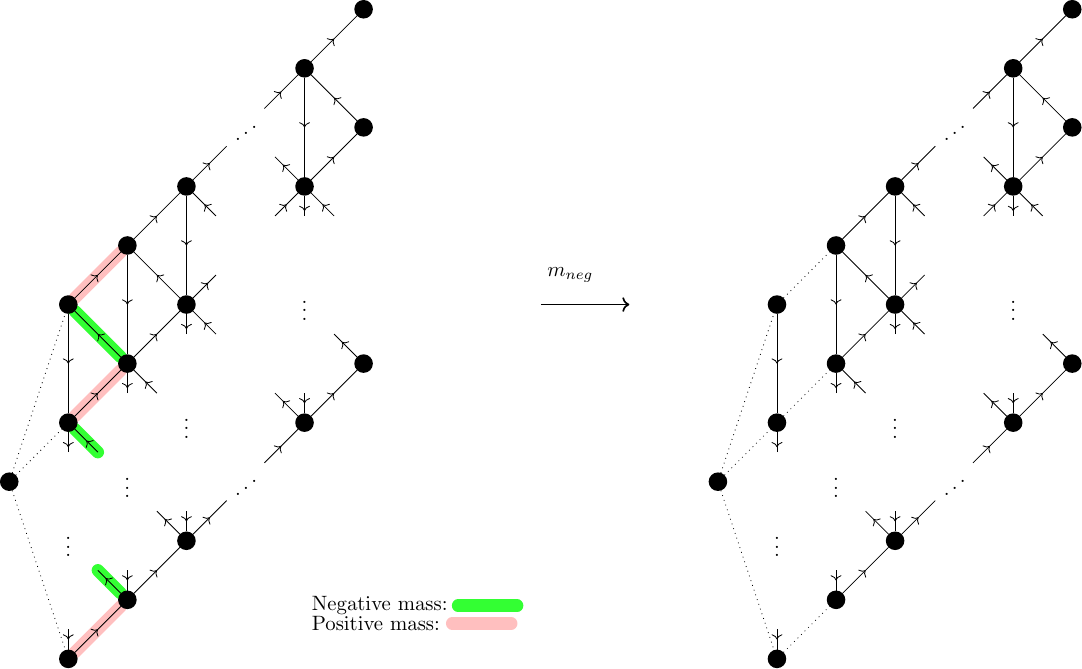}
\end{equation}

Compared to the deformations in Zone 1 presented above, here the quiver does not shrink in size. Instead, some columns of the quiver are now connected to the rest of the quiver only via mixed CS interactions depicted as dotted lines in our compact quiver notation.
As discussed below, such columns do not contribute to the overall topological symmetry of the quiver. Therefore, even if the planar Abelian quiver does not shrink in size, the rank of the global symmetry decreases as more and more columns “detach” from the quiver, as expected when we turn on more and more infinite mass deformations.

Successive $(F-2N-3)$ $\mneg$ deformations act analogously on the bulk of the quiver and we obtain the planar Abelian dual of SQCD in Zone 3 as shown in Equation \ref{eq: Planar_N=4D_k_BF_Col_NegM_Fin}.
\newpage

\begin{equation}
\label{eq: Planar_N=4D_k_BF_Col_NegM_Fin}
    \includegraphics[width=.9\linewidth]{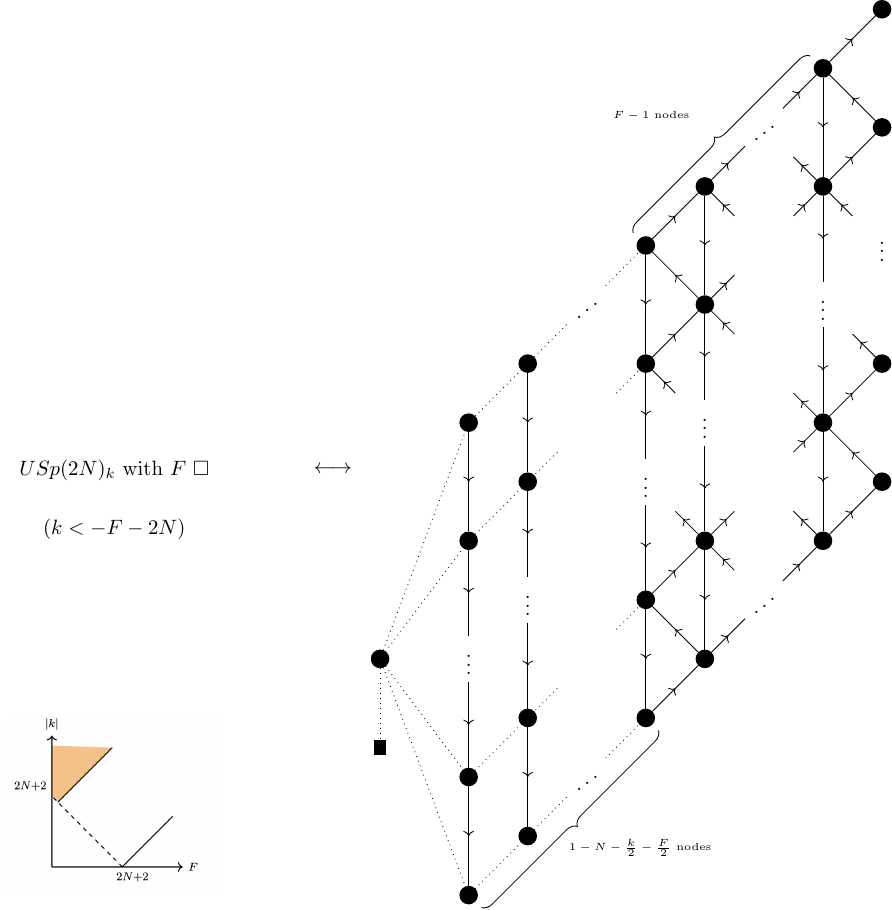}
\end{equation}

Let us discuss some details regarding the monopole interaction in these planar Abelian quiver, as there are some technical differences compared to the theories studied in the previous Sections.
There is a monopole superpotential term for each pair of consecutive nodes in a column with GNO flux $\pm 1$ under the top (bottom) nodes. In the ``bulk" of the quiver, where there are diagonal arrows on both sides of the column, these monopoles are gauge-invariant BPS operators and, hence, enter the superpotential as described before.
For other columns in the quiver these monopoles must be suitably dressed in order to be gauge invariant, in particular:
\begin{itemize}
    \item In the column that separates the ``bulk" of the quiver and the region with BF interactions, as shown in \eqref{eq: mon_W1}, these monopoles are dressed with a vertical chiral field $\phi$ as following:
    \begin{equation}\label{eq: mon_W1}
    \begin{tikzpicture}[baseline=(current bounding box).center, scale=.6]
        \foreach \x in {1,...,3}
        {
            \foreach \y in {1,2}
            {
                \node at (2*\x,-2*\y+\x) (g\x\y) [gaugefill] {$1$};
            }
        \path[->-,draw] (g\x1) -- (g\x2); 
        \path[->-,draw] (g\x1)++(0,1) -- (g\x1);
        \path[-<-,draw] (g\x2)++(0,-1) -- (g\x2);
        }
        
        \path[-<-,draw] (g31) -- (g21) ;
        \path[-<-,draw] (g21) -- (g32);
        \path[-<-,draw] (g32) -- (g22);
        \draw[->-] (g22)++(1,-.5) -- (g22);
        \draw[->-] (g31)++(1,-.5) -- (g31);
        \draw[->-] (g32)++(1,-.5) -- (g32);
        \draw[-<-] (g31)++(1,.5) -- (g31);
        \draw[-<-] (g32)++(1,.5) -- (g32);
        \draw[-<-] (g31)++(-1,.5) -- (g31);
        
        \draw[dotted] (g11)++(-1,-.5) --(g11) -- (g21);
        \draw[dotted] (g12)++(-1,-.5) --(g12) -- (g22);
        \path[] (g11) -- node[midway,left] {$\phi$} (g12);
    \end{tikzpicture}
    \qquad\qquad
    \mathcal{W} \ni \mathfrak M^{\left( \begin{array}{ccc} 0&1&0\\0&-1&0\end{array} \right)} \phi
\end{equation}
    The FIs of the nodes in this type of column are, from top to bottom:
    \begin{equation}
        \text{FIs: } X_i -\tfrac{iQ}{4}, \;\;
                    X_i - \tfrac{iQ}{2}(1-R), \;\;
                    \dots,  X_i - \tfrac{iQ}{2} (1-R), \;\;
                    X_i - \tfrac{iQ}{4}(2 - 3R)
    \end{equation}
    where $X_i$ is a Cartan of the $U(F)$ flavor symmetry and the coefficient in front of $\tfrac{iQ}{2}$ encodes the mixing between the topological and the R-symmetry. We remind the reader that we assign trial R-charge $1-R$ to the fundamental fields of the SQCD.
    One can check that the FIs above are compatible with the monopole superpotential.
\item In the columns with BF interactions on both sides, as shown in \eqref{eq: mon_W2}, the monopoles are dressed with two vertical chiral fields $\phi_1$ and $\phi_2$: 
\begin{equation}
\label{eq: mon_W2}
    \begin{tikzpicture}[baseline=(current bounding box).center, scale=.6]
        \foreach \x in {1,...,3}
        {
            \foreach \y in {1,2}
            {
                \node at (2*\x,-2*\y+\x) (g\x\y) [gaugefill] {$1$};
            }
        \path[->-,draw] (g\x1) -- (g\x2); 
        \path[->-,draw] (g\x1)++(0,1) -- (g\x1);
        \path[-<-,draw] (g\x2)++(0,-1) -- (g\x2);
        }

        \draw[dotted] (g11)++(-1,-.5) --(7,1.5);
        \draw[dotted] (g12)++(-1,-.5) --(7,-.5);
        \path[] (g11) -- node[midway,left] {$\phi_1$} (g12);
        \path[] (g31) -- node[midway,right] {$\phi_2$} (g32);
    \end{tikzpicture}
    \qquad\qquad
    \mathcal{W} \ni \mathfrak M^{\left( \begin{array}{ccc} 0&1&0\\0&-1&0\end{array} \right)} \phi_1 \phi_2
\end{equation}
    The FIs of the nodes in this type of column are, from top to bottom:
    \begin{equation}
        \text{FIs: } \tfrac{iQ}{4}(1-2R), \;\;
                    0, \;\;
                    \dots,  0, \;\;
                    \tfrac{iQ}{4}(2R-1)
    \end{equation}
    One can check that the FIs above are compatible with the monopole superpotential.
\item We further notice that in a column with no diagonal chirals on both sides, the single unbroken topological symmetry can be reabsorbed by a gauge transformation due to the presence of non-zero CS-levels. Therefore, only the columns in the ``bulk" of the quiver and at the edge of the ``bulk" contribute to the Cartan of the global symmetry, thus matching the rank of the global symmetry predicted by the dual SQCD.
\end{itemize}
\newpage

Getting all the way down to $F=2$ we find that the planar Abelian dual has a single set of diagonal arrows between the two right-most columns, denoted as the “saw” throughout this paper. 
A further negative mass deformation leading to SQCD with a single flavor then involves real mass deformations for the diagonal fields in the “saw”.
The mapping of the $\mneg$ deformation on the ``saw" is similar to the previous case \eqref{eq: columns}.
In the resulting quiver all the gauge nodes in the right-most column are $U(1)_{\pm1}$ pure gauge theories and we can perform their path integral exactly.
After this operation we obtain the planar Abelian dual of $USp(2N)$ CS-SQCD with $1$ fundamental chiral shown in Equation \ref{eq: Planar_N=4D_F=1_BF_Col_NegM_Fin}.

\begin{equation}
\label{eq: Planar_N=4D_F=1_BF_Col_NegM_Fin}
    \includegraphics[width=.8\linewidth]{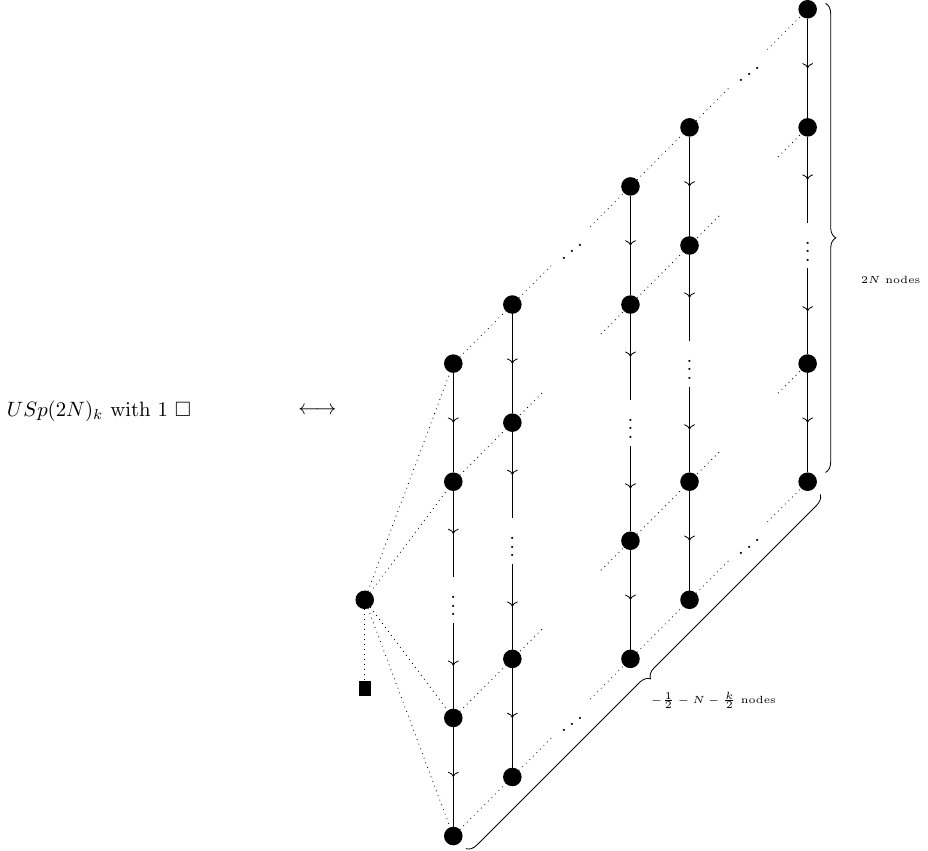}
\end{equation}

\newpage
Here the monopole superpotential terms associated to the last column follow a different pattern compared to the other columns:

\begin{equation}
    \begin{tikzpicture}[baseline=(current bounding box).center,scale=0.6]
    \node[gaugefill] at (0,0) (g11) {};
    \node[gaugefill] at (0,2) (g12) {};
    \node[gaugefill] at (0,4) (g13) {};
    \node[gaugefill] at (0,6) (g14) {};
    \node[gaugefill] at (1.5,1) (g21) {};
    \node[gaugefill] at (1.5,3) (g22) {};
    \node[gaugefill] at (1.5,5) (g23) {};
    \node[gaugefill] at (1.5,7) (g24) {};

    \draw[->-] (g14) -- node[midway,left] {$\phi_1$} (g13);
    \draw[->-] (g13) -- node[midway,left] {$\phi_2$} (g12);
    \draw[->-] (g12) -- node[midway,left] {$\phi_3$} (g11);
    \draw[->-] (g24) -- (g23);
    \draw[->-] (g23) -- (g22);
    \draw[->-] (g22) -- (g21);

    \draw[dotted] (g11)--(g21);
    \draw[dotted] (g12)--(g22);
    \draw[dotted] (g13)--(g23);
    \draw[dotted] (g14)--(g24);

    \draw[->-] (g11) -- ++ (0,-1);
    \draw[->-] (g21) -- ++ (0,-1);
    \draw[-<-] (g14) -- ++ (0,1);
    \draw[-<-] (g24) -- ++ (0,1);

    \draw[dotted] (g11) -- ++ (-1,-.66);
    \draw[dotted] (g12) -- ++ (-1,-.66);
    \draw[dotted] (g13) -- ++ (-1,-.66);
    \draw[dotted] (g14) -- ++ (-1,-.66);
    \end{tikzpicture}
    \qquad
    \begin{gathered}
    \mathcal{W} \ni\;\;
    \mathfrak{M}^{\scalebox{0.6}{$
        \left(
        \begin{array}{cc}
        0 & 1   \\
        0 & 0  \\
        0 & -1   \\
        0 & 0  \\
        \end{array}
        \right)
        $}}
        \phi_1 \phi_2
    +
    \mathfrak{M}^{\scalebox{0.6}{$
        \left(
        \begin{array}{cc}
        0 & 0   \\
        0 & 1  \\
        0 & 0 \\
        0 & -1  \\
        \end{array}
        \right)
        $}}
        \phi_2 \phi_3
    \end{gathered}
\end{equation}
There are $2N-2$ such monopole superpotential terms, therefore there is a $U(1)$ subgroup of the topological symmetries of the last column which is neither gauged nor broken by the superpotential.
This is a proper global symmetry of the planar Abelian quiver and is mapped to the $U(1)$ global symmetry of the SQCD.
Consistently, the FIs of the last column are, from top to bottom:
\begin{equation}
    \text{FIs}: \; iQ \left( \frac{3}{4} - R\right) - X_1, \; \frac{iQ}{2}(1-R) + X_1,\; \frac{iQ}{2} (1-R) - X_1, \dots, \; \frac{iQ}{4} + X_1
\end{equation}
where $X_1$ is the fugacity associated to the flavor $U(1)$ symmetry of the SQCD.

\newpage
Upon integrating out the remaining fundamental chiral field, we obtain pure $USp(2N)_k$ CS-SYM.
On the planar dual side this corresponds to higgsing every other vertical field in the last column, starting from the topmost vertical field.
The resulting planar Abelian dual for $USp(2N)_{k}$ CS-SYM theory is shown in Equation \ref{eq: Planar_N=4D_TQFT_BF_Col_NegM_Fin}. The dual description is not manifestly topological as it contains multiple massless charged fields. 

\begin{equation}
\label{eq: Planar_N=4D_TQFT_BF_Col_NegM_Fin}
    \includegraphics[width=.7\linewidth]{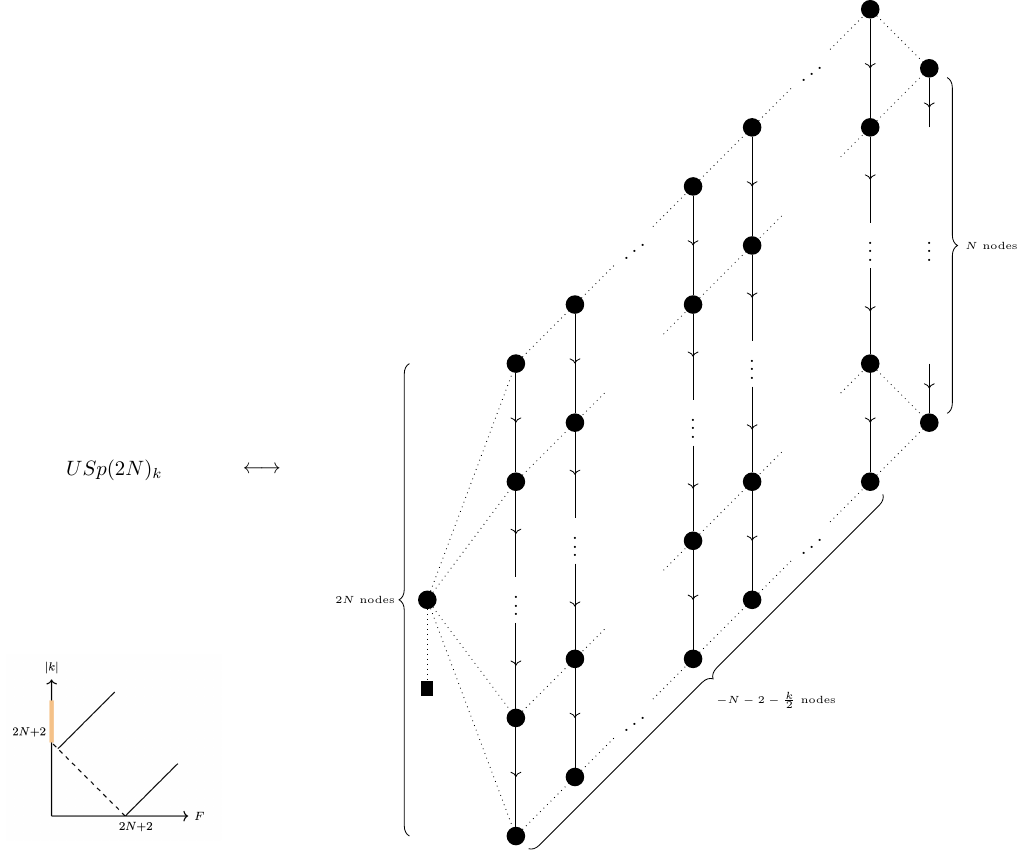}
\end{equation}

Note that the height of all the columns except for the last one is $2N$, while that of the final column is $N$. Furthermore, the vertical chirals in the final column have R-charge $4-4R$, in contrast to the other vertical chirals which have R-charge $2-2R$.
The monopole superpotential terms associated with the last two columns contain the following (dressed) monopoles:
\begin{equation}
    \begin{tikzpicture}[baseline=(current bounding box).center,scale=0.6]
    \node[gaugefill] at (0,0) (g11) {};
    \node[gaugefill] at (0,2) (g12) {};
    \node[gaugefill] at (0,4) (g13) {};
    \node[gaugefill] at (0,6) (g14) {};
    \node[gaugefill] at (1.5,1) (g21) {};
    \node[gaugefill] at (1.5,3) (g22) {};
    \node[gaugefill] at (1.5,5) (g23) {};
    \node[gaugefill] at (1.5,7) (g24) {};
    \node[gaugefill] at (3,2) (g31) {};
    \node[gaugefill] at (3,6) (g32) {};

    \draw[->-] (g14) -- node[midway,left] {$\phi_1$} (g13);
    \draw[->-] (g13) -- node[midway,left] {$\phi_2$} (g12);
    \draw[->-] (g12) -- node[midway,left] {$\phi_3$} (g11);
    \draw[->-] (g24) -- node[midway,left] {$\psi_1$} (g23);
    \draw[->-] (g23) -- node[midway,left] {$\psi_2$} (g22);
    \draw[->-] (g22) -- node[midway,left] {$\psi_3$} (g21);
    \draw[->-] (g32) -- node[midway,left] {$\chi$} (g31);

    \draw[dotted] (g11)--(g21)--(g31);
    \draw[dotted] (g12)--(g22)--(g31);
    \draw[dotted] (g13)--(g23)--(g32);
    \draw[dotted] (g14)--(g24)--(g32);

    \draw[->-] (g11) -- ++ (0,-1);
    \draw[->-] (g21) -- ++ (0,-1);
    \draw[->-] (g31) -- ++ (0,-1);
    \draw[-<-] (g14) -- ++ (0,1);
    \draw[-<-] (g24) -- ++ (0,1);
    \draw[-<-] (g32) -- ++ (0,1);

    \draw[dotted] (g11) -- ++ (-1,-.66);
    \draw[dotted] (g12) -- ++ (-1,-.66);
    \draw[dotted] (g13) -- ++ (-1,-.66);
    \draw[dotted] (g14) -- ++ (-1,-.66);
    \end{tikzpicture}
    \qquad
    \begin{gathered}
    \mathcal{W} \ni\;\;
    \mathfrak{M}^{\scalebox{0.6}{$
        \left(
        \begin{array}{cccc}
        0 & 1 & \multirow{2}{*}{0}  \\
        0 & -1 &  \\
        0 & 0 & \multirow{2}{*}{0}  \\
        0 & 0 &  \\
        \end{array}
        \right)
        $}}
        \phi_1
    +
    \mathfrak{M}^{\scalebox{0.6}{$
        \left(
        \begin{array}{cccc}
        0 & 0 & \multirow{2}{*}{0}  \\
        0 & 0 &  \\
        0 & 1 & \multirow{2}{*}{0}  \\
        0 & -1 &  \\
        \end{array}
        \right)
        $}}
        \phi_3
    \\+
    \mathfrak{M}^{\scalebox{0.6}{$
        \left(
        \begin{array}{cccc}
        0 & 0 & \multirow{2}{*}{0}  \\
        0 & 1 &  \\
        0 & -1 & \multirow{2}{*}{0}  \\
        0 & 0 &  \\
        \end{array}
        \right)
        $}}
        \phi_2\chi
    +
    \mathfrak{M}^{\scalebox{0.6}{$
        \left(
        \begin{array}{cccc}
        0 & 0 & \multirow{2}{*}{1}  \\
        0 & 0 &  \\
        0 & 0 & \multirow{2}{*}{-1}  \\
        0 & 0 &  \\
        \end{array}
        \right)
        $}}
        \psi_1 \psi_2^2 \psi_3
    \end{gathered}
\end{equation}

It would be interesting to study this duality further - particularly its possible connection with level–rank duality. We defer this to future work.

\newpage

\begin{landscape}
\thispagestyle{empty}
    We conclude the analysis of negative mass deformations by discussing another possible flow on the mirror side which is related to the ones discussed above by an (emergent) Weyl transformation.
    In our conventions, a $\mneg$ deformation corresponds to taking a large FI limit for the planar quiver from the left. However, due to invariance under the Weyl of the emergent $U(F)$ flavor symmetry, we could also act on the ``saw" at the end of the quiver. We demonstrate the effect of three $\mneg$ deformations at the end of the quiver in Equation \ref{eq: columns_saw_spicy}. Due to the different pattern of topological symmetries in the last column, the action of a single $\mneg$ deformation is different from the one shown in Equation \ref{eq: columns}. Nevertheless, it can be verified that the IR behavior of the resulting SCFT is independent of the order in which the $X_i$ fugacities are set to $\pm\infty$.
    In the last step we exploit the fact that all the gauge nodes in the last column are $U(1)_{\pm 1}$, which are almost trivial theories.
\begin{equation}
\label{eq: columns_saw_spicy}
    \includegraphics[width=.7\linewidth]{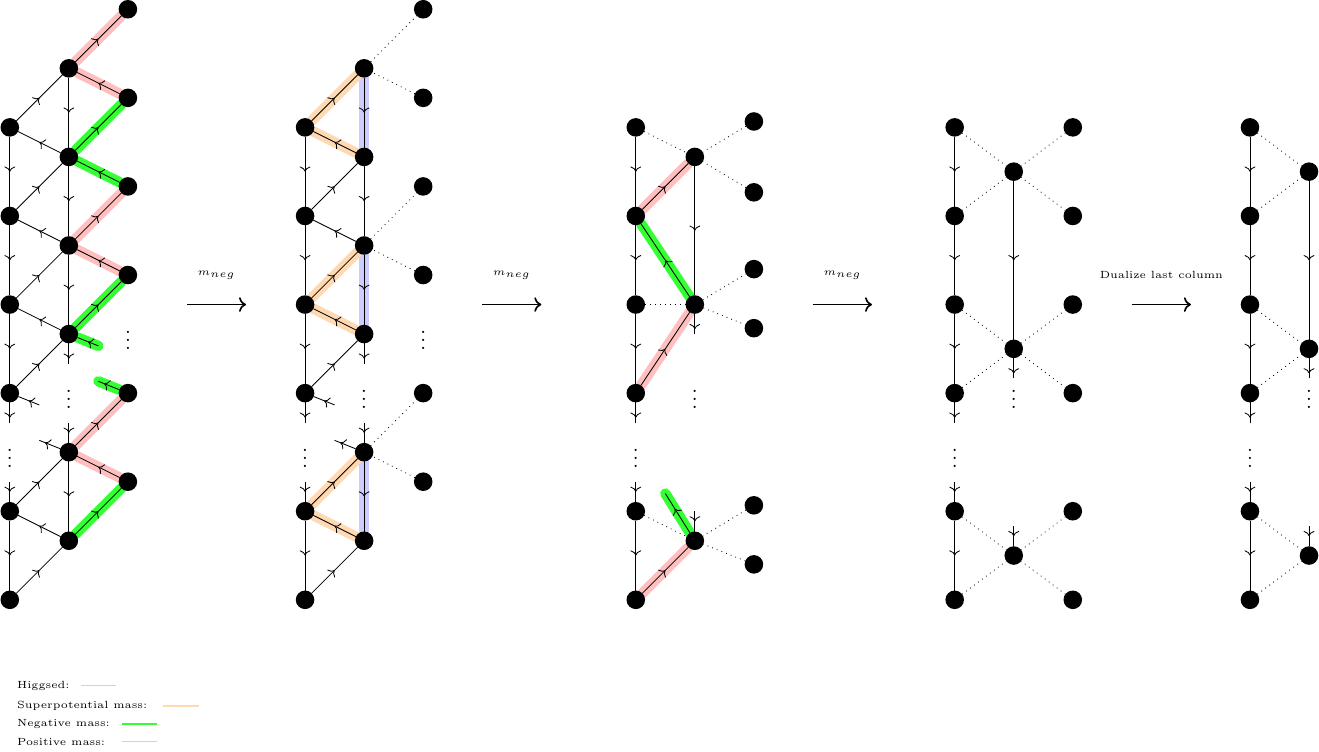}
\end{equation}
\end{landscape}

\subsection{Example: Massive Deformations of \texorpdfstring{USp(4)$_{-1}$ SQCD with $7$ $\Box$}{USp(4)-1 8}}\label{subsec: example}
We now analyze a representative example of such RG flows induced by $\mposneg$ deformations, highlighting how the electric and mirror descriptions encode these deformations. 

As a case study, we examine the RG flow of $USp(4)_{-1}$ CS-SQCD with $7$ fundamental multiplets. As mentioned before, this theory descends from the mirror duality of $\mathcal{N}=4$ $USp$($4$) with $7$ hypermultiplets after adding a suitable real mass deformation. It provides a simple but non-trivial laboratory for testing the mirror construction beyond the parent-fixed CS levels.

We show the resulting $\NN=2$ dual pair in Equation \ref{eq: Level1_C2_k=-1}. As this is our first example, we are pedantic and give many details. The conventions for quiver diagrams follow Section \ref{sec: symplectic}; in particular, we display only the fugacities for the topological symmetries, adopting the shorthand $X_{ij} := X_i - X_j$ for the same. The $\alpha^{\text{th}}$ gauge node in the $I^{\text{th}}$ column of the planar quiver ($\alpha = 1, \dots, |G^{(I)}|$, where $|G^{(I)}|$ denotes the number of $U(1)$ factors in the $I^{\text{th}}$ column of the $\mathcal{N}=2$ quiver) has the FI parameter $X_{i+1}-X_i$.  All other information regarding CS and mixed CS interactions is omitted for brevity and can be reconstructed using the rules outlined earlier. We treat this theory in full generality; hence, we do not assign a value to the R-charge and explicitly display the mixing of the $U(1)_R$ symmetry with the topological symmetries of the mirror theory. Furthermore, the (top-most) bottom-most gauge nodes of each column receive an additional universal contribution of $(-)+ \frac{iQ}{4}R$ to the FI term. We follow these conventions while studying this example. The mixing of the $U(1)_R$ symmetry with the topological symmetries of each column is essential to ensure their enhancement in the infrared.
\newpage
\begin{equation}
\label{eq: Level1_C2_k=-1}
    \includegraphics[]{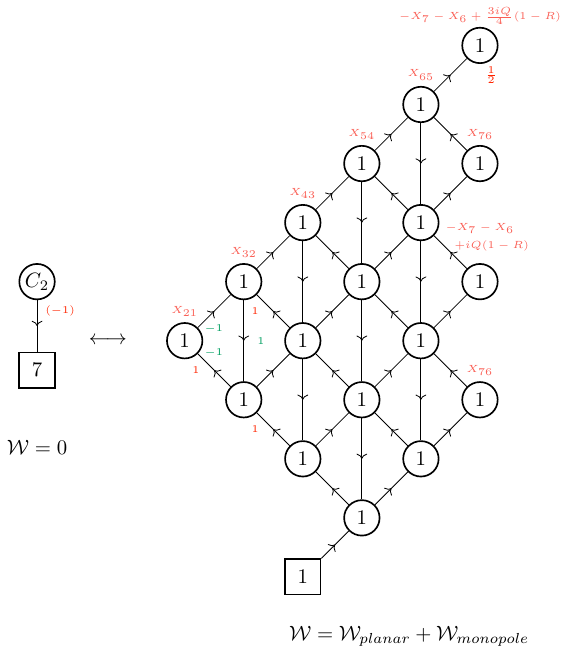}
\end{equation}

\subsubsection*{Dualities with One Real Mass Deformation}
Following the general discussion of Section \ref{subsec: n=4_d}, we study the RG flow induced by $\mposneg$ deformations on the planar dual theory.

\paragraph{\texorpdfstring{$\mpos$ deformations: $USp(4)_{0}$ with $6\,\Box$}{+m USp4 0 w 6}:} We consider the large mass limit $X_7\to-\infty$ in the \textit{electric} SQCD theory. The theory flows to $USp(4)_0$ with $6$ fundamental multiplets in the infrared. At the level of the $\mathbf S^3_b$ partition function, the divergent asymptotic phase generated along this flow is $-2\pi QRX_7$, and we must match this phase once the deformation is mapped to the mirror dual \cite{Benini_2011a, Aharony:2013dha}.

We now consider the large mass limit $X_7\to-\infty$ in the \textit{mirror} partition function. To obtain the right vacuum, corresponding to the divergent asymptotic phase $-2\pi Q R X_7$, we consider shifts in the mirror partition function that implement a Higgs VEV for the fields in the ``saw"-like structure of the quiver (analogous to the setup shown in the bottom left panel of Figure \ref{fig: Def_Planar_Start}). After integrating out the massive fields, we obtain the planar dual of $USp(4)_0$ with $6$ fundamental fields, shown in Equation \ref{eq: Level2_C2_k=0}. We remind the reader that this is also an $\NN=4$ descendant, and this is a non-trivial consistency check of our proposal.

\begin{equation}
\label{eq: Level2_C2_k=0}
    \includegraphics[]{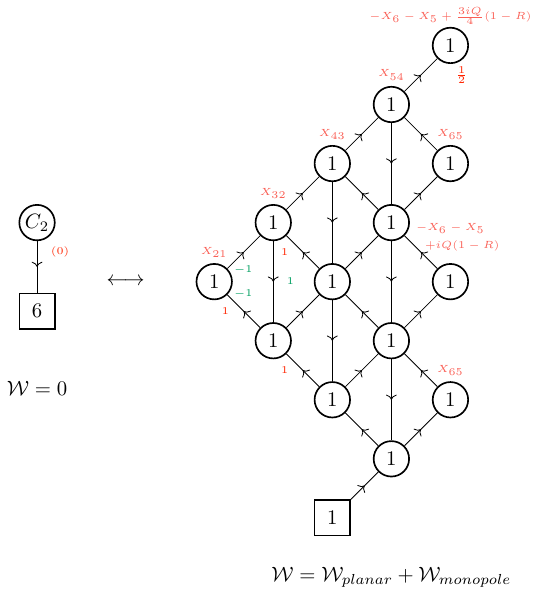}
\end{equation}

The Aharony-like dual of $USp(4)_0$ SQCD with $6$ fundamental multiplets is a Wess--Zumino model with superpotential \cite{Giveon:2008zn}
\[
\mathcal W = \gamma\,\mathrm{Pf}(M),
\]
where $M_{ij}=\tilde Q_i\tilde Q_j$ are the $\binom{6}{2}=15$ mesons of the electric theory, and the singlet field $\gamma$ is dual to the gauge-invariant monopole operator $\mathfrak M$. 

This description predicts the existence of a corresponding gauge-invariant monopole operator in the planar mirror theory, dual to $\gamma$. Explicitly, one finds the operator map
\begin{equation*}
    \mathfrak M \, \leftrightarrow \gamma \, \leftrightarrow \, \mathfrak M^{\begin{pmatrix}
    & & & & + &\\ & & & + & \\& & + & & + \\ & + & & + & \\ + & & + & & 2 \\ & + & &2 \\ & & + & & + \\ & & & 0 & 
    \end{pmatrix}};
\end{equation*}
We generically anticipate the existence of such gauge-invariant higher GNO flux monopoles in the planar mirrors of $USp(2N)_0$ with $2N+2$ fundamental multiplets.
\paragraph{\texorpdfstring{$\mneg$ deformations: $USp(4)_{-2}$ with $6\,\Box$}{-m USp4 -2 w 6}:} We consider the large mass limit $X_1\to+\infty$ in the \textit{electric} SQCD theory. The theory flows to $USp(4)_{-2}$ with $6$ fundamental multiplets in the infrared. At the level of the $\mathbf S^3_b$ partition function, the divergent asymptotic phase generated along this flow is $+i\pi X_1^2$, and we must match this phase once the deformation is mapped to the mirror dual.

We now consider the large mass limit $X_1\to+\infty$ in the \textit{mirror} partition function. To obtain the right vacuum, corresponding to the divergent asymptotic phase $+i\pi X_1^2$, we consider shifts in the mirror partition function that implement large VEVs for the gauge groups along the bottom-left diagonal of the quiver (analogous to the setup shown in the top left panel of Figure \ref{fig: Def_Planar_Start}). After integrating out the heavy fields, we obtain the planar dual of $USp(4)_{-2}$ with $6$ fundamental fields, shown in Equation \ref{eq: Level2_C2_k=-2}. The tail of gauge nodes connected by BF couplings can be confined; however, we do not do this here to highlight the difference in global structure of the resulting theory from the $\NN=4$ descendant. 

\begin{equation}
\label{eq: Level2_C2_k=-2}
    \includegraphics[]{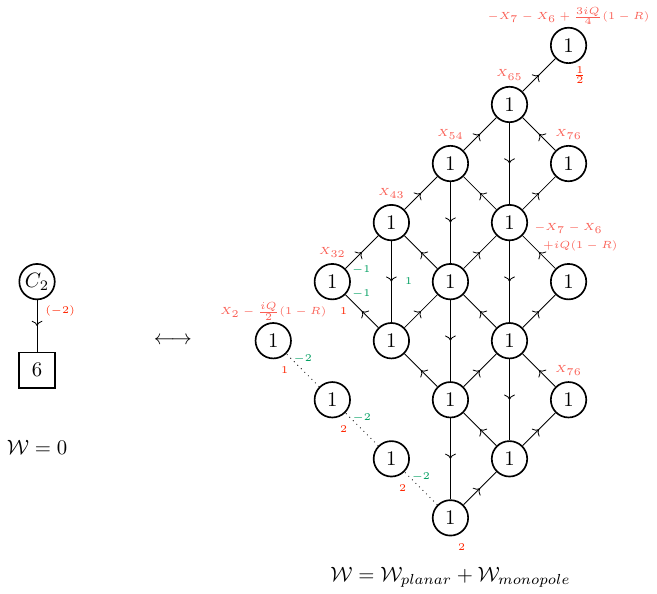}
\end{equation}

\subsubsection*{Dualities with Multiple Real Mass Deformations}
Having established the general framework of how $\mposneg$ deformations are mapped to the planar mirror dual, we now catalog the various dualities generated by $\mposneg$ mass flows. We do not comment on the CS levels and FI terms, which can be recovered from the respective $\mathbf S^3_b$ partition functions, and only display the global topology of the resulting quiver diagrams in Figures \ref{fig: zoology} and \ref{fig: zoology_2}. 

\begin{landscape}
\thispagestyle{empty}
\begin{figure}
    \centering
    \includegraphics[width=.45\linewidth]{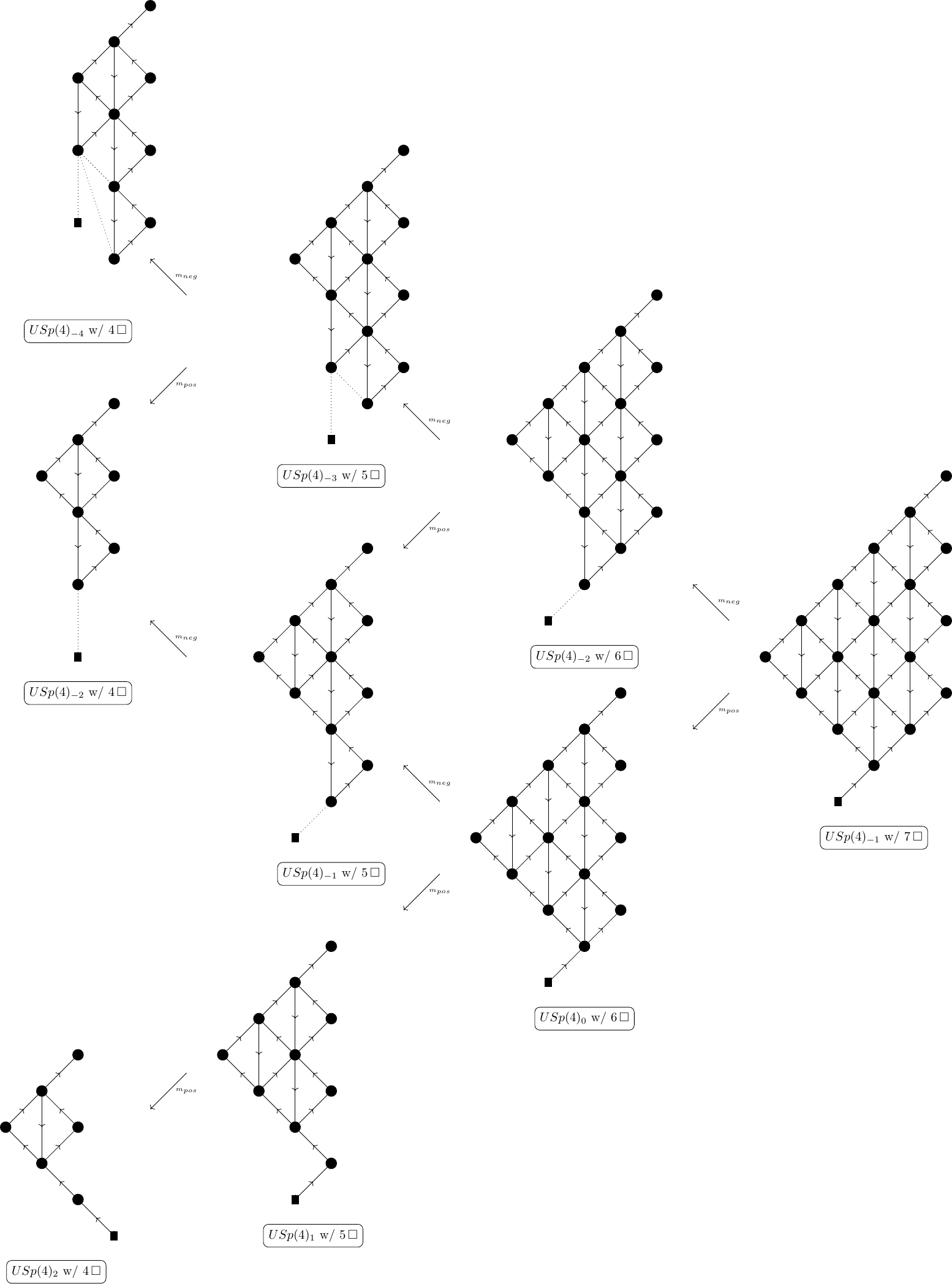}
    \caption{The planar mirror dual of $USp(4)_{-1}$ SQCD with $7$ fundamental multiplets is an $\NN=4$ descendant theory. Starting from this UV theory, we study the RG flow trajectories induced by $\mposneg$ real mass deformations. In this figure, we display the first three levels of this deformation sequence. Specifically, we begin with an electric theory containing $[7]$ fundamental multiplets and follow the flow to theories with $[4]$ fundamental multiplets. We only display the global topology of the resulting quiver diagrams; details of the CS levels and FI terms can be recovered following our usual conventions. }
    \label{fig: zoology}
\end{figure}
\end{landscape}

\begin{landscape}
\thispagestyle{empty}
\begin{figure}
    \centering
    \includegraphics[width=.55\linewidth]{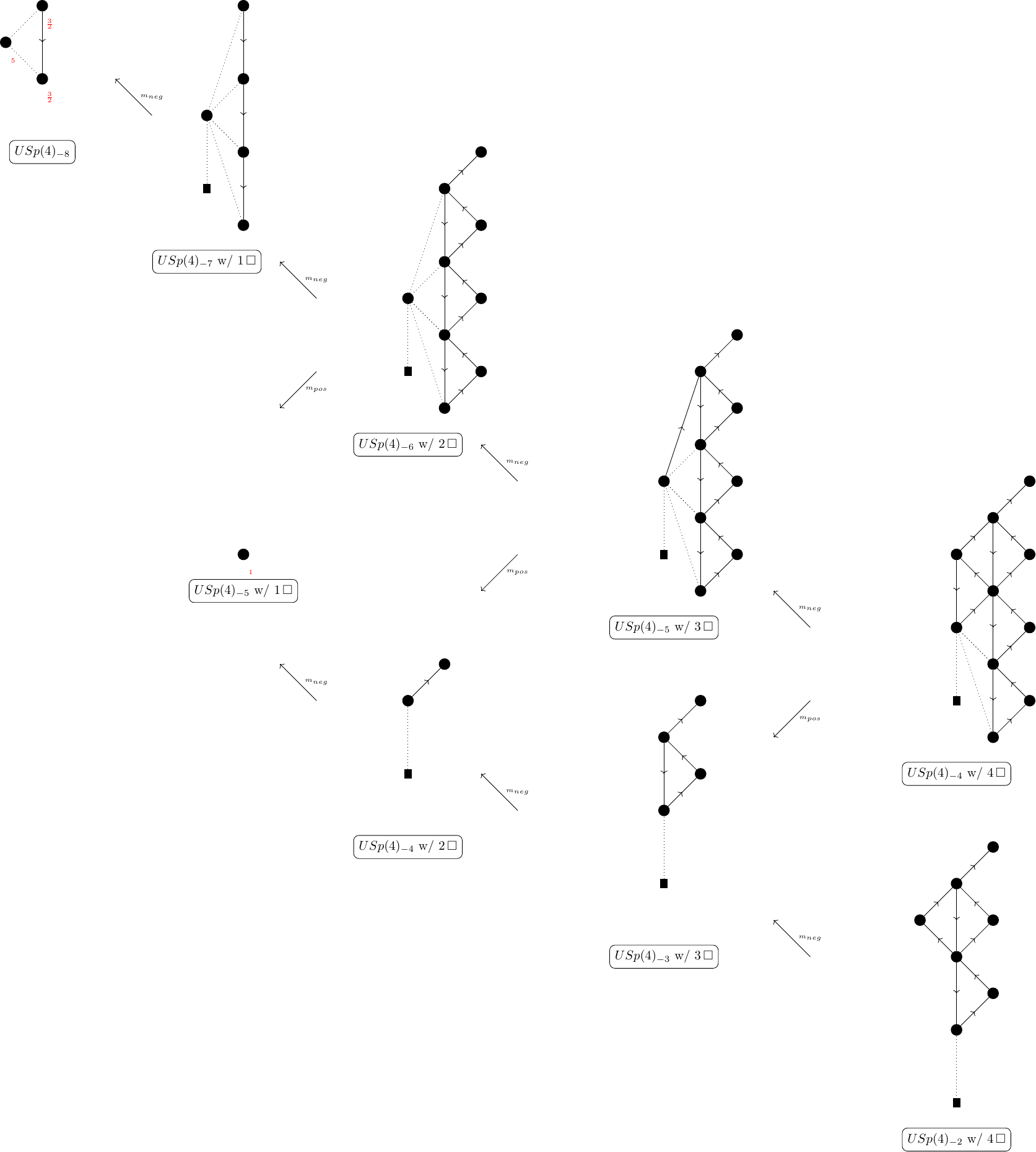}
    \caption{We continue the analysis of the RG flow trajectories induced by $\mposneg$ mass deformations, extending those considered in Figure \ref{fig: zoology}. Among the theories appearing in these flows, only $USp(4)_{-4}$ SQCD and $USp(4)_{-2}$ SQCD with $[4]$ fundamental multiplets admit non-trivial RG flows under $\mposneg$ deformations. Accordingly, we only focus on $USp(4)_{-5}$ SQCD and $USp(4)_{-3}$ SQCD with $[3]$ fundamental multiplets, and follow the flow until all matter is integrated out. As before, details of the CS levels and FI terms can be recovered following our usual conventions.}
    \label{fig: zoology_2}
\end{figure}
\end{landscape}

\subsubsection{S-Confinement}\label{sec: confinement}
Recall that the Aharony-like dual of \textit{electric} $USp(2N)_k$ SQCD with $F$
fundamental multiplets is a \textit{magnetic} $USp(2\tilde N)_{-k}$ SQCD with $F$
antifundamental multiplets, together with $\binom{F}{2}$ gauge-singlet chiral
fields. In the region of parameter space defined by
\[
(F < 2N + 2) \,\cap\, (2N + 2 -F = |k|),
\]
the rank of the magnetic gauge group vanishes, $\tilde N = 0$. As a consequence,
the theory \emph{s-confines} in the deep infrared, and is dual to a Lagrangian that does not involve gauge fields. This behavior
is completely transparent in the Aharony-like magnetic description.

Furthermore,
in the context of Figure~\ref{fig: zoology}, it is known that
$USp(4)_{\pm 1}$ SQCD with $5$ fundamental multiplets and
$USp(4)_{\pm 2}$ SQCD with $4$ fundamental multiplets  both
\textit{s-confine} into a set of free chiral fields. We verify that this expectation is indeed borne out: the
corresponding planar duals of these theories can be made to \textit{s-confine} into free chiral fields
through an appropriate sequence of local dualizations.

We point out that in the corresponding $\mathcal N=4$ parent theories, the deep infrared is characterized by symmetry enhancement that is not visible in the UV Lagrangian \cite{Ferlito:2016grh,Cremonesi_2014, Assel:2018exy,Nawata:2021nse, Nawata:2023rdx}. By contrast, in the planar $\mathcal N=2$ theories obtained after the axial mass deformation, no such enhancement takes place: the UV Abelianized description already accounts for the full Cartan of the infrared symmetry.

\subsubsection*{Insights from Index Spectroscopy}
It is instructive to examine more closely the superconformal index of the \textit{s-confining} theories discussed above. The index of a theory consisting of $F$ free chiral multiplets has a particularly distinctive structure. Assigning a trial R-charge $r$ and a fugacity $\phi_i$ to each free field, the index takes the form
\begin{equation}
\begin{split}
\mathcal{I}_{\rm free}
&= 1 + x^r \left( \sum_{i=1}^F \phi_i \right)
- x^{2-r} \left( \sum_{i=1}^F \phi_i^{-1} \right)
- x^2 \left( \sum_{i,j=1}^F \frac{\phi_i}{\phi_j} \right) + \dots \\
&= 1 + x^r \, \chi_{\Box}(\phi)
- x^{2-r} \, \chi_{\overline{\Box}}(\phi)
- x^2 \, \chi_{\rm adj}(\phi) + \dots \, .
\end{split}
\end{equation}
Beyond the vacuum contribution, the term of order $x^r$ arises from the bosonic primaries of each chiral multiplet and corresponds to the character of the fundamental representation of $U(F)$. The contribution at order $x^{2-r}$ originates from the fermionic primaries and transforms in the antifundamental representation of $U(F)$. Finally, the term of order $x^2$ counts the conserved currents of the theory\footnote{The term at order $x^2$ may in principle receive additional contributions from other composite operators; throughout this discussion, we assume that the trial R-charge $r$ is chosen so as to avoid this possibility.} and correctly reproduces the character of the adjoint representation of $U(F)$.

The salient features of the superconformal index of a free theory can therefore be summarized as follows:
\begin{itemize}
\item There is a bosonic positive contribution at order $x^r$ and a fermionic negative contribution at order $x^{2-r}$; these two terms transform in conjugate representations of the flavor symmetry group.
\item The contribution at order $x^2$ is precisely given by the product of the first bosonic and fermionic terms.
\end{itemize}

These properties must persist even when the full infrared symmetry group is not manifest from an ultraviolet perspective. This is precisely what occurs in the \textit{s-confining} theories discussed here. As an illustrative example, consider the $USp(4)_1$ theory with five chiral multiplets, which is known to confine \cite{Benini_2011a}. In the ultraviolet, the symmetry group is $U(5)$, acting on the five chiral fields, and the ten gauge-invariant mesons transform in the two-index antisymmetric representation $\Lambda^2$ of this group. In the infrared, however, the symmetry is enhanced to $U(10)$ because this theory is dual to $10$ free chiral fields. 
Since the confined theory is free, the two structural properties of the index discussed above must be realized, even though we can only refine the index with respect to the fugacities associated to the Cartan of $U(5)$ visible in the ultraviolet.

The mechanism underlying this symmetry enhancement is highly non-trivial and relies on delicate cancellations among different contributions to the superconformal index, as can be verified by explicit computations. While the mesons and the 25 conserved currents of $U(5)$ are captured by the zero GNO flux sector, the fermionic contribution at order $-x^{2-2r}$, together with the remaining conserved currents, arises from the $+1$ GNO flux sector\footnote{The Weyl group of $USp(2N)$ is $(\mathbb{Z}_2)^N \rtimes S_N$, acting on the GNO flux vector $(m_1,\dots,m_N)$ by permutations and sign inversions. As a result, the $+1$ sector is uniquely characterized by the GNO configuration $(+1,0,\dots,0)$.}. Note that the tensor product $\Lambda^2 \otimes \overline{\Lambda}^2$ always contains the adjoint representation, corresponding to the classical contribution to the current algebra.

It is worth emphasizing that it is quite remarkable that contributions from different GNO flux sectors recombine to reproduce the product structure characteristic of a free theory. This phenomenon results from the interplay between the matter and vector multiplet contributions and the specific value of the Chern-Simons level. It is therefore natural to ask how this structure is modified as the Chern-Simons level is varied. In Table~\ref{tab: close-to-confinement indices}, we report the values of the superconformal index for several choices of the level $k$.

\begin{table}[ht]
\centering
\begin{tabular}{||c|c||} 
 \hline
  $k$ & Superconformal Index \\
 \hline\hline

1 &  
$\begin{array}{l}
1+x^{2r} \chi_{\Lambda^2}(\phi) - x^{2-2r}\chi_{\overline{\Lambda}^2} (\phi) -x^2 \chi_{\Lambda^2}(\phi) \chi_{\overline{\Lambda}^2} (\phi) + \dots
\end{array}$ \\
\hline

2 & 
$\begin{array}{l}
1 + x^{2r} \chi_{\Lambda^2}(\phi) - x^{3-3r} \chi_{\overline{\Lambda}^3} (\phi) -x^2\chi_{\rm adj}(\phi) + \dots
\end{array}$ \\
\hline

3 & 
$\begin{array}{l}
1+ x^{2r} \chi_{\Lambda^2}(\phi) - x^{4-4r} \chi_{\overline{\Lambda}^3}(\phi) -x^2\chi_{\rm adj}(\phi) + \dots
\end{array}$ \\
\hline

4 & 
$\begin{array}{l}
1+x^{2r}\chi_{\Lambda^2}(\phi) - x^{5-5r} \chi_{\overline{\Lambda}^5}(\phi) -x^2\chi_{\rm adj}(\phi) + \dots
\end{array}$ \\
\hline
 
\end{tabular}
\caption{Values of the superconformal index for $USp(4)_{\pm k}$ SQCD$_3$ with five chiral fields are shown here. All fields are assigned a trial R-charge $r$, and $\phi$ denotes the fugacities of the UV flavor symmetry group $U(5)$. The notation $\Lambda^n$ refers to the $n$-index antisymmetric representation of $U(5)$, while $\overline{\Lambda}^n$ denotes its complex conjugate.}

\label{tab: close-to-confinement indices}
\end{table} 
\newpage

As we vary $k$, the rank of the representation appearing in the fermionic contribution changes by one unit, while its R-charge shifts by $1-r$. Confinement is reached when the representation becomes $\overline{\Lambda}^2$. At this point, the coefficient of the $x^2$ term factorizes automatically, in agreement with the expected confining behavior.
\subsubsection*{Confinement via sequential confinement}
We now show that the \textit{s-confinement} of $USp(2N)_{1}$ SQCD with $2N+1$ fundamental multiplets can be understood by performing a sequence of local dualizations on the corresponding planar Abelian dual.

We begin with $\mathcal{N}=2$ $USp(4)_1$ SQCD with $5$ fundamental fields, which is dual to 10 free chiral fields transforming in the antisymmetric representation of the $U(5)$ flavor group — a limiting case of Aharony-like duality. The planar Abelian dual is obtained by performing a positive mass deformation of $USp(4)_0$ with $6$ fundamentals \eqref{eq: Level2_C2_k=0}:
\begin{equation}
    USp(4)_{1} \text{ with 5 }\square 
    \qquad\leftrightarrow\qquad
    \begin{tikzpicture}[baseline=(current bounding box).center, scale=0.6]

        \foreach \x/\y in {1/1, 2/1, 2/2, 3/1, 3/2, 3/3, 4/1, 4/2, 4/3, 4/4}
        {
            \node at (\x,2*\y-\x) (g\x\y) [gaugefill] {};
        }
        \node at (3,-3) (f) [flavorfill] {};

        \foreach \nout/\nin in {f/g41, g41/g31, g31/g21, g21/g11, g42/g32, g32/g22, g43/g33, g31/g42, g21/g32, g32/g43, g11/g22, g22/g33, g33/g44, g22/g21, g33/g32, g32/g31}
        {
            \draw[->-] (\nout) -- (\nin);
        }
    \end{tikzpicture}
\end{equation}

We then perform local dualizations column by column, proceeding from right to left. At each step, the top node of the column confines and produces a free chiral field. The quiver after the first dualization pass is:\newpage
\begin{equation}
        \begin{tikzpicture}[baseline=(current bounding box).center, scale=0.6]

        \foreach \x/\y in {1/1, 2/1, 2/2, 3/1, 3/2, 3/3, 4/1, 4/2, 4/3, 4/4}
        {
            \node at (\x,2*\y-\x) (g\x\y) [gaugefill] {};
        }
        \node at (3,-3) (f) [flavorfill] {};

        \foreach \nout/\nin in {f/g41, g41/g31, g31/g21, g21/g11, g42/g32, g32/g22, g43/g33, g31/g42, g21/g32, g32/g43, g11/g22, g22/g33, g33/g44, g22/g21, g33/g32, g32/g31}
        {
            \draw[->-] (\nout) -- (\nin);
        }
    \end{tikzpicture}
    \qquad \xrightarrow{\text{dualize columns 4,3,2,1}} \qquad
        \begin{tikzpicture}[baseline=(current bounding box).center, scale=0.6]

        \foreach \x/\y in { 2/1, 3/1, 3/2,  4/1, 4/2, 4/3}
        {
            \node at (\x,2*\y-\x) (g\x\y) [gaugefill] {};
        }
        \node at (3,-3) (f) [flavorfill] {};

        \foreach \nout/\nin in {g41/f, g31/f, g21/f, g41/g31, g31/g21,  g42/g32, g31/g42, g21/g32, g32/g43,  g32/g31, g43/g42, g42/g41}
        {
            \draw[->-] (\nout) -- (\nin);
        }
        \draw (6,-4) node {$+$ 4 free chirals};
    \end{tikzpicture}
\end{equation}

The four additional chiral fields produced in this pass correspond to four of the electric mesons. We note that at every step the duality maps a $U(1)$ gauge group either to another $U(1)$ gauge group or to a Wess-Zumino model, so the quiver remains Abelian throughout and can be written in the compact notation adopted in this paper. We refer the reader to \cite{Benvenuti:2026a} for further details on analogous dualization procedures.

The full dualization sequence proceeds as follows:
\begin{equation}
\begin{tikzpicture}[baseline=(current bounding box).center, scale=0.6]

        \foreach \x/\y in { 2/1, 3/1, 3/2,  4/1, 4/2, 4/3}
        {
            \node at (\x,2*\y-\x) (g\x\y) [gaugefill] {};
        }
        \node at (3,-3) (f) [flavorfill] {};

        \foreach \nout/\nin in {g41/f, g31/f, g21/f, g41/g31, g31/g21,  g42/g32, g31/g42, g21/g32, g32/g43,  g32/g31, g43/g42, g42/g41}
        {
            \draw[->-] (\nout) -- (\nin);
        }
        \draw (3,-4) node {$+$ 4 free chirals};
    \end{tikzpicture}
     \qquad \xrightarrow{\text{dualize columns $2, 3,{\color{red}4}, 4, {\color{red}3}, 3, 2, {\color{red}4}, {\color{red}2}, {\color{red}3}, {\color{red}4}$}} \qquad
     \text{ 10 free chirals}
\end{equation}
where the steps highlighted in red are those in which a gauge node confines, producing a free chiral field. At the end of this sequence, the quiver fully confines into 10 free chiral fields. Tracking the charges of these fields confirms that they transform in the antisymmetric representation of $U(5)$, in agreement with the expected confinement of $USp(4)_1$ SQCD with $5$ fundamentals.
\\

By an analogous dualization sequence, one can similarly show that $USp(4)_2$ SQCD with $4$ fundamentals confines into 6 free chiral fields transforming in the antisymmetric representation of $U(4)$, again in agreement with Aharony-like duality. We do not reproduce the individual dualization steps here.\\

The procedure above extends to the general case of $USp(2N)_{1}$ with $F=2N+1$ 
fundamentals, which is known to confine to a set of free chiral fields transforming 
in the antisymmetric representation of $U(2N+1)$.

We proceed by induction, assuming the confinement of $USp(2N-2)_{1}$ with $F=2N-1$ 
fundamentals, with the induction seed given by the case of $USp(4)_1$ with 5 
fundamentals discussed above. We start from the planar Abelian duality:

\begin{equation}    \label{eq:confinement_gen_1}
    USp(2N)_{1} \text{ with $2N+1$ }\square 
    \qquad\leftrightarrow\qquad
    \begin{tikzpicture}[baseline=(current bounding box).center, scale=0.6]
    \newcommand{\tikzprefix}{FIG1} 

        \foreach \x/\y in {1/1, 2/1, 2/2, 3/1, 3/2, 3/3, 4/1, 4/2, 4/3, 4/4, 6/1, 6/5, 6/6, 7/1, 7/5, 7/6, 7/7}
        {
            \node at (\x,2*\y-\x) (\tikzprefix-g\x\y) [gaugefill] {};
        }

        \node at (6,-6) (f) [flavorfill] {};
        \draw[->-] (f) -- (\tikzprefix-g71);

        \foreach \nout/\nin in {g41/g31, g31/g21, g21/g11, g42/g32, g32/g22, g43/g33, g31/g42, g21/g32, g32/g43, g11/g22, g22/g33, g33/g44, g22/g21, g33/g32, g32/g31,
        g44/g43, g43/g42, g42/g41,
        g71/g61, g75/g65, g65/g76, g76/g66, g66/g77,
        g66/g65}
        {
            \ifcsname pgf@sh@ns@\tikzprefix-\nout\endcsname
                \ifcsname pgf@sh@ns@\tikzprefix-\nin\endcsname
                    \draw[->-] (\tikzprefix-\nout) -- (\tikzprefix-\nin);
                \fi
            \fi
        }

        \foreach \n in {g41, g42, g43, g44, g61}
        {
            \ifcsname pgf@sh@ns@\tikzprefix-\n\endcsname
                \draw[->-] (\tikzprefix-\n) -- ++ (.5,0.5);
            \fi
        }
        \foreach \n in {g41, g42, g43, g44}
        {
            \ifcsname pgf@sh@ns@\tikzprefix-\n\endcsname
                \draw[-<-] (\tikzprefix-\n) -- ++ (.5,-0.5);
            \fi
        }
        \foreach \n in {g66, g65, g75}
        {
            \ifcsname pgf@sh@ns@\tikzprefix-\n\endcsname
                \draw[-<-] (\tikzprefix-\n) -- ++ (-.5,-0.5);
            \fi
        }
        \foreach \n in {g65, g61}
        {
            \ifcsname pgf@sh@ns@\tikzprefix-\n\endcsname
                \draw[->-] (\tikzprefix-\n) -- ++ (-.5,0.5);
            \fi
        }
        \foreach \n in { g61}
        {
            \ifcsname pgf@sh@ns@\tikzprefix-\n\endcsname
                \draw[-<-] (\tikzprefix-\n) -- ++ (0,1);
            \fi
        }

        \path (\tikzprefix-g44) -- node[midway] {$\iddots$} (\tikzprefix-g66);
        \path (\tikzprefix-g43) -- node[midway] {$\iddots$} (\tikzprefix-g65);
        \path (\tikzprefix-g65) -- node[midway] {$\vdots$} (\tikzprefix-g61);

        \draw [decorate,decoration={brace,amplitude=5pt,raise=2ex}]
        (\tikzprefix-g11)--(\tikzprefix-g77) node[midway,xshift=-1em,yshift=1.8em]{\tiny$2N$}; 

    \end{tikzpicture}
\end{equation}
Dualizing columns from right to left, we obtain:

\begin{equation}    \label{eq:confinement_gen_2}
    \eqref{eq:confinement_gen_1}\qquad
    \qquad\xrightarrow{\text{dualize columns $2N, 2N-1, \dots, 1$}} \qquad
    \begin{tikzpicture}[baseline=(current bounding box).center, scale=0.6]
    \newcommand{\tikzprefix}{FIG2} 

        \foreach \x/\y in {2/1, 3/1, 3/2, 4/1, 4/2, 4/3, 6/1, 6/5, 7/1, 7/5, 7/6}
        {
            \node at (\x,2*\y-\x) (\tikzprefix-g\x\y) [gaugefill] {};
        }

        \node at (6,-6) (f) [flavorfill] {};
        \draw[->-] (\tikzprefix-g71) -- (f);
        \draw[->-] (\tikzprefix-g61) -- (f);
        \draw[->-] (\tikzprefix-g41) -- (f);
        \draw[->-] (\tikzprefix-g31) -- (f);
        \draw[->-] (\tikzprefix-g21) -- (f);

        \foreach \nout/\nin in {g41/g31, g31/g21, g21/g11, g42/g32, g32/g22, g43/g33, g31/g42, g21/g32, g32/g43, g11/g22, g22/g33, g33/g44, g22/g21, g33/g32, g32/g31,
        g44/g43, g43/g42, g42/g41,
        g71/g61, g75/g65, g65/g76, g76/g66, g66/g77,
        g66/g65,
        g76/g75}
        {
            \ifcsname pgf@sh@ns@\tikzprefix-\nout\endcsname
                \ifcsname pgf@sh@ns@\tikzprefix-\nin\endcsname
                    \draw[->-] (\tikzprefix-\nout) -- (\tikzprefix-\nin);
                \fi
            \fi
        }

        \foreach \n in {g41, g42, g43, g44, g61}
        {
            \ifcsname pgf@sh@ns@\tikzprefix-\n\endcsname
                \draw[->-] (\tikzprefix-\n) -- ++ (.5,0.5);
            \fi
        }
        \foreach \n in {g41, g42, g43, g44}
        {
            \ifcsname pgf@sh@ns@\tikzprefix-\n\endcsname
                \draw[-<-] (\tikzprefix-\n) -- ++ (.5,-0.5);
            \fi
        }
        \foreach \n in {g66, g65, g75}
        {
            \ifcsname pgf@sh@ns@\tikzprefix-\n\endcsname
                \draw[-<-] (\tikzprefix-\n) -- ++ (-.5,-0.5);
            \fi
        }
        \foreach \n in { g61}
        {
            \ifcsname pgf@sh@ns@\tikzprefix-\n\endcsname
                \draw[->-] (\tikzprefix-\n) -- ++ (-.5,0.5);
            \fi
        }
        \foreach \n in { g75}
        {
            \ifcsname pgf@sh@ns@\tikzprefix-\n\endcsname
                \draw[->-] (\tikzprefix-\n) -- ++ (0,-1);
            \fi
        }
        \foreach \n in { g71,g61}
        {
            \ifcsname pgf@sh@ns@\tikzprefix-\n\endcsname
                \draw[-<-] (\tikzprefix-\n) -- ++ (0,1);
            \fi
        }

        \path (\tikzprefix-g43) -- node[midway] {$\iddots$} (\tikzprefix-g65);
        \path (\tikzprefix-g65) -- node[midway] {$\vdots$} (\tikzprefix-g61);

        \draw [decorate,decoration={brace,amplitude=5pt,raise=2ex}]
        (\tikzprefix-g21)--(\tikzprefix-g76) node[midway,xshift=-1em,yshift=1.8em]{\tiny$2N-1$}; 

        \node at (2,-5) (chirals) {$+2N$ free chirals};

    \end{tikzpicture}
\end{equation}

where the $2N$ additional free chiral fields are generated when the top node of each 
column confines. We then perform a ``flip-flip'' dualization pass \cite{Benvenuti:2025a}: 
that is, we dualize columns from left to right, then repeat, stopping at the second-to-last 
column, then the third-to-last, and so on. The resulting quiver is:
\begin{equation}    \label{eq:confinement_gen_3}
\eqref{eq:confinement_gen_2}\qquad
    \xrightarrow{\text{flip-flip}}\qquad
    \begin{tikzpicture}[baseline=(current bounding box).center, scale=0.6]
    \newcommand{\tikzprefix}{FIG3} 

        \foreach \x/\y in {3/1, 4/1, 4/2, 6/1, 6/4,  7/1, 7/4, 7/5}
        {
            \node at (\x,2*\y-\x) (\tikzprefix-g\x\y) [gaugefill] {};
        }

        \node at (6,-6) (f) [flavorfill] {};
        \draw[-<-] (\tikzprefix-g71) -- (f);

        \foreach \nout/\nin in {g41/g31, g31/g21, g21/g11, g42/g32, g32/g22, g43/g33, g31/g42, g21/g32, g32/g43, g11/g22, g22/g33, g33/g44, g22/g21, g33/g32, g32/g31,
        g44/g43, g43/g42, g42/g41,
        g71/g61, g75/g65, g65/g76, g76/g66, g66/g77,
        g66/g65, g64/g75, g74/g64,
        g76/g75, g75/g74}
        {
            \ifcsname pgf@sh@ns@\tikzprefix-\nout\endcsname
                \ifcsname pgf@sh@ns@\tikzprefix-\nin\endcsname
                    \draw[-<-] (\tikzprefix-\nout) -- (\tikzprefix-\nin);
                \fi
            \fi
        }

        \foreach \n in {g41, g42, g43, g44, g61}
        {
            \ifcsname pgf@sh@ns@\tikzprefix-\n\endcsname
                \draw[-<-] (\tikzprefix-\n) -- ++ (.5,0.5);
            \fi
        }
        \foreach \n in {g41, g42, g43, g44}
        {
            \ifcsname pgf@sh@ns@\tikzprefix-\n\endcsname
                \draw[->-] (\tikzprefix-\n) -- ++ (.5,-0.5);
            \fi
        }
        \foreach \n in {g66, g65, g64, g74}
        {
            \ifcsname pgf@sh@ns@\tikzprefix-\n\endcsname
                \draw[->-] (\tikzprefix-\n) -- ++ (-.5,-0.5);
            \fi
        }
        \foreach \n in { g61}
        {
            \ifcsname pgf@sh@ns@\tikzprefix-\n\endcsname
                \draw[-<-] (\tikzprefix-\n) -- ++ (-.5,0.5);
            \fi
        }
        \foreach \n in { g74, g64}
        {
            \ifcsname pgf@sh@ns@\tikzprefix-\n\endcsname
                \draw[-<-] (\tikzprefix-\n) -- ++ (0,-1);
            \fi
        }
        \foreach \n in { g71, g61}
        {
            \ifcsname pgf@sh@ns@\tikzprefix-\n\endcsname
                \draw[->-] (\tikzprefix-\n) -- ++ (0,1);
            \fi
        }

        \path (\tikzprefix-g42) -- node[midway] {$\iddots$} (\tikzprefix-g64);
        \path (\tikzprefix-g64) -- node[midway] {$\vdots$} (\tikzprefix-g61);

        \draw [decorate,decoration={brace,amplitude=5pt,raise=2ex}]
        (\tikzprefix-g31)--(\tikzprefix-g75) node[midway,xshift=-1em,yshift=1.8em]{\tiny$2N-2$}; 

        \node at (4,-7) (chirals) {$+2N + (2N-1)$ free chirals};

    \end{tikzpicture}
\end{equation}

where $2N-1$ additional free chiral fields are generated when the top node of each column 
confines. We then perform another ``flip-flip'' dualization pass up to the second-to-last 
column, obtaining:
\begin{equation}    \label{eq:confinement_gen_4}
    \eqref{eq:confinement_gen_3}\qquad
    \xrightarrow{\text{flip-flip}}
    \begin{tikzpicture}[baseline=(current bounding box).center, scale=0.6]
    \newcommand{\tikzprefix}{FIG4} 

        \foreach \x/\y in {3/1, 4/1, 4/2, 6/1, 6/4,  7/1, 7/2, 7/4, 7/5}
        {
            \node at (\x,2*\y-\x) (\tikzprefix-g\x\y) [gaugefill] {};
        }

        \node at (6,-6) (f) [flavorfill] {};
        \draw[-<-] (\tikzprefix-g71) -- (f);

        \foreach \nout/\nin in {g41/g31, g31/g21, g21/g11, g42/g32, g32/g22, g43/g33, g31/g42, g21/g32, g32/g43, g11/g22, g22/g33, g33/g44, g22/g21, g33/g32, g32/g31,
        g44/g43, g43/g42, g42/g41,
        g71/g61, g75/g65, g65/g76, g76/g66, g66/g77,
        g61/g72,
        g66/g65, g64/g75, g74/g64}
        {
            \ifcsname pgf@sh@ns@\tikzprefix-\nout\endcsname
                \ifcsname pgf@sh@ns@\tikzprefix-\nin\endcsname
                    \draw[->-] (\tikzprefix-\nout) -- (\tikzprefix-\nin);
                \fi
            \fi
        }

        \foreach \n in {g41, g42, g43, g44}
        {
            \ifcsname pgf@sh@ns@\tikzprefix-\n\endcsname
                \draw[->-] (\tikzprefix-\n) -- ++ (.5,0.5);
            \fi
        }
        \foreach \n in {g41, g42, g43, g44}
        {
            \ifcsname pgf@sh@ns@\tikzprefix-\n\endcsname
                \draw[-<-] (\tikzprefix-\n) -- ++ (.5,-0.5);
            \fi
        }
        \foreach \n in {g66, g65, g64, g74}
        {
            \ifcsname pgf@sh@ns@\tikzprefix-\n\endcsname
                \draw[-<-] (\tikzprefix-\n) -- ++ (-.5,-0.5);
            \fi
        }
        \foreach \n in { g61, g72}
        {
            \ifcsname pgf@sh@ns@\tikzprefix-\n\endcsname
                \draw[->-] (\tikzprefix-\n) -- ++ (-.5,0.5);
            \fi
        }
        \foreach \n in {g64}
        {
            \ifcsname pgf@sh@ns@\tikzprefix-\n\endcsname
                \draw[->-] (\tikzprefix-\n) -- ++ (0,-1);
            \fi
        }
        \foreach \n in {g61}
        {
            \ifcsname pgf@sh@ns@\tikzprefix-\n\endcsname
                \draw[-<-] (\tikzprefix-\n) -- ++ (0,1);
            \fi
        }

        \path (\tikzprefix-g42) -- node[midway] {$\iddots$} (\tikzprefix-g64);
        \path (\tikzprefix-g64) -- node[midway] {$\vdots$} (\tikzprefix-g61);

        \draw [decorate,decoration={brace,amplitude=5pt,raise=2ex}]
        (\tikzprefix-g31)--(\tikzprefix-g75) node[midway,xshift=-1em,yshift=1.8em]{\tiny$2N-2$}; 

        \node at (4,-7) (chirals) {$+2N + (2N-1)$ free chirals};

    \end{tikzpicture}
\end{equation}
This pass does not generate any additional free chiral fields. The resulting quiver is 
equivalent to the planar Abelian dual of $USp(2N-2)_{1}$ with $F=2N-1$ fundamentals, 
corresponding to the previous step in the induction \eqref{eq:confinement_gen_1}, which 
confines into $\frac{(2N-1)(2N-2)}{2}$ free chiral fields. Putting everything together, 
the planar Abelian dual of $USp(2N)_{1}$ with $2N+1$ fundamentals confines into 
$\frac{2N(2N+1)}{2}$ free chiral fields. Tracking the charges of these fields under the 
global symmetry confirms that they transform in the antisymmetric representation of 
$U(2N+1)$, completing the induction.

Similarly, one can address \textit{confinement} for $USp(2N)_{|k|}$ with 
$F=2N+2-|k|$ fundamentals for $|k|>0$. All these theories are known to confine into 
free chiral fields transforming in the antisymmetric representation of the flavor group. 
In each case, confinement can be understood by starting from the planar Abelian dual and 
performing an analogous sequence of local Abelian dualizations; we do not reproduce the 
details here.

There is a further family of symplectic SQCD theories known to confine, namely 
$USp(2N)_{0}$ with $F=2N+2$ fundamentals. These theories confine into 
$\frac{(2N+1)(2N+2)}{2}$ chiral fields $M_{ij}$ transforming in the antisymmetric 
representation of $U(2N+2)$, together with a singlet chiral field $Y$ corresponding 
to the monopole operator of the SQCD, interacting via the superpotential
\begin{equation}
    \mathcal{W}_{WZ} = Y\, \mathrm{Pf}(M).
\end{equation}
It would be interesting to understand this confinement via local dualizations of the 
planar Abelian dual. The limiting case of $USp(2)_0 \cong SU(2)_0$ SQCD with 4 
fundamentals was worked out in \cite{Benvenuti:2025a}. For the higher-rank cases, 
we are not aware of a dualization sequence that would establish \textit{s-confinement} 
along these lines, and we leave this to future work.

\acknowledgments

We are grateful to Gabriel Pedde Ungureanu, Sara Pasquetti, and Riccardo Comi for useful conversations. AS thanks Alessandro Piazza, Davide Bason, Leonardo Goller, Davide Morgante, and Marina Moleti for helpful discussions, and Mahak Bhushan for her hospitality (thanks to which a non-trivial part of this paper was written). SB is partially supported by the MUR-PRIN grant No. 2022NY2MXY. SR is supported by the MUR-PRIN grant No. 2022NY2MXY and the POC grant No. 41355/GRFVG.


\appendix

\section{Quiver Diagrams and \texorpdfstring{$\mathbf S^3_b$ Partition Functions}{Notation}}\label{app: notation}
We summarize the conventions (following \cite{Kapustin:2009kz, Hama_2011, Aharony:2013dha, Benini_2011a, Benvenuti:2024mpn, Benvenuti:2025a}) for quiver diagrams that are used throughout this work. A quiver diagram is built with the following fundamental ingredients:
\begin{itemize}
    \item \textbf{Nodes} (or vertices) encode the symmetries of the theory. Gauge symmetries are represented by circular nodes, while global symmetries are denoted by squares. In what follows, we consider classical Lie groups and summarize their respective contributions to the $\mathbf{S}^3_b$ partition function below:
    \begin{enumerate}
        \item the $\mathcal{N}=2$ vector multiplet associated with a \textbf{unitary}\footnote{The contribution of the $\mathcal{N}=2$ vector multiplet associated with a \textbf{special unitary} symmetry can be obtained by a straightforward modification of the integration measure of the unitary symmetry as follows:
        \begin{equation}
            \int\prod_{\alpha=1}^Ndu_{\alpha}\to \int\prod_{\alpha=1}^Ndu_{\alpha}\,\delta\bigg(\sum_{\alpha=1}^Nu_{\alpha}\bigg).
        \end{equation} An $SU(N+1)$ gauge symmetry will be labeled as $A_N$ in nodes of the quiver diagram.} symmetry is reported as follows:

        \begin{equation}
        \begin{tikzpicture}
    
        \node at (-1,.35) (g1) [gauge,black] {$N$};
        \draw[FIcolor] (g1)++(0,.5) node {$^\eta$};
        \draw[CScolor] (g1)++(.8,-.5) node {$_{(k,\,k+\ell N)}$};
        \node at (1,.35) {$\overset{Z_{\mathbf{S}^3_b}}{:=}$};
        \node at (5.5,0) {$\int\prod_{\alpha=1}^N\frac{du_{\alpha}\,e^{-\pi i\big(k\sum_{\alpha} u_{\alpha}^2  \,+\,\ell (\sum_{\alpha} u_{\alpha})^2\big)}\, e^{2\pi i \eta \sum_{\alpha}u_{\alpha}}}{N!\, \prod_{\alpha>\beta}\, s_b\bigg(\frac{iQ}{2}\pm(u_{\alpha}-u_{\beta})\bigg)}$};
        \end{tikzpicture}
        \end{equation}
        where $\{u_{\alpha}\}_{\alpha=1}^N$ are the fugacities associated with the generators of the Cartan subalgebra of the $U(N)$ gauge symmetry, the Chern-Simons level\footnote{A $U(N)$ gauge theory with CS interactions at level $(k, k+N\ell)$ has a Lagrangian of the form 
        \begin{equation*}
            -i\frac{k}{4\pi}\int tr(A\wedge dA) -i\frac{\ell}{4\pi} \int tr(A)\wedge tr(dA) + \text{ supersymmetric completion},
        \end{equation*}
        where $A$ is the gauge potential 1-form of the $U(N)$ gauge symmetry.} is indicated in red below the node, and the fugacity of the topological $U(1)_T\in U(N)$ (denoted by $\eta$) is indicated in orange above the node.

         \item the $\mathcal{N}=2$ vector multiplet associated with a \textbf{symplectic} symmetry is reported as follows:

        \begin{equation}
        \begin{tikzpicture}
        \node at (-.5,.35) (g1) [gauge,black] {$C_N$};
        \draw[CScolor] (g1)++(.5,-.5) node {$_{k}$};
        \node at (1,.35) {$\overset{\mathcal{Z}_{\mathbf{S}^3_b}}{:=}$};
        \node at (4.5,0) {$\int\prod_{\alpha=1}^N\frac{du_{\alpha}\,e^{-\pi i k\sum_{\alpha}u_{\alpha}^2}}{N!\,2^N\, \prod_{\alpha,\beta}\, s_b\bigg(\frac{iQ}{2}\pm u_{\alpha}\pm u_{\beta}\bigg)}$};
        \end{tikzpicture}
        \end{equation}
        where $\{u_{\alpha}\}_{\alpha=1}^N$ are the fugacities associated with the generators of the Cartan subalgebra of the $USp(2N)$ gauge symmetry, and the Chern-Simons level is indicated in red below the node.

         \item the $\mathcal{N}=2$ vector multiplet associated with a \textbf{special orthogonal} symmetry is reported as follows:

        \begin{equation}
        \begin{tikzpicture}
        \node at (-1.5,.35) (g1) [gauge,black] {$B_N$};
        \draw[CScolor] (g1)++(.5,-.5) node {$_{k}$};
        \node at (0,.35) {$\overset{\mathcal{Z}_{\mathbf{S}^3_b}}{:=}$};
        \node at (4.5,0) {$\int\prod_{\alpha=1}^N\frac{du_{\alpha}\,e^{-\pi i k\sum_{\alpha}u_{\alpha}^2}}{N!\,2^N\, \prod_{\alpha<\beta}\, s_b\bigg(\frac{iQ}{2}\pm u_{\alpha}\pm u_{\beta}\bigg)s_b\bigg(\frac{iQ}{2}\pm u_{\alpha}\bigg)}$};
        \end{tikzpicture}
        \end{equation}

        \begin{equation}
        \begin{tikzpicture}\begin{scope}[xshift=-10cm]
        \node at (-1.5,.35) (g1) [gauge,black] {$D_N$};
        \draw[CScolor] (g1)++(.5,-.5) node {$_{k}$};
        \node at (0,.35) {$\overset{\mathcal{Z}_{\mathbf{S}^3_b}}{:=}$};
        \node at (4.5,0) {$\int\prod_{\alpha=1}^N\frac{du_{\alpha}\,e^{-\pi i k\sum_{\alpha}u_{\alpha}^2}}{N!\,2^{N-1}\, \prod_{\alpha<\beta}\, s_b\bigg(\frac{iQ}{2}\pm u_{\alpha}\pm u_{\beta}\bigg)}$};
        \end{scope}
        \end{tikzpicture}
        \end{equation}             
        where $\{u_{\alpha}\}_{\alpha=1}^N$ are the fugacities associated with the generators of the Cartan subalgebra of the $SO(2N+1)$ and $SO(2N)$ gauge symmetries, and the Chern-Simons level is indicated in red below the node.
        
    \end{enumerate}
    \item \textbf{Edges} represent matter fields of the theory. 
    \begin{enumerate}
        \item The bifundamental $\mathcal{Q}$ with R-charge $R$ transforming under various symmetry groups contributes to the $\mathbf{S}^3_b$ partition function as follows:
    \begin{equation}
        \begin{tikzpicture}
            \node at (0,0) (f1) [flavor,black] {$M$};
            \node at (1.5,0) (f2) [flavor,black] {$N$};
            \draw[->-] (f2)--(f1);
            \node at (3,0){$\overset{\mathcal{Z}_{\mathbf{S}^3_b}}{:=}$};

            \node at (7,0) {$\prod_{j=1}^N\prod_{k=1}^Ms_b\bigg(\frac{iQ}{2}(1-R)-X_j + Y_k\bigg)$};
        \end{tikzpicture}   
    \end{equation}

     \begin{equation}
        \begin{tikzpicture}
            \node at (0,0) (f1) [flavor,black] {$M$};
            \node at (1.5,0) (f2) [flavor,black] {$C_N$};
            \draw[->-] (f2)--(f1);
            \node at (3,0){$\overset{\mathcal{Z}_{\mathbf{S}^3_b}}{:=}$};

            \node at (7,0) {$\prod_{j=1}^N\prod_{k=1}^Ms_b\bigg(\frac{iQ}{2}(1-R)\pm X_j + Y_k\bigg)$};
        \end{tikzpicture}   
    \end{equation}
    
    \begin{equation}
        \begin{tikzpicture}
            \node at (0,0) (f1) [flavor,black] {$C_M$};
            \node at (1.5,0) (f2) [flavor,black] {$D_N$};
            \draw[-] (f2)--(f1);
            \node at (3,0){$\overset{\mathcal{Z}_{\mathbf{S}^3_b}}{:=}$};

            \node at (7,0) {$\prod_{j=1}^N\prod_{k=1}^Ms_b\bigg(\frac{iQ}{2}(1-R)\pm X_j \pm Y_k\bigg)$};
        \end{tikzpicture}   
    \end{equation}

    \begin{equation}
        \begin{tikzpicture}
            \node at (0,0) (f1) [flavor,black] {$C_M$};
            \node at (1.5,0) (f2) [flavor,black] {$B_N$};
            \draw[-] (f2)--(f1);
            \node at (3,0){$\overset{\mathcal{Z}_{\mathbf{S}^3_b}}{:=}$};

            \node at (7,0) {$\prod_{j=1}^N\prod_{k=1}^Ms_b\bigg(\frac{iQ}{2}(1-R)\pm X_j \pm Y_k\bigg)$};
            \node at (6.1,-1){$\prod_{k=1}^Ms_b\bigg(\frac{iQ}{2}(1-R)\pm Y_k\bigg)$};
        \end{tikzpicture}   
    \end{equation}
    Note that an edge directed into (out of) a node denotes an anti-fundamental (fundamental) representation.
    \item The presence of an adjoint superfield in the theory is indicated by a closed loop:
    \begin{equation}
        \begin{tikzpicture}
            \node at (0,0) (g1) [flavor,black] {$G_N$};
            \draw[-] (g1) to[out=60,in=0] (0,0.5) to[out=180,in=120] (g1); 
        \end{tikzpicture}
    \end{equation}
    Their contributions to the $\mathbf{S}^3_b$ partition function can be inferred from those of the respective vector multiplets.
    \item The presence of a field with R-charge $R$ transforming in the rank-2 antisymmetric representation of a unitary symmetry (and its conjugate) is indicated as follows:
    \begin{equation}
        \begin{tikzpicture}
            \begin{scope}[xshift=-3.5cm]
            \node at (1.5,0) (ff1) {$\ydiagram{1,1}$};
            \node at (0,0) (g100) [flavor,black] {\tiny$N$};
            
            \draw[->] (g100)++(.3,.1)--(1.2,.1);
            \draw[->] (ff1)++(-.3,-.1)--(.3,-.1);
            \node at (2.5,0) {$\overset{\mathcal{Z}_{\mathbf{S}^3_b}}{:=}$};
            \end{scope}
            \node at (3,0) {$\prod_{i>j}^N\,s_b\bigg(\frac{iQ}{2}(1-R)\pm (X_i+X_j-s)\bigg)$};
        \end{tikzpicture}   
    \end{equation}
    where $\{X_i\}_{i=1}^N$ are the fugacities associated with the generators of the Cartan subalgebra of the $U(N)$ symmetry, and $s$ is the real mass parameter of the rank-2 antisymmetric tensor field.
     \item The presence of a mixed CS interaction at level $k_{XY}$ between the $U(1)\subset U(N)_{\vec{X}}$ and $U(1)\subset U(M)_{\vec{Y}}$ is reported as follows:
     \begin{equation}
        \begin{tikzpicture}
            \node at (0,0) (f1) [flavor,black] {$M$};
            \node at (1.5,0) (f2) [flavor,black] {$N$};
            \draw[dotted,-] (f2)--(f1);
            \draw[BFcolor] node at (.75,.25) {$^{k_{XY}}$};
            \node at (3,0){$\overset{\mathcal{Z}_{\mathbf{S}^3_b}}{:=}$};

            \node at (6,0) {$e^{-\pi i k_{XY}(\sum_j X_j)(\sum_k Y_k)}$};
        \end{tikzpicture}   
    \end{equation}
    
    \end{enumerate}
    \item There is a superpotential interaction associated with each \textbf{cycle} of the quiver diagram. For instance, consider the following quiver diagram:
    \begin{equation}
    \begin{tikzpicture}
        \node at (1,1.5) (g1) [gauge,black] {$G_N$};
        \node at (-1,-.5) (f1) [flavor,black] {$G_F$};
        \node at (3,-.5) (f2) [flavor,black] {$G_A$};
        \draw[->-, thick] (g1)--(f1);
        \draw[->-, thick] (f1)--(f2);
        \draw[->-, thick] (f2)--(g1);
        \node at (-.3,.6) {\tiny$X$};
        \node at (2.3,.6) {\tiny$Y$};
        \node at (1,-.2) {\tiny$Z$};
        \node at (4,.75) {$\longrightarrow$};
        \node at (6,.75) {$\mathcal{W}=XYZ$};
    \end{tikzpicture}
    \end{equation}
   Such interactions are collectively denoted as $\mathcal{W}_{planar}$.  
\end{itemize}
Quiver diagrams are central to this work, as they provide a compact encoding of all matter content and interactions.
When a chiral multiplet takes a large real mass we integrate it out producing an effective CS level. The corresponding operation on the $\mathbf{S}_b^3$ partition function is to take the following limit on the double-sine function $s_b$ encoding the contribution of that chiral multiplet:

\begin{equation*}
    \lim_{\xi\to\pm\infty} s_b(x+\xi) \sim e^{\pm\tfrac{\pi i}{2}(x+\xi)^2},
\end{equation*}

For all mass deformations studied in this paper we checked that the corresponding $\mathbf{S}_b^3$ partition functions factorize into a finite part and a divergent prefactor:
\begin{equation*}
    Z (\vec\mu;\theta)\,\overset{\theta\to\infty}{\longrightarrow}\, e^{\tfrac{\pi i}{2}\Phi(\theta)}\,Z(\vec\mu).
\end{equation*}
and the divergent prefactors match across our dualities when considering dual deformations.
This ensures that the $\mathbf{S}_b^3$ partition functions of the resulting theories agree.

\section{Quantum Numbers of Supersymmetric Monopoles}\label{app: monopole}
Here, we briefly review how to compute the charges of monopole operators (see  \cite{Aharony:1997bx,Benini_2011,Borokhov_2002} for a detailed discussion). Consider a $U(N)_{(k_1,k_2)}$ gauge theory and let $z_i$ be the gauge fugacity associated to the $i$-th Cartan generator. Due to quantum effects, a monopole with GNO flux $\vec{m} = (m_1, \dots,m_N)$ acquires an R-charge. The contribution of a single field $\phi$ with trial R-charge $r$ in representation $\mathcal{R}_\phi$ is given by:

\begin{equation}\label{eq: monopole r contribution}
    \Delta R = -\frac{1}{2} \sum_{\rho \in \mathcal{R}_\phi}(r-1) |\rho(m)|,
\end{equation}  where $\rho$ are the weights of the representation. The monopole is also charged under the gauge group: the contribution of $\phi$ is given by: 

\begin{equation}\label{eq: monopole gauge contribution}
    \Delta Q_{gauge}[\mathfrak{M}^{\vec{m}}] = -\frac{1}{2} \sum_{\rho \in \mathcal{R}_\phi} |\rho(m)| \rho(z).
\end{equation} As a prototypical example of the above formulae, consider $U(N)_{k_1,k_2}$with $[n_f,n_a]$ flavors. The R-charge of a bare monopole operator with GNO flux $\vec m$ is given by:
    \begin{equation}\label{eq: monopole r-charge}
        R[\mathfrak{M}^{\vec{m}}]=-\Lambda\sum_{i=1}^Nm_i+\frac{1}{2}\bigg[n_f(1-R_f)\sum_{i=1}^N|m_i| +n_a(1-R_a)\sum_{i=1}^N|m_i| -\sum_{j\neq i} |m_i-m_j| \bigg],
    \end{equation} 
where $R_f$ ($R_a$) denote the R-charges of the (anti-)fundamental chiral multiplets, while $\Lambda$ parametrizes the mixing between the topological symmetry and the $R$-symmetry. The value of $\Lambda$ can be extracted from the $\mathbf S^3_b$ partition function through the factor $e^{-\pi Q \Lambda\,\vec u}$.

In the same theory, the gauge charge of the given monopole is:

\begin{equation}\label{eq: monopole gauge charge}
    Q_{gauge}[\mathfrak{M}^{\vec{m}}] = k_1\sum_{i=1}^{N} m_iz_i + \ell \left(\sum_{i=1}^{N}m_i\right) \left(\sum_{j=1}^{N} z_j \right) + \frac{1}{2}(n_a-n_f)\sum_{i=1}^{N} |m_i|z_i.
\end{equation} 
Importantly, the above formula uniquely fixes the contribution of a Chern-Simons term to the superconformal index. We define exponentiated gauge fugacities $u_i := e^{i z_i}$; furthermore, assume that in the $\mathbf{S}^3_b$ partition function we have an FI term of the form $2\pi i X \, \sum_i^Nz_i$, so that we may define an exponentiated fugacity for the topological symmetry $\eta := e^{i X}$. The contribution of the Chern-Simons and FI terms to the superconformal index is:

\begin{equation}\label{eq: CS index contribution}
    \mathcal{I}_{CS+FI} = \left(\prod_{i=1}^N u_i^{k_1 m_i}\right) \left( \prod_{i=1}^N u_i \right)^{\ell \sum_{i=1}^N m_i} \, \eta^{-\sum_{i=1}^N m_i}.
\end{equation} Note the minus sign in the FI exponent, which is put for consistency with the partition function. The FI term may be interpreted as a BF coupling between the gauge field and a background field for the topological symmetry. Following our conventions of Appendix \ref{app: notation}, the FI term $2\pi i X \, \sum_{i=1}^Nz_i$ has the opposite sign with respect to other BF terms, thus it should appear with an opposite sign in the superconformal index as well.
As a result, according to our conventions, a monopole with GNO flux $m=+1$ has charge $-1$ under the topological symmetry labelled by $X$.

Another formula we will need in this work applies to the situation in which two $U(1)$ gauge nodes interact both through a BF coupling and through a bifundamental chiral field, as illustrated in Equation~\ref{fig: bifundamental quiver example}. This configuration represents the prototypical building block that appears in planar abelian theories. 

\begin{equation}
    \label{fig: bifundamental quiver example}
    \includegraphics[width=0.4\linewidth]{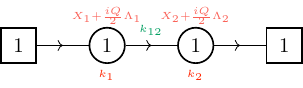}
\end{equation}

Consider a monopole with GNO flux $(m_1, m_2)$ under the two $U(1)$ gauge nodes. We assign trial R-charge $r_1$ to the first flavor, $r_{12}$ to the bifundamental field and $r_2$ to the second flavor. It is convenient to encode the charges carried by $\mathfrak{M}^{(m_1, m_2)}$ in the following polynomial of fugacities:

\begin{equation}\label{eq: monopole fugacity polynomial}
\begin{split}
    \mathcal{Q}[\mathfrak{M}^{(m_1,m_2)}]  &= k_1m_1z_1 + k_2 m_2z_2 + \frac{1}{2} k_{12} \left(m_1 z_2 +m_2 z_1 \right)  \\&+ \frac{1}{2} |m_1|z_1 - \frac{1}{2} |m_1 - m_2|(z_1 - z_2) - \frac{1}{2} |m_2| z_2 \\
    & + \frac{iQ}{2} \left(\frac{1}{2}(1-r_1) |m_1| + \frac{1}{2}(1-r_{12})|m_1-m_2| +  \frac{1}{2}(1-r_2)|m_2| \right)\\
    & -m_1\left( X_1 + \frac{iQ}{2} \Lambda_1\right) -m_2\left( X_2 + \frac{iQ}{2} \Lambda_2\right)
\end{split}
\end{equation} where the coefficient of $iQ/2$ is understood as the R-charge of the monopole. In particular:

\begin{itemize}
 \item the first line is the contribution of CS and BF terms to the gauge charge;
 \item the second line is the contribution of the fields to the gauge charge;
 \item the third line is the contribution of the fields to the R-charge;
 \item the last line is the contribution of the generalized FI terms, which contain both the topological symmetry and R-charge/topological mixing.
\end{itemize}

The key feature of this formula is that there are now two distinct contributions that induce a gauge charge for a monopole under the adjacent gauge node: the BF coupling and the bifundamental chiral field. In planar abelian gauge theories, the gauge invariance of specific monopole operators—namely those that map to mesonic operators on the electric side—is ensured precisely by the cancellation between these two effects.

It is important to stress that the formulae \eqref{eq: monopole r contribution}, \eqref{eq: monopole gauge contribution}, \eqref{eq: monopole r-charge}, \eqref{eq: monopole gauge charge} and \eqref{eq: monopole fugacity polynomial} are valid \emph{only} under the assumption that, in the monopole background, each field contributes through its lowest available modes.

\paragraph{An Example: $U(3)_{(2,2)}$ with $[6,2]$ chiral multiplets.}
As a simple application of the formulae discussed above, we compute the superconformal index of a $U(3)$ gauge theory with Chern--Simons levels $k_1=2$ and $\ell=0$, coupled to six fundamental chiral multiplets $Q_i$ and two antifundamental chiral multiplets $\tilde Q_j$. We assign a common trial R-charge $r=2/5$ to all matter fields.

Using formula~\eqref{eq: monopole gauge charge}, one readily verifies that the monopole operator with GNO flux $\vec m = (1,0,0)$ is gauge invariant:
\begin{equation}\label{eq: gauge monopole example}
   Q[\mathfrak{M}^{(1,0,0)}]
   = 2 z_1 + \frac{(2-6)}{2}\, z_1 = 0 .
\end{equation}
Its R-charge can then be computed from~\eqref{eq: monopole r-charge}, yielding
\begin{equation}\label{eq: r monopole example}
    R[\mathfrak{M}^{(1,0,0)}]
    = \frac{1}{2}\cdot 6\cdot\left(1-\frac{2}{5}\right)
    + \frac{1}{2}\cdot 2\cdot\left(1-\frac{2}{5}\right)
    -2
    = \frac{2}{5}.
\end{equation}
We therefore expect the superconformal index to contain a gauge-invariant monopole contribution at order $x^{2/5}$.

By Weyl symmetry, the GNO fluxes $(1,0,0)$, $(0,1,0)$, and $(0,0,1)$ are indistinguishable, and the above analysis applies equally to each of them. Introducing a topological fugacity $\eta^{-(m_1+m_2+m_3)}$, the superconformal index takes the form
\begin{equation}\label{eq: index example}
   \mathcal{I}
   = 1
   + \frac{1}{\eta}\, x^{2/5}
   + \left(\frac{1}{\eta^2}+12\right) x^{4/5}
   + \frac{12}{\eta}\, x^{6/5}
   + O(x^{8/5}) .
\end{equation}
As expected, the term at order $x^0$ corresponds to the vacuum. At order $x^{2/5}$ we find the gauge-invariant monopole operator, while at order $x^{4/5}$ we recover both its square and the $12$ mesons $Q_i \tilde Q_j$. Finally, the terms at order $x^{6/5}$ arise from monopole--meson composite operators, and higher-order contributions follow similarly.

\paragraph{An Example: A Theory with Multiple Gauge Groups.} As a second example, we compute the superconformal index of the theory shown in Figure~\ref{fig: bifundamental quiver example}. We choose Chern--Simons levels
\(
k_1 = 0,\ k_2 = 0,\ k_{12} = 1
\),
and denote by $\eta_1$ and $\eta_2$ the topological fugacities associated with the two gauge nodes. We further introduce a mixing of the topological symmetries with the $R$-symmetry given by $\Lambda_1 = 1/5$ and $\Lambda_2 = -2/5$.

With this choice of parameters, the monopole operators $\mathfrak{M}^{(-1,0)}$ and $\mathfrak{M}^{(0,1)}$ are gauge invariant. Assigning a common trial R-charge $r = 3/5$ to all matter fields, their charges are found to be:

\begin{equation}\label{eq: bifundamental quiver charges}
\begin{split}
    Q[\mathfrak{M}^{(-1,0)}] &= \frac{1}{2}(-z_2) + \frac{1}{2}z_1 - \frac{1}{2}(z_1-z_2) + \frac{iQ}{2}\left(\frac{1}{2}\cdot 2 \cdot (1-\frac{3}{5})\right) -(-1)\left(X_1 + \frac{iQ}{2} \cdot \frac{1}{5}\right)= \\
    &= X_1 + \frac{iQ}{2} \left( \frac{3}{5}\right) \\
    Q[\mathfrak{M}^{(0,1)}] &= \frac{1}{2}z_1 - \frac{1}{2}(z_1-z_2) - \frac{1}{2}z_2 + \frac{iQ}{2} \left( \frac{1}{2} \cdot 2 \cdot (1- \frac{3}{5}) \right) - \left( X_2 -\frac{iQ}{2} \cdot \frac{2}{5}\right) = \\
    &= -X_2 + \frac{iQ}{2} \left( \frac{4}{5} \right).
\end{split} 
\end{equation} Defining exponentiated fugacities $\eta_i = e^{iX_i}$ and computing the superconformal index explicitly, we find:

\begin{equation}\label{eq: bifundamental index example}
    \mathcal{I}
    = 1 + \eta_1\, x^{3/5} + \frac{x^{4/5}}{\eta_2} + \dots ,
\end{equation}
in agreement with the existence of the two gauge-invariant monopole operators.

\section{Towards Planar Abelian Duals of \texorpdfstring{$\mathcal N=2$}{N=2} Orthogonal SQCD}\label{app: planar bd}

Mirror duals of $\mathcal N=4$ orthogonal and symplectic SQCD can be engineered using \emph{Feng-Hanany} brane setups \cite{Feng:2000eq}, which incorporate orientifold planes in addition to the standard Hanany–Witten configurations. These constructions typically give rise to quivers with alternating orthogonal and symplectic gauge groups. Even in the presence of $\mathcal{N}=4$ supersymmetry, analyzing these theories remains highly nontrivial. This complexity arises from the emergence of various symmetries acting on the moduli space in the deep infrared, which are absent in the UV Lagrangian and hence invisible to supersymmetric localization. A number of tools, such as the \textit{magnetic quivers} paradigm, have been developed to study these theories \cite{Nawata:2021nse, Nawata:2023rdx,Harding:2025vov,Marino:2025ihk}, but a complete and detailed analysis remains intricate. Both $\mathcal{N}=4$ $U(N)$ and $USp(2N)$ SQCD with $F$ hypermultiplets admit RG flows under supersymmetry-breaking deformations to $\mathcal{N}=2$ vacua in which the gauge symmetries are Higgsed to their maximal tori, $U(1)^N$. Finding these vacua is nontrivial, but they form the building blocks of algorithmic planar dual constructions, which proceed by gluing together ``\textit{planarized}" $U(N)$ \cite{Benvenuti:2025a} or $USp(2N)$ \cite{Comi:wip} blocks.

\begin{landscape}
\thispagestyle{empty}
A natural question is whether $SO(2N)$ and $SO(2N+1)$ SQCD with $F$ hypermultiplets also admit such flows. The answer is affirmative: these theories likewise reach $\mathcal{N}=2$ vacua in which the gauge symmetries are Higgsed to $U(1)^N$, as illustrated in Equation \ref{eq:planarBD}. This observation is important because, even if a fully planar Abelianization of orthosymplectic quivers is not yet known, the existence of these planar limits suggests that a proto-algorithm for constructing orthosymplectic planar duals may still be feasible.

    \begin{equation}
    \label{eq:planarBD}
        \includegraphics[width= .6\linewidth]{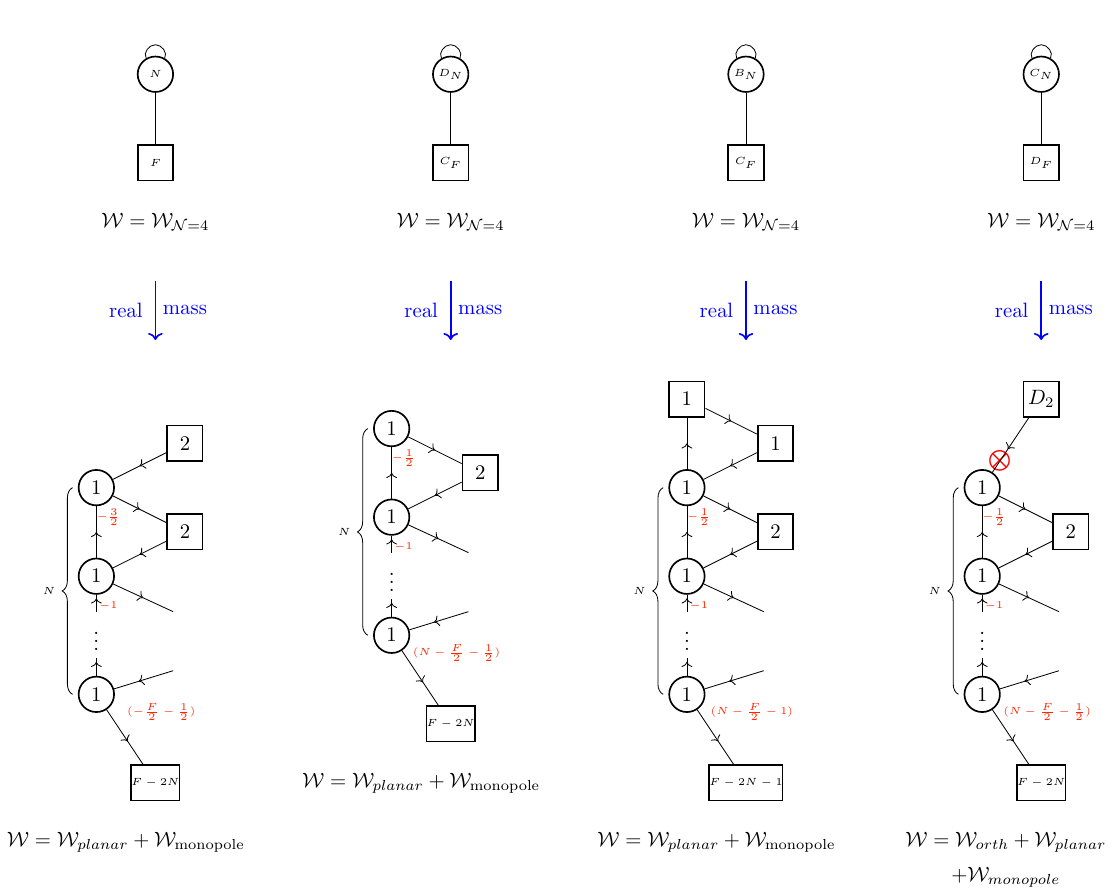}
    \end{equation}
\end{landscape}

\bibliographystyle{JHEP}
\bibliography{References}

\end{document}

%% file: custom_commands.tex
\definecolor{CScolor}{rgb}{1.0, 0.0, 0.0} 
\definecolor{BFcolor} {RGB} {0, 158, 96} 
\definecolor{FIcolor}{RGB}{250, 95, 85} 

\definecolor{bcolor}{RGB}{4, 55, 242} 
\definecolor{fcolor}{RGB}{255, 36, 0} 
\definecolor{labelcolor}{HTML}{808080} 

\newcommand{\NN}{\mathcal{N}}

\def\identityN[#1,#2]{
	_{#1}\mathbb{I}_{#2}^{[N]}
}

\def\identityOne[#1,#2]{
	_{#1}\mathbb{I}_{#2}^{[1^N]}
}

\def\DeltaN[#1]{
	\Delta^{[N]}\left( #1 \right)
}

\def\DeltaOne[#1]{
	\Delta^{[1^N]}\left( #1 \right)
}

\def\mpos{m_{pos}}
\def\mneg{m_{neg}}
\def\mposneg{m_{pos/neg}}

\newcommand{\tikznode}[2]{\relax
\ifmmode%
  \tikz[remember picture,baseline=(#1.base),inner sep=0pt]\node(#1){$#2$};
\else
  \tikz[remember picture,baseline=(#1.base),inner sep=0pt]\node(#1){#2};%
\fi}

\tikzstyle{BFline}=[dashed,black,draw]
\tikzstyle{farrowstyle}=[fcolor,thick,draw]
\tikzstyle{barrowstyle}=[bcolor,thick,draw]

\tikzstyle{flavor}=[rectangle,draw=black,thick,inner sep = 0pt, minimum size = 6mm]
\tikzstyle{flavor2}=[rectangle,draw=red!50,thick,inner sep = 0pt, minimum size = 4mm]
\tikzstyle{manifest}=[rectangle,draw=blue!50,thick,inner sep = 0pt, minimum size = 6mm]
\tikzstyle{gauge}=[circle,draw=black,thick,inner sep = 0pt, minimum size = 6mm] 
\tikzstyle{gauge2}=[circle,draw=black!50,thick,inner sep = 0pt, minimum size = 4.5mm] 
\tikzstyle{gauge3}=[rounded rectangle, draw=black!100, thick, minimum size=5mm] 
\tikzset{->-/.style={decoration={
  markings,
  mark=at position .5 with {\arrow{>}}},postaction={decorate}}}
\tikzset{-<-/.style={decoration={
  markings,
  mark=at position .5 with {\arrow{<}}},postaction={decorate}}}
\tikzstyle{BFline}=[dashed]
  
\tikzstyle{gaugefill}=[circle,draw=black, fill = black, thick,inner sep = 0pt, minimum size = 1.5mm]
\tikzstyle{flavorfill}=[rectangle,draw=black, fill = black,,thick,inner sep = 0pt, minimum size = 1.5mm]

\def\nodeCS(#1,#2)(#3,#4,#5){ 	
	\node at (#1,#2) (#3) [gauge,black] {#4};
	\draw[CScolor] (#3) ++(6pt,-7pt) node[anchor=west] {\tiny#5};
}

\def\flavorCS(#1,#2)(#3,#4,#5){ 	
	\node at (#1,#2) (#3) [flavor,black] {#4};
	\draw[CScolor] (#3) ++(6pt,-8pt) node[anchor=west] {\tiny#5};
}

\def\arrowBF(#1,#2)(#3){ 
	\begin{scope}[every node/.style={auto, outer sep=-1pt}]
	\path (#1) edge[->-] node[BFcolor,midway] {\tiny#3} (#2);
	\end{scope}
}

\def\arrowBFlr(#1,#2)(#3,#4){ 
	\begin{scope}[every node/.style={auto=#4, outer sep=-1pt}]
	\path (#1) edge[->-] node[BFcolor,midway] {\tiny#3} (#2);
	\end{scope}
}

\def\dottedBF(#1,#2)(#3){ 
	\begin{scope}[every node/.style={auto, outer sep=-1pt}]
	\path (#1) edge[BFline] node[BFcolor,midway] {\tiny#3} (#2);
	\end{scope}
}

\def\farrow(#1,#2,#3){
	\begin{scope}[every node/.style={auto=#3, outer sep=-1pt}]
	\path (#1) edge[->-,fcolor] node[fcolor,midway] {\tiny$\psi$} (#2);
	\end{scope}
}

\def\barrow(#1,#2,#3){
	\begin{scope}[every node/.style={auto=#3, outer sep=-1pt}]
	\path (#1) edge[->-,bcolor] node[bcolor,midway] {\tiny$\phi$} (#2);
	\end{scope}
}

\def\barrowBF(#1,#2,#3,#4){
	\begin{scope}[every node/.style={auto=#3, outer sep=-1pt}]
	\path (#1) edge[->-,bcolor] node[bcolor,midway] {\tiny$\phi$}  node[BFcolor,midway,swap] {\tiny#4} (#2);
	\end{scope}
}

\def\farrowBF(#1,#2,#3,#4){
	\begin{scope}[every node/.style={auto=#3, outer sep=-1pt}]
	\path (#1) edge[->-,fcolor] node[fcolor,midway] {\tiny$\psi$}  node[BFcolor,midway,swap] {\tiny#4} (#2);
	\end{scope}
}

\def\fQED(#1,#2){ 
\begin{tikzpicture}{baseline=(current bounding box).center}
	\node at (0,0) (g) [gauge,black] {$1$};
	\draw[CScolor] (g.south east) ++(-3pt,3pt) node[anchor=north west] {\tiny$#2$};
	\node at (1.5,0) (f) [flavor,black] {$#1$};
	\farrow(g,f,left)
\end{tikzpicture}
}

\def\sQED(#1,#2){ 
\begin{tikzpicture}{baseline=(current bounding box).center}
	\node at (0,0) (g) [gauge,black] {$1$};
	\draw[CScolor] (g.south east) ++(-3pt,3pt) node[anchor=north west] {\tiny$#2$};
	\node at (1.5,0) (f) [flavor,black] {$#1$};
	\barrow(g,f,left)
\end{tikzpicture}
}
\def\bQED(#1,#2){ \sQED(#1,#2) }

\def\KFplane(#1)(#2,#3){ 
\begin{scope}[scale=#2]
\ifthenelse{\equal{#1}{1}}{
    \fill[fill=#3] (3.5,0)--(6,0)--(6,2.5)--cycle;
}{
\ifthenelse{\equal{#1}{1t}}{
    \fill[fill=#3] (4,0)--(2,0)--(2,2)--cycle;
}{
\ifthenelse{\equal{#1}{2old}}{
    \fill[fill=#3] (4,0)--(6,2)--(6,6)--(4,4)--cycle;
}{
\ifthenelse{\equal{#1}{2t}}{
    \fill[fill=#3] (4,0)--(4,4)--(2,2)--cycle;
}{
\ifthenelse{\equal{#1}{3old}}{
    \fill[fill=#3] (4,4)--(6,6)--(2,6)--cycle;
}{
\ifthenelse{\equal{#1}{3t}}{
    \fill[fill=#3] (0,4)--(2,2)--(4,4)--(2,6)--(0,6)--cycle;
}{
\ifthenelse{\equal{#1}{2}}{
    \fill[fill=#3] (3.5,0)--(6,2.5)--(3,5.9)--(.3,3.2)--cycle;
}{
\ifthenelse{\equal{#1}{34}}{
    \fill[fill=#3] (4,4)--(6,6)--(2,6)--cycle;
     \fill[fill=#3] (0,4)--(2,2)--(4,4)--(2,6)--(0,6)--cycle;
}
\ifthenelse{\equal{#1}{3}}{
    \fill[fill=#3] (2,2)--(6,6)--(2,6)--cycle;
}{
\ifthenelse{\equal{#1}{4}}{
    \fill[fill=#3] (.3,3.2)--(3,5.9)--(0,6)--(0,3.5)--cycle;
}{
}}}}}}}}}

\draw[->,thick] (0,0) -- (6.5,0) node[right] {$\scriptstyle F$};
\draw[->,thick] (0,0) -- (0,6.5) node[above] {$\scriptstyle |k|$};

\node[left] at (0,4) {$\scriptstyle 2N+2$};
\node[below] at (4,0) {$\scriptstyle 2N+2$};

\draw[dashed,thick] (3.5,0) -- (0,3.5);
\draw[thick,black] (.3,3.2) -- (3,5.9);  
\draw[thick,black] (3.5,0) -- (6,2.5);

\ifthenelse{\equal{#1}{N4}}{
    \draw[#3,line width=3pt]  (3.5,0) -- (6,2.5);
}{
\ifthenelse{\equal{#1}{k0}}{
    \draw[#3,line width=3pt] (3.5,0) -- (6,0);
}{
\ifthenelse{\equal{#1}{mchi}}{
    \draw[#3,line width=3pt] (.3,3.2) -- (3,5.9);
}{
\ifthenelse{\equal{#1}{F0}}{
    \draw[#3,line width=3pt] (0,3.5) -- (0,5.5);
}{
\ifthenelse{\equal{#1}{NOSUSYbdyLeft}}{
    \draw[#3,line width=3pt] (2,0)--(2,2);
}{
\ifthenelse{\equal{#1}{NOSUSYbdyTop}}{
    \draw[#3,line width=3pt] (2,2)--(0,4);
}{
\ifthenelse{\equal{#1}{cusp}}{
    \draw[#3,fill=#3] (2,2) circle (.2cm);
}{
\ifthenelse{\equal{#1}{bdy1t2}}{
    \draw[#3,line width=3pt] (2,2)--(4,0);
}{
}}}}}}}}
\end{scope}
}

\tikzset{cross/.style={cross out, draw=black, minimum size=5*(#1-\pgflinewidth), inner sep=0pt, outer sep=0pt},
cross/.default={2pt}}

\tikzset{snake it/.style={decorate, decoration=snake}}

\tikzset{mid arrow/.style={postaction={decorate,decoration={
        markings,
        mark = at position .55 with {\arrow[#1]{Straight Barb[width=5pt]}}
      }}}}
      
\tikzset{mid arrowsm/.style={postaction={decorate,decoration={
        markings,
        mark = at position .55 with {\arrow[#1]{Straight Barb[width=3pt]}}
      }}}}
      
\tikzset{middx arrowsm/.style={postaction={decorate,decoration={
        markings,
        mark = at position .7 with {\arrow[#1]{Straight Barb[width=3pt]}}
      }}}}
      
\tikzset{midsx arrowsm/.style={postaction={decorate,decoration={
        markings,
        mark = at position .4 with {\arrow[#1]{Straight Barb[width=3pt]}}
      }}}}


%% file: References.bib
@article{Hanany:2005ve,
    author = "Hanany, Amihay and Kennaway, Kristian D.",
    title = "{Dimer models and toric diagrams}",
    eprint = "hep-th/0503149",
    archivePrefix = "arXiv",
    reportNumber = "MIT-CTP-3613",
    month = "3",
    year = "2005"
}

@article{Franco:2005rj,
    author = "Franco, Sebastian and Hanany, Amihay and Kennaway, Kristian D. and Vegh, David and Wecht, Brian",
    title = "{Brane dimers and quiver gauge theories}",
    eprint = "hep-th/0504110",
    archivePrefix = "arXiv",
    reportNumber = "MIT-CTP-3619",
    doi = "10.1088/1126-6708/2006/01/096",
    journal = "JHEP",
    volume = "01",
    pages = "096",
    year = "2006"
}

@article{Benvenuti:2005cz,
    author = "Benvenuti, Sergio and Kruczenski, Martin",
    title = "{Semiclassical strings in Sasaki-Einstein manifolds and long operators in $\mathcal{N}=1$ gauge theories}",
    eprint = "hep-th/0505046",
    archivePrefix = "arXiv",
    doi = "10.1088/1126-6708/2006/10/051",
    journal = "JHEP",
    volume = "10",
    pages = "051",
    year = "2006"
}

@article{Benvenuti:2005ja,
    author = "Benvenuti, Sergio and Kruczenski, Martin",
    title = "{From Sasaki-Einstein spaces to quivers via BPS geodesics: $L^{p,q|r}$}",
    eprint = "hep-th/0505206",
    archivePrefix = "arXiv",
    doi = "10.1088/1126-6708/2006/04/033",
    journal = "JHEP",
    volume = "04",
    pages = "033",
    year = "2006"
}

@article{Franco:2005sm,
    author = "Franco, Sebastian and Hanany, Amihay and Martelli, Dario and Sparks, James and Vegh, David and Wecht, Brian",
    title = "{Gauge theories from toric geometry and brane tilings}",
    eprint = "hep-th/0505211",
    archivePrefix = "arXiv",
    reportNumber = "MIT-CTP-3646, CERN-PH-TH-2005-084, HUTP-05-A0027",
    doi = "10.1088/1126-6708/2006/01/128",
    journal = "JHEP",
    volume = "01",
    pages = "128",
    year = "2006"
}

@article{Butti:2005sw,
    author = "Butti, Agostino and Forcella, Davide and Zaffaroni, Alberto",
    title = "{The Dual superconformal theory for $L^{p,q,r}$ manifolds}",
    eprint = "hep-th/0505220",
    archivePrefix = "arXiv",
    reportNumber = "BICOCCA-FT-05-11",
    doi = "10.1088/1126-6708/2005/09/018",
    journal = "JHEP",
    volume = "09",
    pages = "018",
    year = "2005"
}

@article{Benvenuti:2024seb,
    author = "Benvenuti, Sergio and Comi, Riccardo and Pasquetti, Sara and Pedde Ungureanu, Gabriel and Rota, Simone and Shri, Anant",
    title = "{Planar Abelian Mirror Duals of $\mathcal{N}=2$ SQCD$_3$}",
    journal = "Phys. Rev. D",
    doi = "10.1103/k5b3-lrnz",
    volume = "112",
    issue = "6",
    pages = "L101703",
    numpages = "6",
    year = "2025",
    eprint = "2411.05620",
    archivePrefix = "arXiv",
    primaryClass = "hep-th",
}

@article{Benvenuti:2025a,
     author = "Benvenuti, Sergio and Comi, Riccardo and Pasquetti, Sara and Pedde Ungureanu, Gabriel and Rota, Simone and Shri, Anant",
    title = "{A Chiral-Planar Dualization Algorithm for $3d$ $\mathcal{N}=2$ Chern-Simons-Matter Theories}",
    journal = "JHEP",
    year = "2025",
    doi = "10.1007/JHEP10(2025)211",
    volume = "10",
    pages = "211",
    eprint = "2505.02913 ",
    archivePrefix = "arXiv",
    primaryClass = "hep-th"
}

@article{Benvenuti:2026a,
    author = "Benvenuti, Sergio and Comi, Riccardo and Pedde Ungureanu, Gabriel and Rota, Simone and Shri, Anant",
    title = "{Universal Planar Abelian Duals for 3d $\mathcal{N}=2$ Unitary CS-SQCD}",
    eprint = "2603.08842",
    archivePrefix = "arXiv",
    primaryClass = "hep-th",
    month = "3",
    year = "2026"
}

@article{Hanany:1996ie,
    author = "Hanany, Amihay and Witten, Edward",
    title = "{Type IIB superstrings, BPS monopoles, and three-dimensional gauge dynamics}",
    eprint = "hep-th/9611230",
    archivePrefix = "arXiv",
    reportNumber = "IASSNS-HEP-96-121",
    doi = "10.1016/S0550-3213(97)00157-0",
    journal = "Nucl. Phys. B",
    volume = "492",
    pages = "152--190",
    year = "1997"
}

@article{Borokhov_2002,
   title={Monopole Operators and Mirror Symmetry in Three Dimensions},
   volume={2002},
   ISSN={1029-8479},
   url={http://dx.doi.org/10.1088/1126-6708/2002/12/044},
   DOI={10.1088/1126-6708/2002/12/044},
   number={12},
   journal={Journal of High Energy Physics},
   publisher={Springer Science and Business Media LLC},
   author={Borokhov, Vadim and Kapustin, Anton and Wu, Xinkai},
   year={2002},
   month=dec, pages={044–044} }

@article{Giveon:2008zn,
    author = "Giveon, Amit and Kutasov, David",
    title = "{Seiberg Duality in Chern-Simons Theory}",
    eprint = "0808.0360",
    archivePrefix = "arXiv",
    primaryClass = "hep-th",
    doi = "10.1016/j.nuclphysb.2008.09.045",
    journal = "Nucl. Phys. B",
    volume = "812",
    pages = "1--11",
    year = "2009"
}

@article{Imamura:2011su,
    author = "Imamura, Yosuke and Yokoyama, Shuichi",
    title = "{Index for three dimensional superconformal field theories with general R-charge assignments}",
    eprint = "1101.0557",
    archivePrefix = "arXiv",
    primaryClass = "hep-th",
    reportNumber = "UT-11-01, TIT-HEP-607",
    doi = "10.1007/JHEP04(2011)007",
    journal = "JHEP",
    volume = "04",
    pages = "007",
    year = "2011"
}

@article{Kapustin:2011jm,
    author = "Kapustin, Anton and Willett, Brian",
    title = "{Generalized Superconformal Index for Three Dimensional Field Theories}",
    eprint = "1106.2484",
    archivePrefix = "arXiv",
    primaryClass = "hep-th",
    reportNumber = "68-2840",
    month = "6",
    year = "2011"
}

@article{Aharony:1997bx,
    author = "Aharony, Ofer and Hanany, Amihay and Intriligator, Kenneth A. and Seiberg, N. and Strassler, M. J.",
    title = "{Aspects of $\mathcal{N}=2$ supersymmetric gauge theories in three-dimensions}",
    eprint = "hep-th/9703110",
    archivePrefix = "arXiv",
    reportNumber = "RU-97-10A, IASSNS-HEP-97-18",
    doi = "10.1016/S0550-3213(97)00323-4",
    journal = "Nucl. Phys. B",
    volume = "499",
    pages = "67--99",
    year = "1997"
}

@article{Kapustin:1998fa,
    author = "Kapustin, Anton",
    title = "{$D_n$ quivers from branes}",
    eprint = "hep-th/9806238",
    archivePrefix = "arXiv",
    reportNumber = "IASSNS-HEP-98-62",
    doi = "10.1088/1126-6708/1998/12/015",
    journal = "JHEP",
    volume = "12",
    pages = "015",
    year = "1998"
}

@phdthesis{vandebult,
title = "{Hyperbolic hypergeometric functions}",
author = {van de Bult, F.J.},
year = {2007},
month = {November},
address = {Amsterdam, Netherlands}, 
school = {Universiteit van Amsterdam},
note = {Available at \url{https://pure.uva.nl/ws/files/4364984/53985_vdbult_tekst.pdf}},
}

@article{Tong:2000ky,
    author = "Tong, David",
    title = "{Dynamics of $\mathcal{N}=2$ supersymmetric Chern-Simons theories}",
    eprint = "hep-th/0005186",
    archivePrefix = "arXiv",
    reportNumber = "MIT-CTP-2985, KCL-TH-00-28",
    doi = "10.1088/1126-6708/2000/07/019",
    journal = "JHEP",
    volume = "07",
    year = "2000"
}

@article{Gaiotto_2009,
   title="{$\mathcal{S}$-duality of boundary conditions in $\mathcal{N}=4$ Super Yang-Mills theory}",
   volume={13},
   ISSN={1095-0753},
   url={http://dx.doi.org/10.4310/ATMP.2009.v13.n3.a5},
   DOI={10.4310/atmp.2009.v13.n3.a5},
   number={3},
   journal={Advances in Theoretical and Mathematical Physics},
   publisher={International Press of Boston},
   author={Gaiotto, David and Witten, Edward},
   year={2009},
   pages={721–896} }

@article{Benini_2011a,
   title="{Comments on 3d Seiberg-like dualities}",
   volume={2011},
   ISSN={1029-8479},
   url={http://dx.doi.org/10.1007/JHEP10(2011)075},
   DOI={10.1007/jhep10(2011)075},
   number={10},
   journal={Journal of High Energy Physics},
   publisher={Springer Science and Business Media LLC},
   author={Benini, Francesco and Closset, Cyril and Cremonesi, Stefano},
   year={2011},
   month={10} }

@article{Benini_2011,
   title="{Quantum moduli space of Chern-Simons quivers, wrapped D6-branes and $AdS_4/CFT_3$}",
   volume={2011},
   ISSN={1029-8479},
   url={http://dx.doi.org/10.1007/JHEP09(2011)005},
   DOI={10.1007/jhep09(2011)005},
   number={9},
   journal={Journal of High Energy Physics},
   publisher={Springer Science and Business Media LLC},
   author={Benini, Francesco and Closset, Cyril and Cremonesi, Stefano},
   year={2011},
   month={9} }

@article{Kapustin:1999ha,
    author = "Kapustin, Anton and Strassler, Matthew J.",
    title = "{On mirror symmetry in three-dimensional Abelian gauge theories}",
    eprint = "hep-th/9902033",
    archivePrefix = "arXiv",
    reportNumber = "IASSNS-HEP-99-15",
    doi = "10.1088/1126-6708/1999/04/021",
    journal = "JHEP",
    volume = "04",
    year = "1999"
}

@article{Intriligator_2013,
   title="{Aspects of 3d $ \mathcal{N}=2 $ Chern-Simons-Matter theories}",
   volume={2013},
   ISSN={1029-8479},
   url={http://dx.doi.org/10.1007/JHEP07(2013)079},
   DOI={10.1007/jhep07(2013)079},
   number={7},
   journal={Journal of High Energy Physics},
   publisher={Springer Science and Business Media LLC},
   author={Intriligator, Kenneth and Seiberg, Nathan},
   year={2013},
   month={7} }

@article{Intriligator_1996,
   title={Mirror symmetry in three dimensional gauge theories},
   volume={387},
   ISSN={0370-2693},
   url={http://dx.doi.org/10.1016/0370-2693(96)01088-X},
   DOI={10.1016/0370-2693(96)01088-x},
   number={3},
   journal={Physics Letters B},
   publisher={Elsevier BV},
   author={Intriligator, K. and Seiberg, N.},
   year={1996},
   month={10}, pages={513–519} }

@article{Cremonesi_2014,
   title="{Monopole operators and Hilbert series of Coulomb branches of 3d $ \mathcal{N} $ = 4 gauge theories}",
   volume={2014},
   ISSN={1029-8479},
   url={http://dx.doi.org/10.1007/JHEP01(2014)005},
   DOI={10.1007/jhep01(2014)005},
   number={1},
   journal={Journal of High Energy Physics},
   publisher={Springer Science and Business Media LLC},
   author={Cremonesi, Stefano and Hanany, Amihay and Zaffaroni, Alberto},
   year={2014},
   month={1} }

@article{Hama_2011,
   title="{SUSY Gauge Theories on Squashed Three-Spheres}",
   volume={2011},
   ISSN={1029-8479},
   url={http://dx.doi.org/10.1007/JHEP05(2011)014},
   DOI={10.1007/jhep05(2011)014},
   number={5},
   journal={Journal of High Energy Physics},
   publisher={Springer Science and Business Media LLC},
   author={Hama, Naofumi and Hosomichi, Kazuo and Lee, Sungjay},
   year={2011},
   month={5} }

@article{Polyakov1977,
    author = "Polyakov, Alexander M.",
    title = "{Quark Confinement and Topology of Gauge Groups}",
    reportNumber = "NORDITA-76/33",
    doi = "10.1016/0550-3213(77)90086-4",
    journal = "Nucl. Phys. B",
    volume = "120",
    pages = "429--458",
    year = "1977"
}

@article{Comi:2022aqo,
    author = "Comi, Riccardo and Hwang, Chiung and Marino, Fabio and Pasquetti, Sara and Sacchi, Matteo",
    title = "{The SL(2, \ensuremath{\mathbb{Z}}) dualization algorithm at work}",
    eprint = "2212.10571",
    archivePrefix = "arXiv",
    primaryClass = "hep-th",
    reportNumber = "CTPU-PTC-22-28",
    doi = "10.1007/JHEP06(2023)119",
    journal = "JHEP",
    volume = "06",
    pages = "119",
    year = "2023"
}

@article{witten2003sl2zactionthreedimensionalconformal,
      author = "Witten, Edward",
    title = "{$SL(2,\mathbb{Z})$ action on three-dimensional conformal field theories with Abelian symmetry}",
    booktitle = "{From Fields to Strings: Circumnavigating Theoretical Physics: A Conference in Tribute to Ian Kogan}",
    eprint = "hep-th/0307041",
    archivePrefix = "arXiv",
    pages = "1173--1200",
    month = "7",
    year = "2003"
}

@article{Aharony:2013dha,
    author = "Aharony, Ofer and Razamat, Shlomo S. and Seiberg, Nathan and Willett, Brian",
    title = "{3d dualities from 4d dualities}",
    eprint = "1305.3924",
    archivePrefix = "arXiv",
    primaryClass = "hep-th",
    reportNumber = "WIS-04-13-APR-DPPA",
    doi = "10.1007/JHEP07(2013)149",
    journal = "JHEP",
    volume = "07",
    pages = "149",
    year = "2013"
}

@article{Kapustin:2009kz,
    author = "Kapustin, Anton and Willett, Brian and Yaakov, Itamar",
    title = "{Exact Results for Wilson Loops in Superconformal Chern-Simons Theories with Matter}",
    eprint = "0909.4559",
    archivePrefix = "arXiv",
    primaryClass = "hep-th",
    reportNumber = "CALT-68-2750",
    doi = "10.1007/JHEP03(2010)089",
    journal = "JHEP",
    volume = "03",
    pages = "089",
    year = "2010"
}

@article{Benvenuti:2024mpn,
    author = "Benvenuti, Sergio and Comi, Riccardo and Pasquetti, Sara",
    title = "{Star-triangle dualities and supersymmetric improved bifundamentals}",
    eprint = "2410.19049",
    archivePrefix = "arXiv",
    primaryClass = "hep-th",
    month = "10",
    year = "2024"
}

@article{Benvenuti:2025qnq,
    author = "Benvenuti, Sergio and Comi, Riccardo and Pasquetti, Sara and Pedde Ungureanu, Gabriel and Rota, Simone and Shri, Anant",
    title = "{Planar Abelian Duals of Chern-Simons QCD}",
    eprint = "hep-th/2506.05465",
    archivePrefix = "arXiv",
    primaryClass = "hep-th",
    month = "6",
    year = "2025"
}

@article{Porrati:1996xi,
    author = "Porrati, Massimo and Zaffaroni, Alberto",
    title = "{M-Theory origin of mirror symmetry in three-dimensional gauge theories}",
    eprint = "hep-th/9611201",
    archivePrefix = "arXiv",
    reportNumber = "NYU-TH-96-11-01, IASSNS-HEP-96-117",
    doi = "10.1016/S0550-3213(97)00061-8",
    journal = "Nucl. Phys. B",
    volume = "490",
    pages = "107--120",
    year = "1997"
}

@article{Hanany:1999sj,
    author = "Hanany, Amihay and Zaffaroni, Alberto",
    title = "{Issues on orientifolds: On the brane construction of gauge theories with SO(2n) global symmetry}",
    eprint = "hep-th/9903242",
    archivePrefix = "arXiv",
    reportNumber = "CERN-TH-99-82, MIT-CTP-2845",
    doi = "10.1088/1126-6708/1999/07/009",
    journal = "JHEP",
    volume = "07",
    pages = "009",
    year = "1999"
}

@article{Giveon:1998sr,
    author = "Giveon, Amit and Kutasov, David",
    editor = "Bais, F. A. and Bergshoeff, E. A. and de Wit, B. and Dijkgraaf, R. and Schellekens, A. N. and Verlinde, Erik P. and Verlinde, Herman L.",
    title = "{Brane Dynamics and Gauge Theory}",
    eprint = "hep-th/9802067",
    archivePrefix = "arXiv",
    reportNumber = "RI-2-98, EFI-98-06",
    doi = "10.1103/RevModPhys.71.983",
    journal = "Rev. Mod. Phys.",
    volume = "71",
    pages = "983--1084",
    year = "1999"
}

@article{Feng:2000eq,
    author = "Feng, Bo and Hanany, Amihay",
    title = "{Mirror symmetry by O3 planes}",
    eprint = "hep-th/0004092",
    archivePrefix = "arXiv",
    reportNumber = "MIT-CTP-2959, NSF-ITP-00-20",
    doi = "10.1088/1126-6708/2000/11/033",
    journal = "JHEP",
    volume = "11",
    pages = "033",
    year = "2000"
}

@article{Hori:1997zj,
    author = "Hori, Kentaro and Ooguri, Hirosi and Vafa, Cumrun",
    title = "{Non-Abelian conifold transitions and N=4 dualities in three-dimensions}",
    eprint = "hep-th/9705220",
    archivePrefix = "arXiv",
    reportNumber = "HUTP-97-A024, LBL-40349, LBNL-40349, UCB-PTH-97-27",
    doi = "10.1016/S0550-3213(97)00529-4",
    journal = "Nucl. Phys. B",
    volume = "504",
    pages = "147--174",
    year = "1997"
}

@article{Comi:wip,
    author = "Comi, Riccardo and Garavaglia, Sebastiano and Giacomelli, Simone and Pasquetti, S.",
    title = "{Algorithmic Dualization of USp
    Good and Bad Quivers}",
    journal = "in progress",
    year = "2026"
}

@article{Nawata:2021nse,
    author = "Nawata, Satoshi and Sperling, Marcus and Wang, Hao Ellery and Zhong, Zhenghao",
    title = "{Magnetic quivers and line defects {\textemdash} On a duality between 3d $ \mathcal{N} $ = 4 unitary and orthosymplectic quivers}",
    eprint = "2111.02831",
    archivePrefix = "arXiv",
    primaryClass = "hep-th",
    doi = "10.1007/JHEP02(2022)174",
    journal = "JHEP",
    volume = "02",
    pages = "174",
    year = "2022"
}

@article{Assel:2018exy,
    author = "Assel, Benjamin and Cremonesi, Stefano",
    title = "{The Infrared Fixed Points of 3d $\mathcal{N}=4$ $USp(2N)$ SQCD Theories}",
    eprint = "1802.04285",
    archivePrefix = "arXiv",
    primaryClass = "hep-th",
    reportNumber = "CERN-TH-2018-031",
    doi = "10.21468/SciPostPhys.5.2.015",
    journal = "SciPost Phys.",
    volume = "5",
    number = "2",
    pages = "015",
    year = "2018"
}

@article{Nawata:2023rdx,
    author = "Nawata, Satoshi and Sperling, Marcus and Wang, Hao Ellery and Zhong, Zhenghao",
    title = "{3d $\mathcal{N}=4$ mirror symmetry with 1-form symmetry}",
    eprint = "2301.02409",
    archivePrefix = "arXiv",
    primaryClass = "hep-th",
    doi = "10.21468/SciPostPhys.15.1.033",
    journal = "SciPost Phys.",
    volume = "15",
    number = "1",
    pages = "033",
    year = "2023"
}

@article{Dey:2014tka,
    author = "Dey, Anindya and Hanany, Amihay and Koroteev, Peter and Mekareeya, Noppadol",
    title = "{Mirror Symmetry in Three Dimensions via Gauged Linear Quivers}",
    eprint = "1402.0016",
    archivePrefix = "arXiv",
    primaryClass = "hep-th",
    reportNumber = "UTTG-36-13, TCC-030-13, CERN-PH-TH-2013-279",
    doi = "10.1007/JHEP06(2014)059",
    journal = "JHEP",
    volume = "06",
    pages = "059",
    year = "2014"
}

@article{Dey:2014dwa,
    author = "Dey, Anindya",
    editor = "Donagi, Ron and Douglas, Michael R. and Kamenova, Ljudmila and Rocek, Martin",
    title = "{Mirror Symmetry in Flavored Affine D-type Quivers}",
    eprint = "1401.6462",
    archivePrefix = "arXiv",
    primaryClass = "hep-th",
    doi = "10.1090/pspum/088/01466",
    journal = "Proc. Symp. Pure Math.",
    volume = "88",
    pages = "259--270",
    year = "2014"
}

@article{Ferlito:2016grh,
    author = "Ferlito, Giulia and Hanany, Amihay",
    title = "{A tale of two cones: the Higgs Branch of Sp(n) theories with 2n flavours}",
    eprint = "1609.06724",
    archivePrefix = "arXiv",
    primaryClass = "hep-th",
    month = "9",
    year = "2016"
}

@article{Goddard:1976qe,
    author = "Goddard, P. and Nuyts, J. and Olive, David I.",
    title = "{Gauge Theories and Magnetic Charge}",
    reportNumber = "CERN-TH-2255",
    doi = "10.1016/0550-3213(77)90221-8",
    journal = "Nucl. Phys. B",
    volume = "125",
    pages = "1--28",
    year = "1977"
}

@book{Tachikawa:2013kta,
    author = "Tachikawa, Yuji",
    title = "{$\mathcal N=2$ supersymmetric dynamics for pedestrians}",
    eprint = "1312.2684",
    archivePrefix = "arXiv",
    primaryClass = "hep-th",
    reportNumber = "UT-13-42, IPMU-13-0234, UT-13-42, IPMU-13-0234",
    doi = "10.1007/978-3-319-08822-8",
    volume = "890",
    month = "12",
    year = "2013"
}

@article{Cabrera:2018ann,
    author = "Cabrera, Santiago and Hanany, Amihay",
    title = "{Quiver Subtractions}",
    eprint = "1803.11205",
    archivePrefix = "arXiv",
    primaryClass = "hep-th",
    doi = "10.1007/JHEP09(2018)008",
    journal = "JHEP",
    volume = "09",
    pages = "008",
    year = "2018"
}

@article{Bennett:2026gtm,
    author = "Bennett, Sam and Hanany, Amihay and Kumaran, Guhesh",
    title = "{Quotient Quiver Subtraction -- Classical Groups}",
    eprint = "2603.08774",
    archivePrefix = "arXiv",
    primaryClass = "hep-th",
    reportNumber = "Imperial/TP/26/AH/02",
    month = "3",
    year = "2026"
}

@article{Bennett:2025edk,
    author = "Bennett, Sam and Hanany, Amihay and Kumaran, Guhesh",
    title = "{Orthosymplectic quotient quiver subtraction. Part II. Framed quivers}",
    eprint = "2503.19954",
    archivePrefix = "arXiv",
    primaryClass = "hep-th",
    reportNumber = "Imperial/TP/25/AH/03",
    doi = "10.1007/JHEP12(2025)046",
    journal = "JHEP",
    volume = "12",
    pages = "046",
    year = "2025"
}

@article{Hanany:2023tvn,
    author = "Hanany, Amihay and Kalveks, Rudolph and Kumaran, Guhesh",
    title = "{Quotient quiver subtraction}",
    eprint = "2308.05853",
    archivePrefix = "arXiv",
    primaryClass = "hep-th",
    doi = "10.1016/j.nuclphysb.2024.116731",
    journal = "Nucl. Phys. B",
    volume = "1009",
    pages = "116731",
    year = "2024"
}

@article{Bennett:2024llh,
    author = "Bennett, Sam and Hanany, Amihay and Kumaran, Guhesh",
    title = "{Orthosymplectic quotient quiver subtraction}",
    eprint = "2409.15419",
    archivePrefix = "arXiv",
    primaryClass = "hep-th",
    reportNumber = "Imperial/TP/24/AH/02, Imperial/TP/24/AH/02",
    doi = "10.1007/JHEP12(2024)063",
    journal = "JHEP",
    volume = "12",
    pages = "063",
    year = "2024"
}

@article{Harding:2025vov,
    author = "Harding, William and Mekareeya, Noppadol and Zhong, Zhenghao",
    title = "{Orthosymplectic quivers: indices, Hilbert series, and generalised symmetries}",
    eprint = "2505.03875",
    archivePrefix = "arXiv",
    primaryClass = "hep-th",
    doi = "10.1007/JHEP09(2025)212",
    journal = "JHEP",
    volume = "09",
    pages = "212",
    year = "2025"
}

@article{Marino:2025ihk,
    author = {Marino, Fabio and Moura Soys{\"u}ren, Sinan and Sperling, Marcus},
    title = "{Orthosymplectic Chern-Simons Matter Theories: Global Forms, Dualities, and Vacua}",
    eprint = "2509.11733",
    archivePrefix = "arXiv",
    primaryClass = "hep-th",
    month = "9",
    year = "2025"
}
